\documentclass[12pt,a4wide]{article}
\usepackage{afterpage}
\usepackage{epsfig}
\usepackage[a4paper]{geometry}
\usepackage{float}
\usepackage[stable]{footmisc}
\usepackage[utf8x]{inputenc}
\usepackage{a4wide}
\usepackage{placeins}
\usepackage{amsmath,amssymb,amstext,amsthm,amssymb}
\usepackage{amsfonts}
\usepackage{mathrsfs}
\usepackage{framed}
\usepackage{graphicx}
\usepackage{bbold}
\usepackage{seqsplit}
\usepackage{slashed} 
\usepackage{tensor} 
\usepackage{braket} 
\usepackage{array}
\usepackage{graphicx}
\usepackage{enumerate}
\usepackage{setspace}
\usepackage{hyperref}
\usepackage{overpic}
\usepackage{bbm}
\usepackage{cite}
\usepackage{color}
\usepackage{lipsum}
\usepackage{bm}
\usepackage{tikz}

\DeclareSymbolFontAlphabet{\mathbb}{AMSb}

\newcommand{\del}[0]{\partial}

\let\baraccent=\=
\renewcommand{\=}[1]{\stackrel{#1}{=}}
\newcommand{\id}[0]{\mathbb{I}}

\newcommand{\eps}[0]{\varepsilon}

\begin{document}

\pagestyle{plain}

\makeatletter
\@addtoreset{equation}{section}
\makeatother
\renewcommand{\theequation}{\thesection.\arabic{equation}}
\pagestyle{empty}

\vspace{0.5cm}

\begin{center}
{\LARGE \bf{Small Cosmological Constants}}\\
\vskip 9pt
{\LARGE \bf{in String Theory}}\\[15mm]
\end{center}

\begin{center}
\scalebox{0.95}[0.95]{{\fontsize{14}{30}\selectfont Mehmet Demirtas,$^{a,b}$ Manki Kim,$^{a,c}$ Liam McAllister,$^{a}$}} \vspace{0.35cm}
\scalebox{0.95}[0.95]{{\fontsize{14}{30}\selectfont Jakob Moritz,$^{a}$ and Andres Rios-Tascon$^{a}$}}
\end{center}

\begin{center}
\vspace{0.25 cm}
\textsl{$^{a}$Department of Physics, Cornell University, Ithaca, NY 14853, USA}\\
\textsl{$^{b}$Department of Physics, Northeastern University, Boston, MA 02115, USA}\\
\textsl{$^{c}$Center for Theoretical Physics, MIT,
Cambridge, MA 02139, USA}\\

	 \vspace{1cm}
	\normalsize{\bf Abstract} \\[8mm]
\end{center}
\begin{center}
	\begin{minipage}[h]{15.0cm}
	
We construct supersymmetric  $\mathrm{AdS}_4$ vacua of type IIB string theory in
compactifications on orientifolds of Calabi-Yau threefold hypersurfaces.
We first find explicit orientifolds and quantized fluxes for which
the superpotential takes the form proposed by Kachru, Kallosh, Linde, and Trivedi.
Given very mild assumptions on the numerical values of the Pfaffians, these compactifications admit vacua in which all moduli are stabilized at weak string coupling.
By computing high-degree Gopakumar-Vafa invariants we give strong evidence that the $\alpha'$ expansion is likewise well-controlled.
We find extremely small cosmological constants, with magnitude $ < 10^{-123}$ in Planck units.
The compactifications are large, but not exponentially so, and hence these vacua manifest hierarchical scale-separation, with the AdS length exceeding the Kaluza-Klein length by a factor of a googol.

	\end{minipage}
\end{center}
\newpage
\setcounter{page}{1}
\pagestyle{plain}
\renewcommand{\thefootnote}{\arabic{footnote}}
\setcounter{footnote}{0}
%
%
\setcounter{tocdepth}{2}
\tableofcontents
\newpage
\section{Introduction}

To understand the effects of the quantization of gravity in cosmology, one can search for cosmological solutions of string theory.
A first step is to characterize isolated vacua in well-controlled settings, such as the four-dimensional $\mathcal{N}=1$
supergravity theories that arise in compactifications of type IIB string theory on Calabi-Yau orientifolds.

Our goal in this work is to find supersymmetric $\mathrm{AdS}_4$ vacua of the type proposed by Kachru, Kallosh, Linde, and Trivedi (KKLT) \cite{Kachru:2003aw}.
Three conditions are necessary for such vacua to exist.  First, the expectation value of the classical flux superpotential
must be exponentially small.  Second, the nonperturbative superpotential must contain
at least $h^{1,1}$ independent terms for the $h^{1,1}$ K\"ahler moduli.  Third, there must exist a point inside the K\"ahler cone at which the F-terms for all the K\"ahler moduli vanish, and the $\alpha'$ and $g_s$ expansions are well-controlled.  An important open problem is to determine how widely these requirements are
fulfilled in ensembles of flux compactifications on Calabi-Yau orientifolds.

To compute the nonperturbative superpotential in such a compactification, one needs to identify the seven-brane gauge groups that generate gaugino condensates, and also find all the rigid divisors that support Euclidean D3-brane superpotential terms.
The leading superpotential terms then take the form
\begin{equation}\label{kkltw}
  W = W_{\mathrm{flux}}(z,\tau) + \sum_D \mathcal{A}_D(z,\tau)\,\mathrm{exp}\Bigl(-\tfrac{2\pi}{c_D} T_D\Bigr)\,.
\end{equation}
Here $W_{\mathrm{flux}}(z,\tau)$ is the classical Gukov-Vafa-Witten flux superpotential \cite{Gukov:1999ya}, which depends on the complex structure moduli $z$ and the axiodilaton $\tau$.  The sum runs over nonperturbative contributions supported on divisors $D$ with complexified volumes $T_D$, either from Euclidean D3-branes when $D$ is suitably rigid, or from strong gauge dynamics on a stack of seven-branes wrapping $D$.  In the former case $c_D=1$, while in the latter $c_D$ is the dual Coxeter number of the gauge theory.  The Pfaffian prefactors $\mathcal{A}_D(z,\tau)$ in general depend on the complex structure moduli and the axiodilaton.

Computing the dependence of the Pfaffians $\mathcal{A}_D(z,\tau)$  on the moduli remains challenging, but for divisors
whose uplifts to F-theory have trivial intermediate Jacobian,
the $\mathcal{A}_D$ are sections of the trivial bundle over moduli space \cite{Witten:1996hc,Blumenhagen:2010ja}.
This constancy with respect to the moduli simplifies the study of moduli stabilization (cf.~e.g.~\cite{Blumenhagen:2010ja}), and for this reason we will ensure that
every Pfaffian occurring in our studies is constant.

Recent advances have made it possible to find quantized fluxes for which\footnote{To avoid writing $|W_0|$ throughout, we have defined $W_0$ to be \emph{positive}, and we write instead $\langle W_{\mathrm{flux}} \rangle$ for the rare cases where phase information is relevant.} $W_0 := \langle |W_{\mathrm{flux}}| \rangle \ll 1$ \cite{Demirtas:2019sip,Alvarez-Garcia:2020pxd,Demirtas:2020ffz,Honma:2021klo,Marchesano:2021gyv}.
However, the problem of finding such fluxes is Diophantine in character, and the computation becomes extremely expensive for $h^{2,1} \gg 1$.  Prior to the present work, examples had been found only for $h^{2,1}=2$ and $3$, but one learns from the Kreuzer-Skarke list \cite{Kreuzer:2000xy} that
the \emph{smallest} value of $h^{1,1}$ for a Calabi-Yau threefold hypersurface with $h^{2,1} \leq 3$ is 21.\footnote{We
have found new solutions with $h^{2,1}$ as large as $7$, but $h^{1,1}$ remains large: in the examples detailed in \S\ref{sec:examples}, $h^{1,1} \ge 51$.}

Thus, in the cases where one can find fluxes yielding exponentially small values of the flux superpotential, the K\"ahler moduli space is high-dimensional.  As a result, in seeking supersymmetric $\mathrm{AdS}_4$ vacua in such geometries one encounters certain challenges.  First, one needs to construct explicit orientifolds at $h^{1,1} \gg 1$.
Second, one needs to count fermion zero-modes on Euclidean D3-branes in such orientifolds, and find cases in which there are enough nonperturbative superpotential terms.
Third, one needs to actually find supersymmetric vacua in this high-dimensional moduli space, consisting of exponentially many chambers of the K\"ahler cone.  Finally, establishing control of the $\alpha'$ expansion in such vacua involves computing Gopakumar-Vafa invariants of curves at large $h^{1,1}$.

In this paper we overcome these obstacles.  We exhibit compactifications in which the superpotential takes the form \eqref{kkltw}, containing at least $h^{1,1}$ independent nonperturbative terms, all with constant Pfaffians, and with $W_0$ as small as $10^{-95}$.  The examples are explicit orientifolds of Calabi-Yau threefold hypersurfaces with $4 \le h^{2,1} \le 7$ and $51 \le h^{1,1} \le 214$, in which all tadpoles are cancelled.
We show that with very mild assumptions about the numerical values of the Pfaffians, these compactifications admit supersymmetric $\mathrm{AdS}_4$ vacua.  All closed string moduli are explicitly stabilized, near weak string coupling, large complex structure, and large Einstein-frame volumes.  All seven-branes occur in $\mathfrak{so}(8)$ stacks, and we argue that the seven-brane moduli are therefore automatically stabilized in the presence of three-form fluxes. By computing the genus-zero Gopakumar-Vafa invariants to high degree, we give strong evidence that the leading worldsheet instanton corrections to the K\"ahler potential are well-controlled.

Because our constructions
unite a number of nontrivial components, the critical reader may wonder which components are most likely to `fail', i.e.~which are the least understood, or the most vulnerable to higher-order corrections of some form.  To address this question, we briefly summarize the status of our examples.
The orientifolds, and the classical flux vacua with $W_0 \ll 1$, are extremely well-controlled.  In particular, we have computed the type IIA worldsheet instanton corrections to the prepotential up to curves of degree much higher than those that generate the racetrack of \cite{Demirtas:2019sip}, and have verified that the omitted terms are indeed negligible.
The D7-brane gauge sectors are all $\mathfrak{so}(8)$ stacks with well-understood low-energy dynamics.
The Euclidean D3-branes in our examples are straightforward: they wrap prime toric divisors $D$ that are rigid, i.e.~with $h^{\bullet}(\mathcal{O}_D)=(1,0,0)$,
and intersect the O7-planes transversely, so that the counting of fermion zero-modes is standard, see e.g.~\cite{Blumenhagen:2009qh}.
Moreover, these divisors uplift to divisors $\widehat{D}$ in fourfolds with $h^{2,1}(\widehat{D})=0$, so that
the M5-brane partition function is a section of the trivial bundle, i.e.~a pure constant, with no dependence on the complex structure moduli \cite{Witten:1996hc}.
In sum, the superpotential is very well-characterized: we have shown explicitly that it takes the form proposed in \cite{Kachru:2003aw},
and the only presently-unknown parameters in the leading superpotential data are the constant prefactors $\mathcal{A}_D$ of the nonperturbative terms, which we term Pfaffian numbers.

We lack a theory of the Pfaffian numbers, but will show that as long as they are not exponentially large or small, our compactifications admit supersymmetric $\mathrm{AdS}_4$ vacua.  Setting all the $\mathcal{A}_D\rightarrow 1$ leads to a relative error in the expectation values of the K\"ahler moduli that is of order $\log(\mathcal{A}_D)/\log(W_0)$, and so for sufficiently small $W_0$, as in our examples, the numerical values of the Pfaffians become irrelevant.

Crucially, because $W_0$ is exponentially small, the string coupling $g_s$ is stabilized at very weak coupling.  Perturbative corrections in $g_s$, and the effects of Euclidean D(-1)-branes, can then be neglected.
We further argue that because of the smallness of $g_s$,
and correspondingly the weakness of the $\mathcal{N}=2 \rightarrow \mathcal{N}=1$ breaking effects of fluxes and D-branes,
the leading corrections in the $\alpha'$ expansion are \emph{not} perturbative\footnote{Except for the famous term at order $\alpha'^3$ \cite{Becker:2002nn}, which we show is negligible.} corrections,
but are instead $\mathcal{N}=2$ corrections from worldsheet instantons wrapping curves.

In sum, of all the possible corrections to the vacuum structure that one obtains from the superpotential \eqref{kkltw} and the leading-order K\"ahler potential, we find that the most significant ones are the contributions of worldsheet instantons to the K\"ahler potential.  Evaluating such worldsheet instantons in a Calabi-Yau threefold is conceptually straightforward: one need only compute the genus-zero Gopakumar-Vafa invariants of curves, for example by means of mirror symmetry \cite{Hosono:1994ax}.  In practice, however, systematic computation of Gopakumar-Vafa invariants in compact Calabi-Yau threefolds with many K\"ahler moduli has not yet been achieved, to the best of our knowledge: except in special cases, threefolds with $h^{1,1} \gg 10$ have remained inaccessible.\footnote{See for example the recent work \cite{Carta:2021sms}.} Yet in our ensemble of vacua, $h^{1,1}$ is no smaller than 51, and in fact $h^{1,1} \sim \mathcal{O}(100)$ in many examples.  In order to ensure convergence of the $\alpha'$ expansion in our solutions, we apply improved methods --- to appear in \cite{compmirror} --- for computing genus-zero Gopakumar-Vafa invariants in compact Calabi-Yau threefold hypersurfaces.
We compute these invariants systematically, to rather high degree, and we apply specialized techniques to identify and study the smallest curves that are not collapsible.  With the aid of these new computational tools, we establish control of the worldsheet instanton corrections, and \emph{a fortiori} of the (largely unknown) perturbative-in-$\alpha'$ corrections that are suppressed by one or more additional powers of $g_s \ll 1$.

The construction of flux vacua employed here has been shown \cite{Alvarez-Garcia:2020pxd,Demirtas:2020ffz} to be compatible with the existence of near-conifold regions, including Klebanov-Strassler throat regions \cite{Klebanov:2000hb} that could plausibly host supersymmetry-breaking anti-D3-branes \cite{Kachru:2002gs}.
However, establishing the validity of the supergravity approximation in such regions, for the K\"ahler moduli expectation values obtained in our vacua, will require separate treatment.  Moreover, introducing supersymmetry breaking leads to a further host of issues.  The search for de Sitter vacua based on our solutions is therefore left as a task for the future.

The pioneering works \cite{Denef:2004dm,Denef:2005mm,Lust:2006zg} already presented evidence for the existence of supersymmetric $\mathrm{AdS}_4$ vacua of KKLT type, so we should explain what has been gained in our approach.
First of all, in \cite{Denef:2004dm,Denef:2005mm,Lust:2006zg} the methods for finding flux vacua with $W_0 \ll 1$ were less powerful, and the smallest value obtained was $\mathcal{O}(10^{-2})$, whereas we have found examples with $W_0$ as small as\footnote{ See \S\ref{sec:discussion} for a comparison of our findings to the statistical predictions of \cite{Denef:2004ze}.} $10^{-95}$.
Second, the constructions of \cite{Denef:2004dm,Denef:2005mm,Lust:2006zg} relied on special structures: a key example in  \cite{Denef:2004dm} stabilized the complex structure moduli on the locus invariant under the Greene-Plesser symmetry \cite{Greene:1990ud}, which presents certain subleties; a very high degree of symmetry among the various divisors is crucial in \cite{Denef:2005mm}; and the approach of \cite{Lust:2006zg} is restricted to a class of resolved orbifolds that generalize \cite{Denef:2005mm}.
Finally, and for us most significantly, the constructions of \cite{Denef:2004dm,Denef:2005mm,Lust:2006zg} required considerable insights into the detailed properties of a few examples.  Our approach, building on the software package {\tt{CYTools}} \cite{CYTools}, amounts to a \emph{general method} that can be applied to the entire Kreuzer-Skarke database, and in principle generate vast numbers of vacua.  In this work we have presented only an initial harvest at $h^{2,1} \le 7$, but extending our findings to larger $h^{2,1}$ is a purely computational task.

\subsection{Plan of the paper}\label{roadmap}

The organization of this paper is as follows.  In \S\ref{sec:setup} we set our notation, explain how we construct orientifolds, and review
how we select quantized fluxes that yield small $W_0$, following \cite{Demirtas:2019sip}.  We find classical solutions in which the F-terms of the complex structure moduli and axiodilaton vanish, and these moduli are stabilized at weak string coupling and large complex structure.  At this level the K\"ahler moduli remain unstabilized.
Then, in \S\ref{ss:ed3}, we explain how we identify orientifolds in which there are at least $h^{1,1}$ nonperturbative superpotential terms from Euclidean D3-branes or strong gauge dynamics on rigid prime toric divisors.  We further detail how we select geometries in which the Pfaffian prefactor of each such term is a constant.

At this point we have proved that the effective superpotential for the K\"ahler moduli in our ensemble of compactifications takes the form\footnote{The ellipsis in \eqref{kkltwsimple} denotes subleading corrections: from Euclidean D(-1)-brane contributions to the flux superpotential,
and from further nonperturbative contributions to the superpotential for the K\"ahler moduli, resulting in particular from Euclidean D3-branes on autochthonous divisors.
These corrections are shown to be negligible in \S\ref{ss:fluxvacua} and \S\ref{sec:autochthonous}, respectively.}
\begin{equation}\label{kkltwsimple}
  W = W_0 + \sum_{D_I} \mathcal{A}_{D_I}\,\mathrm{exp}\Bigl(-\tfrac{2\pi}{c_{D_I}} T_{D_I}\Bigr)+\ldots\,,
\end{equation}
with $W_0 \ll 1$.
Here the $D_I$ are the $h^{1,1}+4$
prime toric divisors of the Calabi-Yau threefold hypersurface, and the $\mathcal{A}_{D_I}$ are \emph{constants}, at least $h^{1,1}$ of which are nonzero, according to standard counting of fermion zero-modes.  We parameterize the K\"ahler moduli by the complexified volumes
$T_i$ of a basis $\{D_i\}$ of $h^{1,1}$ prime toric divisors for which $\mathcal{A}_{D_i} \neq 0$.

In order to find supersymmetric vacua, we must then find points in the K\"ahler moduli space at which the F-terms for the K\"ahler moduli vanish.
Such points take the form\footnote{Throughout this paper, $\mathrm{log}$ denotes the natural logarithm.}
\begin{equation}\label{schema}
\mathrm{Re}(T_{i}) \approx \frac{c_i}{2\pi}\,\mathrm{log}(W_0^{-1})+\ldots\,,
\end{equation}
where the ellipsis denotes corrections that will be computed in \S\ref{sec:control}.
The Einstein-frame volumes of the basis divisors are then large, because $W_0$ is exponentially small.

Demonstrating that one or more points obeying (the appropriately corrected form of) \eqref{schema} are in fact inside the K\"ahler moduli space is the subject of \S\ref{sec:control} and \S\ref{sec:comp}.  First, in \S\ref{sec:control}, we examine perturbative and nonperturbative corrections to the K\"ahler potential for the K\"ahler moduli,
and argue that because of the smallness of $g_s$ in our vacua, the leading effects occur at string tree level, and result from worldsheet instantons wrapping small curves.
Then, in \S\ref{ss:fflat} we give an algorithm for finding vacua at large $h^{1,1}$.
In order to explicitly include the aforementioned worldsheet instanton effects, and more generally to ensure control of the $\alpha'$ expansion, we compute the Gopakumar-Vafa invariants of a vast set of curves in our examples (\S\ref{ss:GV}), and then determine the
radius of convergence (\S\ref{ss:convergence}).

In \S\ref{sec:examples} we  give the details of a few examples that result from applying this procedure to the Kreuzer-Skarke list, for $h^{2,1} \le 7$. We discuss the implications of our findings in \S\ref{sec:discussion}, and we conclude, traditionally, in \S\ref{sec:conclusions}.  Appendix \ref{app:ds} contains comments on the prospects for an uplift to de Sitter space.  A brief summary and discussion of our results appears in the companion paper \cite{shortpaper}.

\section{Classical flux vacua}\label{sec:setup}

In this section we set our notation and terminology, and then show how we find orientifolds and classical flux vacua.

\subsection{Setup}\label{ss:class}

Let $X$ be a Calabi-Yau threefold, and denote by $\widetilde{X}$ its mirror threefold. Let $\{\omega^i\}_{i=1}^{h^{1,1}(X)}$ be a basis of $H^4(X,\mathbb{Z})$, and let $\{\omega_i\}_{i=1}^{h^{1,1}(X)}$ be a dual basis of $H^2(X,\mathbb{Z})$, with $\int_X\omega^i\wedge  \omega_j={\delta^i}_j$.
We adopt a notation where a $p$-form class and its Poincar\'e-dual cycle class are denoted by the same symbol, to be understood from the context.

Let $J$ be the string-frame K\"ahler class of $X$, taking values in the K\"ahler cone $\mathcal{K}_X\subset H^{1,1}(X,\mathbb{R})$.
The Mori cone $\mathcal{M}(X) \subset H_{2}(X,\mathbb{R})$ is the cone dual to $\mathcal{K}_X$.
We may expand
\begin{equation}
J=\sum_i t^i\,\omega_i\,
\end{equation}
in terms of K\"ahler parameters $\{t^i\}_{i=1}^{h^{1,1}(X)}$.

Let $\{\alpha^A,\beta_A\}_{A=0}^{h^{2,1}(X)}$ be a symplectic basis of the middle cohomology $H^3(X,\mathbb{Z})$, with $\int_X \alpha^A\wedge \beta_B={\delta^A}_B$, and let $\Omega$ be the holomorphic three-form of $X$. Then, it is useful to represent $\Omega$ by a \textit{period vector}
\begin{equation}
\vec{\Pi}:=\begin{pmatrix}
\int_{X}\Omega\wedge \beta_A \\
\int_{X}\Omega\wedge \alpha^A
\end{pmatrix}=\begin{pmatrix}
\mathcal{F}_A\\
z^A
\end{pmatrix}\, ,
\end{equation}
and more generally to represent closed three-forms via $(2h^{2,1}(X)+2)$-dimensional vectors.
Furthermore,
we introduce a symplectic pairing $\Sigma:=\begin{pmatrix}
0 & \id\\
-\id & 0
\end{pmatrix}$.

Locally, in a suitable patch, the periods $z^A$ serve as \textit{homogeneous} complex coordinates on the complex structure moduli space of $X$, and away from the locus $z^0=0$ we may normalize $\Omega$ such that $z^0=1$. Henceforth,
we do so and let $a=1,...,h^{2,1}(X)$. The dual periods $\mathcal{F}_a$ are determined in terms of the $z^a$ by the \textit{prepotential} $\mathcal{F}(z)$ via  $\mathcal{F}_a(z)=\del_{z^a} \mathcal{F}(z)$, and
$\mathcal{F}_0 =2\mathcal{F}-z^a\del_{z^a}\mathcal{F}$. In this paper we will restrict ourselves to the large complex structure (LCS) patch, where
\begin{equation}\label{eq:prepotential}
\mathcal{F}(z)=\mathcal{F}_{\text{poly}}(z)+\mathcal{F}_{\text{inst}}(z)\,,
\end{equation}
with
\begin{equation}
\mathcal{F}_{\text{poly}}(z)=-\frac{1}{3!}\widetilde{\kappa}_{abc}z^az^bz^c+\frac{1}{2}\tilde{a}_{ab}z^az^b+\frac{1}{24}\tilde{c}_a z^a+\frac{\zeta(3)\chi(\widetilde{X})}{2(2\pi i)^3}\, .
\end{equation}
Here $\widetilde{\kappa}_{abc}$ are the triple intersection numbers of the mirror threefold $\widetilde{X}$, and
\begin{equation}
\tilde{c}_a=\int_{\widetilde{X}}c_2(\widetilde{X})\wedge \tilde{\beta}_a\, ,\quad
\tilde{a}_{ab}\equiv \frac{1}{2}\begin{cases}
\widetilde{\kappa}_{aab} & a\geq b\\
\widetilde{\kappa}_{abb} & a<b
\end{cases}\, , \quad \text{and} \quad \chi(\widetilde{X})=\int_{\widetilde{X}} c_3(\widetilde{X})\, ,
\end{equation}
where $\{\tilde{\beta}_a\}_{a=1}^{h^{2,1}(X)}$ is a basis of $H^2(\widetilde{X},\mathbb{Z})$ mirror dual to the set of three-forms $\beta_a\in H^3(X,\mathbb{Z})$, and $c_2(\widetilde{X})$ and $c_3(\widetilde{X})$ are the second and third Chern classes, respectively, of $\widetilde{X}$.  The type IIA worldsheet instanton corrections are given by
\begin{equation}\label{eq:finst}
\mathcal{F}_{\text{inst}}(z)=-\frac{1}{(2\pi i)^3}\sum_{\tilde{\mathbf{q}}\in \mathcal{M}(\widetilde{X})}\mathscr{N}_{\tilde{\mathbf{q}}}\,\text{Li}_3\Bigl(e^{2\pi i\,\tilde{\mathbf{q}}\cdot \mathbf{z}}\Bigr)\, ,
\end{equation}
where $\text{Li}_k(q):=\sum_{n=1}^\infty q^n/n^k$ is the polylogarithm, the $\tilde{\mathbf{q}}$ represent effective curve classes in $H^4(\widetilde{X},\mathbb{Z})\simeq H_2(\widetilde{X},\mathbb{Z})$ expressed in a basis $\left\{\tilde{\alpha}^a\right\}_{a=1}^{h^{2,1}(X)}$ mirror dual to the set of three-forms $\alpha^a\in H^3(X,\mathbb{Z})$,
and $\mathscr{N}_{\tilde{\mathbf{q}}}$ are the genus-zero Gopakumar-Vafa invariants of
$\widetilde{X}$.

Type IIB string theory compactified on $X$ gives four-dimensional $\mathcal{N}=2$ supergravity coupled to $h^{2,1}(X)$ vector multiplets and $h^{1,1}(X)+1$ hypermultiplets. Throughout this paper we will consider orientifold projections of O3/O7 type, defined by holomorphic involutions $\mathcal{I}$. The induced action of $\mathcal{I}$ on cohomology groups $H^{p,q}(X,\mathbb{Q})$ allows us to define the even/odd eigenspaces $H^{p,q}_{\pm}(X,\mathbb{Q})$, and we will make the additional restriction on $\mathcal{I}$ that $h^{2,1}_+(X)=h^{1,1}_-(X)=0$, so that all the
geometric moduli
survive the projection.
This results in an effective $\mathcal{N}=1$ supergravity theory coupled to $h^{2,1}(X)$ complex structure moduli, the axiodilaton $\tau$, and $h^{1,1}(X)$ K\"ahler moduli, all in chiral multiplets. Their scalar components can be parameterized by the $z^a$ and $\tau:=C_0+ie^{-\phi}$, and the K\"ahler parameters $t^i$ and four-form axions $\int_{X}C_4\wedge \omega_i$,
where $C_4$ is the self-dual four-form of type IIB string theory. We postpone a discussion of the proper choice of \textit{holomorphic} coordinates to \S\ref{sec:control}.

Each of our compactifications contains some number $N_{\mathrm{O7}}$ of O7-planes wrapping mutually non-intersecting divisors $D_{\alpha}^{\mathrm{O7}}$, $\alpha=1,\ldots,N_{\mathrm{O7}}$.
We choose to cancel the D7-brane charge tadpole of the O7-planes \textit{locally}, by placing four D7-branes on top of each O7-plane. This gives rise to seven-brane stacks with gauge algebras $\mathfrak{so}(8)$. As usual, potential Freed-Witten
anomalies on seven-branes \cite{Freed:1999vc} are cancelled by turning on half-integral worldvolume fluxes on the D7-branes,
\begin{equation}
\frac{1}{2\pi}F_\alpha=\frac{1}{2}\imath^*_{\alpha}[D^{\mathrm{O7}}_\alpha]\, ,\quad \alpha=1,...,N_{\mathrm{O7}}\,,
\end{equation} where $\imath^{*}_{\alpha}$ denotes the pullback to $D_{\alpha}^{\mathrm{O7}}\subset X$.
The gauge-invariant field strengths $\frac{1}{2\pi}\mathcal{F}_{\alpha}=\frac{1}{2\pi} F_{\alpha}-\imath^*_\alpha B_2$ can then be set to zero by choosing a $\frac{1}{2}\mathbb{Z}$-valued $B_2$ background
\begin{equation}\label{eq:Bfield}
B_2=\frac{1}{2}\sum_{\alpha}[D_{\alpha}]\in H^2\bigl(X,\mathbb{Z}/2\bigr)\, ,
\end{equation}
and for later reference we define
\begin{equation}
b^i:=\int_X B_2\wedge \omega^i\, ,\quad \gamma^i:=2b^i\in \mathbb{Z}\, .
\end{equation}
The configuration described so far carries a net D3-brane charge $Q^{\mathrm{D3}}=-\frac{1}{4}\chi_f$, where $\chi_f$ is the Euler characteristic of the fixed locus of $\mathcal{I}$ in $X$. This tadpole can be cancelled by including $N_{\mathrm{D3}}\geq 0$ mobile D3-branes and turning on quantized three-form fluxes $(F_3,H_3)$, represented by integer vectors $(\vec{f},\vec{h})$, so that\footnote{In our conventions, a D3-brane stuck on an orientifold plane has D3-brane charge $1/2$.}
\begin{equation}
N_{\mathrm{D3}}+\frac{1}{2}\int_X H_3\wedge F_3=N_{\mathrm{D3}}+\frac{1}{2}\vec{f}^t\,\Sigma\,\vec{h}=\frac{1}{4}\chi_f\, .
\end{equation}
The classical superpotential, which is perturbatively exact in the $g_s$ and $\alpha'$ expansions \cite{PhysRevLett.57.2625,Burgess:2005jx},
is entirely determined by the pair $(F_3,H_3)$ \cite{Gukov:1999ya,Giddings:2001yu},
\begin{equation}
W_{\text{flux}}(\tau,z^a)=\sqrt{\tfrac{2}{\pi}}\int_X (F_3-\tau H_3)\wedge \Omega(z)=\sqrt{\tfrac{2}{\pi}}\,\vec{\Pi}^t\,\Sigma\, (\vec{f}-\tau \vec{h})\, ,
\end{equation}
but receives nonperturbative corrections from Euclidean D(-1)-branes,
\begin{equation}\label{eq:edm}
W_{\text{flux}}^{\mathrm{ED(-1)}} = \sum_{k=1}^\infty B_k(z)e^{2\pi i k \tau}\, ,
\end{equation}
which can be computed in F-theory, where they are naturally thought of as part of the flux superpotential \cite{Gukov:1999ya}.
Throughout this work we can consistently omit the terms \eqref{eq:edm}: see \eqref{eq:negled1} below.
The tree-level K\"ahler potential reads \cite{Grimm:2004uq}
\begin{equation}\label{eq:kis}
\mathcal{K}_{\text{tree}}=-2\log\,\Bigl(2^{\frac{3}{2}}\mathcal{V}_E\Bigr)-\log\,\Bigl(-i(\tau-\bar{\tau})\Bigr)-\log\left(-i\int_X \Omega\wedge \overline{\Omega}\right)\, ,
\end{equation}
with
\begin{equation}
\mathcal{V}_E:=\frac{1}{6}\text{Im}(\tau)^{\frac{3}{2}}\kappa_{ijk}t^it^jt^k\, \quad \text{and} \quad \int_X\Omega \wedge \overline{\Omega}=\vec{\Pi}^\dagger \, \Sigma \, \vec{\Pi}\,.
\end{equation}
The nonperturbative superpotential for the K\"ahler moduli is given by \eqref{kkltw},
\begin{equation}\label{eq:wnp}
W_{\text{np}}=\sum_D \mathcal{A}_D(z,\tau)\,\mathrm{exp}\Bigl(-\tfrac{2\pi}{c_D} T_D\Bigr)\,,
\end{equation}
up to higher-order corrections that are proportional to products of terms appearing in \eqref{eq:wnp}, and can be safely neglected in this work.

\subsection{Orientifolds of Calabi-Yau hypersurfaces}\label{sec:CYhypersurfaces_orientifolds}
The Calabi-Yau threefolds considered in this paper are hypersurfaces $X$ in toric fourfolds $V$ whose toric fans $\Sigma$ arise from triangulating four-dimensional reflexive polytopes $\Delta^\circ$; all such polytopes have been enumerated by Kreuzer and Skarke in \cite{Kreuzer:2000xy}.
Specifically, we consider regular, star triangulations of $\Delta^\circ$ in which points interior to facets are omitted, but each point not interior to a facet is a vertex of a simplex in the triangulation.
Such triangulations define partial desingularizations of $V$
in which a generic hypersurface $X$ is smooth. Let $\Sigma(1)$ be the set of edges (one-dimensional cones) of $\Sigma$, and denote by $\{x_1,...,x_n\}$ the homogeneous coordinates associated with the edges: these are the generators of the Cox ring. We have $h^{1,1}(V)=n-4$, and we define
\begin{equation}
\mathscr{D}_I:=\{x_I=0\}\subset V\, .
\end{equation}
The \textit{prime toric divisors} $\mathscr{D}_I$ generate $H_6(V,\mathbb{Z})$, and over $\mathbb{Z}_+$ they generate the cone of effective divisors on $V$.
The \textit{inherited prime toric divisors} of $X$,
\begin{equation}
D_I:=\mathscr{D}_I\cap X\, ,
\end{equation}
are effective divisors in $H_4(X,\mathbb{Z})$.
The $D_I$ generate all of $H_4(X,\mathbb{Z})$ --- i.e.~the Picard group of $X$ is inherited from the Picard group of $V$, so that $h^{1,1}(X)=h^{1,1}(V)$ --- if and only if $\Delta^\circ$ has the property that
every 2-face of $\Delta^\circ$ that has interior points is dual to a 1-face of the polar dual polytope $\Delta$ that has no interior points.  We call a model with this property $\Delta^\circ$\textit{-favorable}.
Even in cases that are not $\Delta^\circ$-favorable, the $h^{1,1}(X)+4$ irreducible components of the $D_I$,
which we will call the \textit{prime toric divisors} of $X$, furnish an overcomplete set of effective generators of $H_4(X,\mathbb{Z})$.
We note also that in general there can exist effective divisor classes on $X$ that cannot be written as non-negative linear combinations of the prime toric divisors.  Such divisor classes are called \textit{autochthonous}.

A (partial) triangulation of the polar dual polytope $\Delta$ defines a toric variety $\widetilde{V}$
whose Calabi-Yau hypersurface $\widetilde{X}$ is the \textit{mirror} of $X$: in particular, $h^{2,1}(X)=h^{1,1}(\widetilde{X})$. We call a model $\Delta$\textit{-favorable} if $h^{2,1}(X)=h^{1,1}(\widetilde{V})$, which occurs if and only if
every 2-face of $\Delta$ with interior points is dual to a 1-face of $\Delta^{\circ}$ without interior points.

Each orientifold model will be constructed using
a holomorphic involution $\mathcal{I}:\, X\rightarrow X$ that can be defined via restricting an involution $\widehat{\mathcal{I}}:\, V\rightarrow V$ to the hypersurface $X$, and tuning the hypersurface such that $\text{im}(\widehat{\mathcal{I}}|_X)=X$.
The subgroup of the automorphism group $\text{Aut}(V,\mathbb{C})$ that is connected to the identity, $\text{Aut}^0(V,\mathbb{C})$, is obtained by mapping the homogeneous coordinates $x_I$ to general sections of $\mathcal{O}_V(\mathscr{D}_I)$ \cite{Cox}. For simplicity we restrict to involutions $\widehat{\mathcal{I}}\in \text{Aut}^0(X,\mathbb{C})$, as it is these that lead to
$h^{1,1}_-(V)=0$. A general $\mathbb{Z}_2$ conjugacy class of $\text{Aut}^0(V,\mathbb{C})$ can be represented by negating a subset $\{x_{I_1},...,x_{I_k}\}\subset \{x_1,...,x_n\}$ of the homogeneous coordinates $x_I$,
\begin{equation}
\mathcal{I}:\,x_{I_\alpha}\mapsto -x_{I_\alpha}\, , \quad \alpha =1,\ldots,k\, .
\end{equation}
As stated earlier, for simplicity we will restrict to involutions for which $h^{2,1}_+(X)=0$, a very large class of which can be found systematically \cite{trilayer}.  More general orientifold models will be discussed in \cite{orientifoldKS}.

\subsection{Flux vacua}\label{ss:fluxvacua}
We now construct classical flux vacua
with exponentially small $W_0=\langle |W_{\text{flux}}|\rangle $, following \cite{Demirtas:2019sip}.
We make use of the form $\mathcal{F}=\mathcal{F}_{\text{poly}}+\mathcal{F}_{\text{inst}}$ of the prepotential near LCS,\footnote{For recent work on flux compactifications in this regime see e.g.~\cite{Grimm:2019ixq}.}
which was explained below \eqref{eq:prepotential}, and write
\begin{equation}\label{eq:wdecomp}
W_{\mathrm{flux}} = W_{\mathrm{flux}}^{(\mathrm{pert})} + W_{\mathrm{flux}}^{(\mathrm{inst})}\,,
\end{equation}
where the first term is obtained by approximating $\mathcal{F}_a$ by $\del_{z^a} \mathcal{F}_{\text{poly}}$,
and the second term is the correction to this approximation from $\mathcal{F}_{\text{inst}}$.
We now seek to solve
\begin{equation}\label{eq:pert_Fterms}
D_{\tau,z^a}W_{\mathrm{flux}}^{(\mathrm{pert})}(\tau,z^a)=0\, ,
\end{equation}
which is possible provided that we can find flux choices $(\vec{f},\vec{h})$ --- which we write as
\begin{equation}\label{eq:PFVfluxes}
\vec{f}=\left(\frac{c_a}{24}M^a,a_{ab}M^b,0,M^a\right)\, ,\quad \vec{h}=(0,K_a,0,0)\, ,
\end{equation}
in terms of a pair $(\mathbf{M},\mathbf{K})\in \mathbb{Z}^{h^{2,1}}\times \mathbb{Z}^{h^{2,1}}$ --- that fulfill the following constraints:
\begin{enumerate}\label{eq:pert.flat.vacua}
  \item $0\leq -\frac{1}{2}\mathbf{M} \cdot \mathbf{K} \leq \frac{\chi_f}{4}$, \text{i.e.~the D3-brane charge tadpole constraint;}
  \item $p^a:=(\widetilde{\kappa}_{abc}M^c)^{-1}K_b \in \mathcal{K}_{\widetilde{X}}$, \text{i.e.~}$\mathbf{p}$ lies in the K\"ahler cone of the mirror threefold;
  \item $\mathbf{K} \cdot \mathbf{p} = 0$\,.
\end{enumerate}
Such solutions, termed \textit{perturbatively flat vacua}, have a few key properties: along the one-complex-dimensional valley $\mathbf{z}=\mathbf{p}\tau$,
the F-flatness condition \eqref{eq:pert_Fterms} is satisfied, and $W_{\mathrm{flux}}^{(\mathrm{pert})}(\tau,\mathbf{p}\tau)\equiv 0$, and the type IIA worldsheet instanton corrections to the flux superpotential,
which take the form
\begin{equation}\label{eq:weff}
W_{\text{flux}}(\tau)=-\zeta\,\sum_{\tilde{\mathbf{q}}\in  \mathcal{M}(\widetilde{X})} \mathbf{M}\cdot\tilde{\mathbf{q}}\,\, \mathscr{N}_{\tilde{\mathbf{q}}}\,\text{Li}_2 \Bigl(e^{2\pi i \tau\,\tilde{\mathbf{q}}\cdot \mathbf{p}}\Bigr)\, ,
\end{equation}
become exponentially suppressed at large $\text{Im}(\tau)$.
In \eqref{eq:weff} we have defined the useful constant
\begin{equation}\label{eq:zeta}
\zeta:=\frac{1}{2^{3/2}\pi^{5/2}}\,.
\end{equation}

Now suppose one finds a pair $(\tilde{\mathbf{q}}_1,\tilde{\mathbf{q}}_2)$ of generators of the semigroup of effective curves on $\widetilde{X}$, such that:
\begin{enumerate}\setcounter{enumi}{3}\label{pairconditions}
  \item $\mathbf{p}\cdot\tilde{\mathbf{q}}_1 <1~~\text{and}~~\mathbf{p}\cdot\tilde{\mathbf{q}}_2<1$;
  \item $0<\epsilon:=\mathbf{p}\cdot(\tilde{\mathbf{q}}_2-\tilde{\mathbf{q}}_1)< 1$;
  \item at large $\text{Im}(\tau)$ along $\mathbf{z}=\mathbf{p}\tau$, the instanton terms from $\tilde{\mathbf{q}}_1$ and $\tilde{\mathbf{q}}_2$ in \eqref{eq:weff}
are parametrically larger than all other terms in \eqref{eq:weff}.
\end{enumerate}
Using condition (f), at large $\text{Im}(\tau)$ along $\mathbf{z}=\mathbf{p}\tau$ we have
\begin{equation}\label{eq:wraceeff}
W_{\text{flux}}(\tau)\approx -\zeta\,\Biggl(\mathbf{M}\cdot\tilde{\mathbf{{q}}}_1\,\, \mathscr{N}_{\tilde{\mathbf{q}}_1}\,e^{2\pi i \tau\,{\tilde{\mathbf{q}}_1\cdot \mathbf{p}}}+
\mathbf{M}\cdot\tilde{\mathbf{{q}}}_2\,\, \mathscr{N}_{\tilde{\mathbf{q}}_2}\,e^{2\pi i \tau\,{\tilde{\mathbf{q}}_2\cdot \mathbf{p}}}\Biggr)\,.
\end{equation}
Now if furthermore the pair $(\tilde{\mathbf{q}}_1,\tilde{\mathbf{q}}_2)$ has a suitable hierarchy between the superpotential coefficients,
\begin{equation}
\delta:=-\frac{(\mathbf{M}\cdot\tilde{\mathbf{q}}_1)(\mathbf{p}\cdot \tilde{\mathbf{q}}_1)\, \mathscr{N}_{\tilde{\mathbf{q}}_1} }{(\mathbf{M}\cdot\tilde{\mathbf{q}}_2)(\mathbf{p}\cdot \tilde{\mathbf{q}}_2)\, \mathscr{N}_{\tilde{\mathbf{q}}_2}}\, ,\quad |\delta|<1\, ,
\end{equation}
then \eqref{eq:wraceeff} is a \textit{racetrack superpotential} with a minimum at weak string coupling and large complex structure.

Specifically, setting the F-terms of the complex structure moduli and the axiodilaton to zero, we find
\begin{equation}\label{eq:tauform}
\left\langle e^{2\pi i \tau}\right\rangle\approx \delta^{\frac{1}{\epsilon}}\ll 1\, ,
\end{equation}
where we have approximated the F-term $D_{\tau}W$ by $\del_{\tau}W$, which becomes accurate at small $g_s$.
In the vacuum we have
\begin{equation}\label{eq:W0_formula}
W_0 = \bigl\langle |W_{\text{flux}}| \bigr\rangle \approx \zeta\,\Bigl|\mathbf{M}\cdot \tilde{\mathbf{q}}_1\,\mathscr{N}_{\tilde{\mathbf{q}}_1}^0 \,\delta^{\mathbf{p}\cdot \tilde{\mathbf{q}}_1/\epsilon}+\mathbf{M}\cdot \tilde{\mathbf{q}}_2\,\mathscr{N}_{\tilde{\mathbf{q}}_2}^0\,\delta^{\mathbf{p}\cdot \tilde{\mathbf{q}}_2/\epsilon}\Bigr|\,,
\end{equation} and so
\begin{equation}\label{eq:W0_formulasimpl}
W_0  \sim \delta^{\mathbf{p}\cdot \tilde{\mathbf{q}}_{1}/\epsilon}\sim \delta^{\mathbf{p}\cdot \tilde{\mathbf{q}}_2/\epsilon}\ll 1\,.\end{equation}
Viewed as functions of the quantized parameters --- namely, the three-form fluxes $\mathbf{M}$ and $\mathbf{K}$; the homology classes of curves in $\widetilde{X}$, $\tilde{\mathbf{q}}_1$ and $\tilde{\mathbf{q}}_2$; and the Gopakumar-Vafa invariants of these curves, $\mathscr{N}_{\tilde{\mathbf{q}}_1}$ and $\mathscr{N}_{\tilde{\mathbf{q}}_2}$ ---
the string coupling $g_s = 1/\mathrm{Im}(\tau)$ is \textit{polynomially} small, while the flux superpotential is \textit{exponentially} small.

In obtaining \eqref{eq:tauform} and \eqref{eq:W0_formula} we have consistently omitted the effects of other worldsheet instantons, by virtue of the condition
(f) that we imposed above on the pair $(\tilde{\mathbf{q}}_1,\tilde{\mathbf{q}}_2)$.
We have also omitted the effects of Euclidean D(-1)-branes, which from \eqref{eq:edm} give contributions to the superpotential of order $\mathrm{exp}(-2\pi/g_s)$.
Let us now explain why this is justified.
The worldsheet instanton terms in \eqref{eq:wraceeff} have actions
$S_i=2\pi {\tilde{\mathbf{q}}_i}\cdot{\mathbf{p} /g_s}$ for $i=1,2$, and in our flux vacua one has $e^{-S_i}\sim W_0$.
In terms of the parameter
\begin{equation}\label{eq:ctau}
c_{\tau}^{-1} :=g_s\frac{\log (W_0^{-1})}{2\pi} =\mathbf{p}\cdot \tilde{\mathbf{q}}_2 +\mathcal{O}(\epsilon)= \mathbf{p}\cdot\tilde{\mathbf{q}}_1+\mathcal{O}(\epsilon)\,,
\end{equation}
we can write
\begin{equation}\label{eq:negled1}
e^{-2\pi/g_s}=\left(e^{-S_i}\right)^{\frac{1}{\tilde{\mathbf{q}}_i\cdot \mathbf{p}}}\sim (W_0)^{c_{\tau}}\ll W_0\,,
\end{equation} where we have used the condition (d) that was imposed above.
Thus, Euclidean D(-1)-brane effects are parametrically sub-leading in comparison to the terms in \eqref{eq:wraceeff} that determine the vacuum structure.

The conditions for a perturbatively flat vacuum in \eqref{eq:pert.flat.vacua} are Diophantine in nature, and so are difficult to solve in general.
Nevertheless, in practice we have been able to find solutions to the constraints when $h^{2,1}$ is relatively small.

\section{Nonperturbative superpotential}\label{ss:ed3}

\subsection{Rigid divisors}\label{ss:rigid}

A Euclidean D3-brane (ED3) wrapped on an effective divisor is half-BPS and can thus contribute to the superpotential provided the only exact fermion zero-modes are the two universal modes, i.e.~the goldstini associated to the breaking of half the supercharges. In the absence of worldvolume flux and bulk three-form flux, the zero-modes take values in the cohomology groups $H_{\pm}^{\bullet}(D,\mathcal{O}_D)$, and a superpotential term is therefore generated if $D$ is smooth and
\begin{equation}\label{eqn:rigid1}
h_+^\bullet(D)=(1,0,0)\, ,\quad \text{and}\quad h_-^\bullet(D)=(0,0,0)\,.
\end{equation}
We call a divisor $D$ that satisfies \eqref{eqn:rigid1}
a \emph{rigid} divisor.\footnote{More generally, worldvolume flux and bulk three-form flux will generically lift the zero-modes associated with $h^{2}_{\pm}(D,\mathcal{O}_D)$ and $h^{1}_-(D,\mathcal{O}_D)$ \cite{Bianchi:2011qh,Bianchi:2012pn,Bianchi:2012kt}, but we will not rely on such lifting.}

Equivalently, one can consider the dual F-theory compactification on an elliptically fibered Calabi-Yau fourfold $\pi_{\Bbb{E}}:Y_4\rightarrow B_3,$ where $B_3\simeq X/\mathcal{I}$ is the base manifold of the elliptic fibration.
A Euclidean D3-brane on a divisor $D$ uplifts to a Euclidean M5-brane wrapped on a vertical divisor $\widehat{D}\subset Y_4.$
The fermion zero-modes take values in $H^\bullet(\widehat{D},\mathcal{O}_{\widehat{D}})$, so the F-theory version of the rigidity condition \eqref{eqn:rigid1}
is
\begin{equation}\label{eqn:rigidM}
h^\bullet(\widehat{D},\mathcal{O}_{\widehat{D}})=(1,0,0,0)\,.
\end{equation}
A divisor $\widehat{D}$ that satisfies \eqref{eqn:rigidM} is likewise called a rigid divisor, and a \emph{smooth} rigid divisor $\widehat{D}$ contributes to the superpotential \cite{Witten:1996bn}. In this paper it will suffice to study \eqref{eqn:rigid1} and \eqref{eqn:rigidM}
for prime toric divisors of Calabi-Yau hypersurfaces, for which smoothness is guaranteed.\footnote{At a generic point in the complex structure moduli space of a smooth Calabi-Yau threefold $X$, the prime toric divisors $D_I$ are smooth, because their strata are inherited from the strata of $X$ \cite{danilov1986newton,Batyrev:1994hm,Braun:2017nhi}.}

We now turn our attention to non-abelian gauge theories on seven-branes. In the Calabi-Yau orientifolds considered in this paper, most O7-planes wrap rigid divisors, and
as explained in \S\ref{ss:class}, we cancel the D7-brane tadpole locally by placing four D7-branes on each O7-plane. As a result, for each $\mathfrak{so}(8)$ stack on a rigid divisor $D$ we obtain pure $\mathcal{N}=1$ $\mathfrak{so}(8)$ super Yang-Mills (SYM) theory, whose holomorphic gauge coupling is given by $T_D/4\pi$ at high energy.
At low energies the gauginos of pure $\mathcal{N}=1$ SYM condense and generate a nonperturbative superpotential term $\mathcal{A}_D(z,\tau) e^{-2\pi T_D/c_D}$.

In some of our compactifications, a small subset of
the O7-planes wrap divisors $D_N$ that have normal bundle deformations, with $h^{\bullet}(D_N)=(1,0,1)$ or, in rare cases, $h^{\bullet}(D_N)=(1,0,2)$.
As the volume of $D_N$ is typically many times larger than that of the
$h^{1,1}$ smallest rigid prime toric divisors, stabilization of the K\"ahler moduli does not require, and is negligibly affected by, potential
gaugino condensation in the $\mathfrak{so}(8)$ stack on $D_N$, which can occur if fluxes lift all the normal bundle deformations.

Indeed, we expect that normal bundle deformations of the seven-branes on $D_N$ \textit{are} stabilized by background three-form fluxes. To rigidify a D7-brane on a divisor $D$ one can turn on the worldvolume flux $\mathcal{F}_2=[C]-[C']$ \cite{Bianchi:2011qh} on the D7-brane, where $C\subset D$ is a rigid holomorphic curve and $C'$ is its orientifold image. This procedure cannot be applied to rigidify the $\mathfrak{so}(8)$ stack, because every holomorphic curve $C$ in an O7-plane is pointwise invariant under the orientifold action, and hence $[C]=[C'].$
However, in a nontrivial $H_3$ background, where locally we can write $H_3=dB_2$, upon displacing a D7-brane away from an O7-plane on $D_N$, the $B_2$ field induced on the displaced D7-brane grows, which eventually leads to D7-brane monodromies \cite{Hebecker:2014eua,Marchesano:2014mla,Bielleman:2016olv,Kim:2018vgz}. As a result, the displaced D7-brane feels a monodromy potential
\begin{equation}
V_{D7}= \frac{2\pi}{\ell_s^8}\int_{D_N}d^4 y\,  e^{-\phi}\sqrt{\det(g_{D_N}+\imath^*_{D_N}B_2)}\, .
\end{equation}
The minimum of this potential lies at the $\mathfrak{so}(8)$ stack configuration, where $B_2=0$: the O7-plane is a fixed locus of the orientifold involution $\mathcal{I}$, and the orientifold acts as $B_2(x)\mapsto -B_2(\mathcal{I}(x))$. The displacement of the D7-brane also induces D3-brane charge on its worldvolume, and thus by conservation of D3-brane charge the bulk D3-brane charge density from fluxes, and their energy density, gets reduced within the region swept out by the moving D7-brane. The overall potential is positive semi-definite, and vanishes if and only if the induced $\imath^*_{D_N}B_2$ happens to be self-dual on $D_N$ \cite{Marchesano:2014mla,Kim:2018vgz}.\footnote{This is related to the non-generic situations described in \cite{Bianchi:2012kt} in which the normal bundle deformations of a Euclidean D3-brane do \textit{not} get lifted by fluxes.} In this paper we will not check this condition explicitly, but note that the anti-self dual part generically does \textit{not} vanish, and seven-branes should be stabilized automatically. Moreover, even if they \textit{do} turn out to be unstabilized at this level, they either stay exact moduli after inclusion of all perturbative and non-perturbative corrections --- which seems implausible but would in any case not endanger the existence of our vacua
--- or they receive a potential from those corrections. In the latter case, due to the orientifold plane boundary conditions, the potential would have to be minimized or maximized in the $\mathfrak{so}(8)$ configuration, and thus the D7-branes would be stabilized there by virtue of the unbroken supersymmetry.\footnote{In our AdS vacua the potential could have a maximum with negative mass$^2$ above the Breitenlohner-Freedman (BF) bound.}

Next, let us remark that the superpotential terms from gaugino condensation on confining seven-brane gauge theories have a simple M/F-theory description, again described by a Euclidean M5-brane.  Consider a smooth F-theory uplift $\overline{\Delta}_{\Bbb{E}}$ of an irreducible discriminant locus $\Delta_{\Bbb{E}}$ of the elliptic fibration. We assume that a  gauge group $G$ is engineered on $\Delta_{\Bbb{E}}$, and for simplicity we assume that there is no curve $\gamma$ in $\Delta_{\Bbb{E}}$ where the gauge group $G$ is enhanced to a larger group $G'.$
Then, $\pi^{-1}(\Delta_{\Bbb{E}})=\overline{\Delta}_{\Bbb{E}}$ is topologically equivalent to a union of $\mathbb{P}^1$'s (corresponding to the Dynkin nodes of $G$) times $\Delta_{\Bbb{E}}$. It follows that the fermion zero-modes of an M5-brane wrapped on $\overline{\Delta}_{\Bbb{E}}$ are counted by $h^\bullet(\overline{\Delta}_{\Bbb{E}},\mathcal{O}_{\overline{\Delta}_{\Bbb{E}}})=c_2(G) h^\bullet(\Bbb{P}^1\times \Delta_{\Bbb{E}},\mathcal{O}_{\Bbb{P}^1\times \Delta_{\Bbb{E}}}).$ For a rigid $\Delta_{\Bbb{E}},$ a Euclidean M5-brane wrapped on a Dynkin node $\Bbb{P}^1\times D$ has the right number of zero-modes to contribute to the superpotential. Under the projection $\pi_{\Bbb{E}},$ the Dynkin node maps to a fractional divisor class $[\Delta_{\Bbb{E}}]/c_2(G).$ Hence, we conclude again that an $\mathfrak{so}(8)$ stack on a rigid divisor $D$ generates a nonperturbative superpotential term $\mathcal{A}_D(z,\tau) e^{-2\pi T_D/6}.$

Finally, we would like to briefly comment on the matching between the zero-mode counting in the type IIB picture and the dual M/F-theory picture. Consider a blowdown of the elliptic fibration $\pi_{\Bbb{E}}':Y_4'\rightarrow B_3$ such that the elliptic fiber, $\Bbb{E},$ develops singularities at the discriminant locus.
We note that odd-dimensional cycles of $\Bbb{E}$ can be viewed as orientifold-odd and even-dimensional cycles of $\Bbb{E}$ as orientifold-even, due to the $-1\in SL(2,\mathbb{Z})$ monodromy picked up under encircling an $\mathfrak{so}(8)$ stack. To compute the Hodge numbers of the blowdown of $\widehat{D}$, denoted $\widehat{D}'$, one can count orientifold-even cycles of $D\times \Bbb{E}$: we have
$h^{i,0}(D\times \Bbb{E})=h^{i,0}_+(D)\times h^{0,0}(\Bbb{E})+h^{i-1,0}_-(D)\times h^{1,0}(\Bbb{E})= h^{i,0}_+(D)+h^{i-1,0}_-(D).$ Because the blowup of the elliptic fiber along the discriminant locus $\Delta_{\Bbb{E}}|_D$ does not change $h^{i,0}(\widehat{D}'),$ we arrive at the identification\footnote{In the special case that $B_3$ is smooth, it was shown in \cite{Grassi:1997mr} that $h^i(\widehat{D},\mathcal{O}_{\widehat{D}})=h^i (D_B,\mathcal{O}_{D_B})+h^{i-1}(D_B,-\Delta_{\Bbb{E}}|_{D_B})$ for a divisor $D_B\subset B_3.$  This formula is equivalent to \eqref{eqn:identification} under the identifications $h_-^{i,0}(D)\equiv h^i(D_B,-\Delta_{\Bbb{E}}|_{D_B})$ and $h_+^{i,0}(D)\equiv h^i(D_B,\mathcal{O}_{D_B}).$}
\begin{equation}\label{eqn:identification}
 h^i(\widehat{D},\mathcal{O}_{\widehat{D}})=h^{i,0}_+(D)+h^{i-1,0}_-(D).
\end{equation}

\subsection{Pfaffian prefactors}

The Pfaffian prefactor $\mathcal{A}_D(z,\tau)$ of a nonperturbative superpotential term is related to the partition function of the corresponding M5-brane worldvolume theory in the F-theory uplift $\widehat{D}$ of the divisor $D$, or equivalently to the partition function of the $O(1)$ D3-brane worldvolume theory on $D$ in the type IIB orientifold. In general,  $\mathcal{A}_D(z,\tau)$ can be a section of a non-trivial line bundle on moduli space,\footnote{The moduli space in question is the complex structure moduli space of the fourfold, or equivalently the axiodilaton and complex structure moduli space of $X$.} so it can have zeros along divisors $\mathfrak{D}$ in moduli space.
Along such $\mathfrak{D}$, the corresponding nonperturbative superpotential term no longer contributes to the potential for the K\"ahler moduli, while in the immediate neighborhood of $\mathfrak{D}$ the contribution is suppressed; either case could impact the vacuum structure.

The stabilization of the complex structure moduli and axiodilaton by fluxes leads to expectation values $\langle z \rangle$, $\langle \tau \rangle$, and the generic situation is that these expectation values do not lie on $\mathfrak{D}$, or exponentially near $\mathfrak{D}$.  If we now define
\begin{equation}
\mathcal{A}_D^{\mathrm{vac}} := \mathcal{A}_D\bigl(\langle z \rangle,\langle \tau \rangle\bigr)\,,
\end{equation}
then the expectation values $\mathcal{A}_D^{\mathrm{vac}}$ are simply (unknown) complex numbers, and the vacuum configuration for the K\"ahler moduli could be expressed in terms of their values.

Even so, one could worry that a conspiracy might cause some of the $\mathcal{A}_D^{\mathrm{vac}}$ to be exponentially small in the classical flux vacua of \S\ref{ss:fluxvacua} that yield $W_0 \ll 1$.  This would jeopardize a search for AdS vacua.\footnote{We are not aware of any reasoning that predicts that such a conspiracy should actually occur in string theory, but we can predict that the solutions presented here would be criticized on this basis if the possibility were not strictly excluded.}
For the avoidance of doubt, we will ensure that zeros of the Pfaffian cannot arise in our examples, by selecting compactifications in which the Pfaffians of all leading rigid prime toric divisors are pure numbers, i.e.~sections of the trivial line bundle on moduli space.  Let us now explain how this can be achieved.

\subsubsection{General properties of the Pfaffian}

Around LCS and weak string coupling, the $\mathcal{A}_D(z,\tau)$ enjoy an expansion
\begin{equation}\label{pfaffexp}
\mathcal{A}_D(z,\tau)=\sum_{k=0}^\infty\mathcal{A}_D^{(k)}(z)e^{2\pi i k \tau}=\sum_{\tilde{\mathbf{q}}\in \mathcal{M}(\tilde{X})}\sum_{k=0}^\infty \mathcal{A}_D^{(\tilde{\mathbf{q}},k)} e^{2\pi i (\tilde{\mathbf{q}}\cdot \mathbf{z}+k \tau)} \, .
\end{equation}
Here $\mathcal{A}_D^{(k)}(z)$ is the Pfaffian of a Euclidean D3-brane with a fixed gauge bundle of instanton number $k$, and each of the $\mathcal{A}_D^{(k)}(z)$ enjoys its own expansion around LCS.\footnote{In general, the expansion of the $\mathcal{A}_D^{(k)}(z)$ around LCS may contain terms that are polynomial in the $z^a$, which are not displayed in \eqref{pfaffexp}.  Such terms are absent along perturbatively flat vacua due to the unbroken discrete shift symmetry preserved by the fluxes of \eqref{eq:PFVfluxes} \cite{Demirtas:2019sip}.}  The $\mathcal{A}_D^{(\tilde{\mathbf{q}},k)}$ are a priori unknown
complex
numbers. We note that the above expansion can be reinterpreted via mirror symmetry as a poly-instanton expansion including Euclidean D2-branes and worldsheet instantons in a type IIA O6 orientifold.

It is convenient to write the leading terms in \eqref{pfaffexp} in the more schematic form
\begin{equation}\label{pfaffexpsimp}
\mathcal{A}_D(z,\tau) = \mathcal{A}_D^{(0,0)} + \mathcal{A}_D^{(1,0)} e^{2\pi i\tau} +\mathcal{A}_D^{(0,1)} e^{2\pi i z} + \ldots
\end{equation}
The type IIB zero-mode counting --- i.e.,
the rigidity condition imposed on $D$ in \S\ref{ss:rigid} --- implies that $\mathcal{A}_D(z,\tau)$ does not vanish identically.
However, at this stage one cannot exclude that $\mathcal{A}_D^{(0,0)}= 0$.\footnote{This situation would be mirror dual to a single Euclidean D2-brane on a special Lagrangian three-cycle with non-trivial Betti number $b^1$.}
In such a case, $\mathcal{A}_D^{\mathrm{vac}}$ would be extremely small, as our flux vacua occur
at weak string coupling and near LCS.

To avoid this outcome, we will ensure that the following three relations hold:
\begin{enumerate}\label{eq:pfaffzero}
  \item $\mathcal{A}_D  (z,\tau)$~~\text{is not identically zero};
  \item $\mathcal{A}_D^{(j,k)}=0~\forall j>0$\,,
  \item $\mathcal{A}_D^{(j,k)}=0~\forall k>0$\,,
\end{enumerate} which together imply that $\mathcal{A}_D^{(0,0)}\neq 0$.  In sum, by enforcing (a), (b), and (c) we guarantee that $\mathcal{A}_D^{\mathrm{vac}}$ is not systematically suppressed near weak coupling and LCS.

In fact we have already imposed condition (a), by insisting that $\widehat{D}$ must be rigid.

To impose (b), we recall from standard zero-mode counting that
the $\mathcal{A}_D^{(k)}(z)$ are nonzero for gauge bundles $\mathcal{F}\in H^{1,1}_-(D,\mathbb{Z})$
that do \emph{not} descend via restriction from nontrivial classes in $H^2(X,\mathbb{Z})$.
In particular, if $h^{1,1}_-(D)=0$ then only $\mathcal{A}_D^{(0)}(z)$ is non-vanishing, and thus $h^{1,1}_-(D)=0$ implies condition (b).\footnote{Note that condition (b) is not actually necessary for ensuring that $\mathcal{A}_D^{(0,0)}\neq 0$: a rigid $O(1)$ Euclidean D3-brane has Pfaffian $\mathcal{A}_D^{(0)}(z)$ which, via zero-mode counting, is not identically zero.
If $\mathcal{A}_D^{(0)}(z)$ is also $z$-independent then it already follows that $\mathcal{A}_D^{(0,0)}\neq 0$, even if $h^{1,1}_-(D)\neq 0$. We are imposing (b) here purely to simplify the reasoning that leads to $\mathcal{A}_D^{(0,0)}\neq 0$.}

Finally, to impose (c), we will arrange that $\mathcal{A}_D^{(0)}(z)$ is actually \textit{independent} of $z$. We are aware of two mechanisms for ensuring this, which we now discuss in turn.

\subsubsection{Pure rigid divisors}

In \cite{Witten:1996hc} it was shown that the partition function of the worldvolume theory of an M5-brane wrapping a divisor $\widehat{D}$ is an elliptic theta function of the complex structure moduli of the intermediate Jacobian $\mathcal{J}:=H^3(\widehat{D},\Bbb{R})/H^3(\widehat{D},\Bbb{Z}).$ As the complex structure moduli of $\mathcal{J}$ depend on
the complex structure moduli of $Y_4,$ in principle by computing $\mathcal{J}$ one can determine $\mathcal{A}_D(z,\tau).$
In particular, if $h^{2,1}(\widehat{D})=0$, then the corresponding M5-brane partition function is a section of the \emph{trivial} line bundle over the complex structure moduli space of the fourfold, and so the Pfaffian $\mathcal{A}_{D}$ is a pure (complex) number \cite{Witten:1996hc}.  We will call a rigid divisor $\widehat{D}$ with $h^{2,1}(\widehat{D})=0$ a \emph{pure rigid divisor}. By extension, if $D\subset X$ descends from a pure rigid divisor $\widehat{D}$ in the fourfold, we will call $D$ a pure rigid divisor.

To impose the condition of pure rigidity, the first step is to compute the dimension of $\mathcal{J}$ for a vertical divisor $\widehat{D}$ in an elliptic Calabi-Yau fourfold. For each Calabi-Yau orientifold compactification considered in this paper, we have constructed the dual elliptic Calabi-Yau fourfold, by first obtaining the base $B_3:=X/\mathcal{I}$ as a hypersurface in $V/\widehat{\mathcal{I}}$, and further defining the Calabi-Yau fourfold as a codimension-two complete intersection in a toric sixfold $V_6$ given by a toric twofold fibration over $V/\widehat{\mathcal{I}}$.\footnote{See e.g.~\cite{Collinucci:2008zs}.} Next, one can generalize the results of the early works \cite{danilov1986newton,Batyrev:1994hm,borisov1993towards,Batyrev:1994pg,Klemm:1996ts} to obtain combinatorial formulas for the Hodge numbers of prime toric divisors in $Y_4$ \cite{hodge}.
Equipped with these formulas, one can compute $h^{2,1}(\widehat{D})$ \cite{hodge,danilov1986newton}.

Let us briefly explain the type IIB perspective on $h^{2,1}(\widehat{D}).$ We again consider a blowdown of the elliptic fibration $\pi_\Bbb{E}:Y_4'\rightarrow B_3$.  Then $h^{2,1}(\widehat{D}')$ is counted by
\begin{equation}
h^{2,1}(\widehat{D}')=h^{1,0}_+(D)  b^{2}(\Bbb{E})+h^{2,1}_+(D) b^{0}(\Bbb{E})+h^{2,0}_-(D) \frac{b^{1}(\Bbb{E})}{2}+h_-^{1,1}(D) \frac{b^{1}(\Bbb{E})}{2}\overset{D\,\, \text{rigid}}{=} h^{1,1}_-(D)\, ,
\end{equation}
and indeed we had concluded above that the Pfaffian $\mathcal{A}_D(z,\tau)$ is $\tau$-independent if $h^{1,1}_-(D)=0$. As $h^{1,1}_-(D)$ is insensitive to the intersection locus with seven-branes, a natural interpretation is that $h^{1,1}_-(D)$ encodes the dependence of the Pfaffian on the \textit{bulk} complex structure moduli of $X/\mathcal{I}$ and the dilaton $\tau$, though we will not rely on such an interpretation in our models. Upon blowing up along the discriminant locus of the elliptic fibration, $h^{2,1}(\widehat{D})$ can in general be larger than $h^{2,1}(\widehat{D}')$. Thus, we interpret the difference $h^{2,1}(\widehat{D})-h^{2,1}(\widehat{D}')$ as encoding the dependence of the Pfaffian on the D7-brane moduli.\footnote{For related work see \cite{Blumenhagen:2010ja}.}
For this paper, however, we will compute $h^{2,1}(\widehat{D})$ directly in F-theory.

\subsubsection{Inflexible rigid divisors}

The condition $h^{2,1}(\widehat{D})=0$ is \textit{sufficient} to imply property (c) above, and indeed (b) at the same time, because the axiodilaton $\tau$ is of course a complex structure modulus of the fourfold.
In every example presented in this paper, all of the leading contributions to the nonperturbative superpotential come from pure rigid prime toric divisors, with $h^{2,1}(\widehat{D})=0$ and $h^{\bullet}(\widehat{D},\mathcal{O}_{\widehat{D}})=(1,0,0,0)$.

However, a condition that can be checked directly in the type IIB compactification implies (c) but not (b), as we now explain.
Though we will not make use of it here, in future model-building this alternative may be useful, as the uplift to F-theory is not always easy to analyze.

One can forbid $z$-dependence of the Pfaffian
by imposing that $D$ has no complex structure deformations: in terms of the tangent bundle $\mathscr{T}_D$ of $D$, one requires
that $h^{1}(\mathscr{T}_D)=0$.
In this case, the complex structure of $D$ is necessarily independent of the bulk complex structure moduli $z$, and thus the partition function of the Euclidean D3-brane worldvolume theory cannot depend on $z$.   We call a divisor $D$ obeying $h^{1}(\mathscr{T}_D)=0$ \emph{inflexible}.

The constraint $h^{1}(\mathscr{T}_D)=0$ is satisfied by a considerable majority of prime toric divisors $D$ in Calabi-Yau threefold hypersurfaces $X$ with $h^{1,1}(X) \gg 1$.
At large $h^{1,1}(X)$ almost all prime toric divisors of $X$ arise from points interior to 1-faces and 2-faces of the reflexive polytope $\Delta^\circ$. It is straightforward to see that divisors from points interior to 2-faces are toric surfaces themselves, and thus trivially satisfy $h^1(\mathscr{T}_D)=0$. In general, divisors from points interior to 1-faces are $\mathbb{P}^1$-fibrations over curves of genus $g$, where $g$ is determined by the number of points interior to the dual 2-face in the dual polytope $\Delta$. As explained in \S\ref{sec:CYhypersurfaces_orientifolds}, for convenience we impose $\Delta$-favorability in our models, and in particular we have $g=0$ for all 1-face divisors.
Because $\mathbb{P}^1$ fibrations over $\mathbb{P}^1$ are likewise toric, we again find $h^1(\mathscr{T}_D)=0$ for all divisors from points interior to 1-faces.
The only divisors that need to be checked case by case are those arising from vertices of $\Delta^\circ$. Their complex structure deformations are inherited from those of $X$, i.e.~$h^1(\mathscr{T}_D)=h^1(\mathscr{T}_X|_D)$, so all one needs to do is inspect the anti-canonical polynomial $f$ along $D$. The surviving monomials correspond to the points of the facet of $\Delta$ dual to the vertex, and rigidity of $D$ is in one-to-one correspondence with absence of interior points in the facet. After setting to zero the toric coordinate associated to $D$ one can still use the action of an algebraic torus $(\mathbb{C}^*)^3$ to gauge fix three coefficients of $f|_D$, and finally use the freedom of overall rescaling of $f|_D$.
Thus, $h^1(\mathscr{T}_D)=0$ if and only if there are exactly four points in the dual facet, i.e.~if and only if the facet is a simplex.

In summary, prime toric divisors $D_f$ corresponding to points interior to 2-faces $f$ of $\Delta^\circ$ are always rigid and inflexible,
and in $\Delta$-favorable models, prime toric divisors $D_e$ corresponding to points interior to 1-faces $e$ are likewise always rigid and inflexible.
A prime toric divisor $D_v$ corresponding to a vertex $v$ is inflexible if and only the dual facet is a simplex, and is rigid if and only if the dual facet has no interior points.
Equipped with these results, we are able to check the condition $h^1(\mathscr{T}_D)=0$ in our models.

\subsubsection{Pfaffian numbers}

The complex numbers\footnote{Note that as we can neglect Euclidean D3-brane corrections to the K\"ahler potential, nothing is lost by using the axion shift symmetries $T_i\rightarrow T_i+\delta T_i$ with $\delta T_i\in i\mathbb{R}$ to absorb the complex phases in $h^{1,1}$ Pfaffian numbers, but in some examples we find $h^{1,1}+1$, $h^{1,1}+2$, or $h^{1,1}+3$ leading contributions, and in such cases there are one, two, or three phases remaining, respectively.} $\mathcal{A}_{D_I} \equiv \mathcal{A}_{D_I}^{(0,0)}$ associated to pure rigid prime toric divisors $D_I$
are the \emph{only} unknowns in the leading data of the effective supergravity theories studied in this paper. A few comments are in order regarding their properties.

Although in all our models we have proved that the $\mathcal{A}_{D_I}$ are \emph{numbers}, one could worry that one or more of them is actually the number 0, which after all is a famous section of the trivial line bundle.\footnote{See e.g.~\cite{zerotablet,zero}.}
We have excluded the main physical reasons for such a zero --- namely, integrals over moduli space, extra fermion zero modes, and cancellations \cite{Beasley:2003fx,Cvetic:2011gp,Cvetic:2012ts} --- and so the $\mathcal{A}_{D_I}$ are \emph{generically nonzero} by the usual standards of instanton calculus.  Even so, computing their values directly, perhaps along the lines of \cite{Sen:2021tpp,Alexandrov:2021shf}, would be worthwhile.

Moreover, one might wonder whether the $\mathcal{A}_{D_I}$ could be hierarchical, because similar prefactors are often related to BPS state counts, which can in principle involve large numbers.  However, changing the $\mathcal{A}_{D_I}$ leads to relative corrections in the vevs of the K\"ahler moduli of order $\log(\mathcal{A}_{D_I})/\log(W_0)$.
In our examples, $W_0^{-1}$ far exceeds any number that could reasonably appear in a BPS state count at low degrees, and so we expect our approximation to be excellent.
Even so, after finding supersymmetric vacua for the reference value $\mathcal{A}_{D_I}=1~\forall I$, we have repeated our analysis with $\mathcal{A}_{D_I} \in \{10^{-4},10^{4}\}$, and recovered the existence of vacua.

Let us further point out that because the $\mathcal{A}_{D_I}$ remain unchanged as we select fluxes to explore vacua with smaller and smaller $W_0$, there is no possibility of a conspiracy in which the $\mathcal{A}_{D_I}$ become ill-behaved as $W_0 \to 0$ and thus destroy the vacuum structure.\footnote{Likewise, the mass of the perturbatively-flat direction $\mathbf{z}=\mathbf{p}\tau$ is comparable to the masses of the K\"ahler moduli \cite{Demirtas:2019sip}, and if some of the
$\mathcal{A}_{D_I}$ were to vary along $\mathbf{z}=\mathbf{p}\tau$, the stabilization of the perturbatively-flat direction and of the K\"ahler moduli would be entangled.  Because the $\mathcal{A}_{D_I}$ are constant this possibility does not arise in our examples.}

Finally, we remark that thus far we have ensured that the $\mathcal{A}_{D_I}$ do not depend on the closed string moduli and the seven-brane moduli.
However, in some of our compactifications, mobile D3-branes will be present, and all of the $\mathcal{A}_{D_I}$ do \textit{necessarily} depend on all the D3-brane position moduli.
As a D3-brane approaches a rigid divisor $D_I$, the corresponding nonperturbative superpotential term tends to zero: linearly in the separation for Euclidean D3-branes, and with fractional power $c_{D_I}^{-1}$ for gaugino condensation \cite{Ganor:1996pe,Berg:2004ek,Baumann:2006th,Koerber:2007xk,Braun:2018fdp}.
Thus, the F-flat configuration for the D3-brane position moduli has the D3-branes stabilized \textit{away} from the vanishing loci $\mathcal{A}_{D_I}=0$ \cite{DeWolfe:2007hd}.
In the following it will be understood that the Pfaffians $\mathcal{A}_{D_I}$ are evaluated at the F-flat minimum for the D3-brane moduli.\footnote{Note that if $\frac{1}{2}\int_X H_3\wedge F_3 \in \tfrac{1}{2}\mathbb{Z}$ then one needs to introduce a `half' D3-brane stuck either on one of the seven-brane stacks or on one of the O3-planes. When this is necessary, we will place the half D3-brane on a seven-brane stack on a divisor $D$ that is not relevant for K\"ahler moduli stabilization, thus adding a chiral multiplet (or a half-hypermultiplet if $D=K3$) in the \textbf{8} of $SO(8)$ and two neutral chiral multiplets (a hypermultiplet if $D=K3$) parameterizing the position of the half D3-brane along the seven-branes: see e.g.~\cite{Vafa:1997nx}.}

\subsection{Autochthonous divisors}\label{sec:autochthonous}

The most obvious corrections to the superpotential of \eqref{kkltw} come from multi-instantons or, potentially, from Euclidean D3-branes wrapped on divisors that can be written as non-negative linear combinations of two or more prime toric divisors. At points in moduli space where the one-instanton and gaugino condensation terms that we have already incorporated in \eqref{kkltw} are small, such corrections are parametrically sub-leading.

However, as we now explain, another class of Euclidean D3-brane contributions deserves a more detailed analysis: as recalled in \S\ref{sec:CYhypersurfaces_orientifolds}, a Calabi-Yau threefold hypersurface $X$ in a toric variety $V$ inherits effective divisors $D$ from divisors $\mathscr{D}$ of $V$ via intersection with $X$, i.e.~$D = \mathscr{D} \cap X$.
Effective divisors on $X$ that are \emph{not} inherited are termed autochthonous.\footnote{Through the inclusion of $X$ in $V$, an autochthonous divisor $D_A$ on $X$ corresponds to an effective subvariety of complex dimension two in $V$, but unlike an inherited effective divisor, this subvariety is not the intersection $\mathscr{D} \cap X$ for any effective divisor $\mathscr{D}$ on $V$.}
Methods for identifying the classes of autochthonous divisor will be presented elsewhere \cite{autochthonousinprogress}.  For the present work, it suffices to remark that computing \emph{all} effective divisor classes in a Calabi-Yau threefold with large $h^{1,1}$ is not currently feasible, and so we will study the nonperturbative superpotential terms that result from Euclidean D3-branes that wrap inherited divisors, which are very easy to identify from toric data.

One might then ask whether rigid autochthonous divisors could support Euclidean D3-brane superpotential terms that alter the vacuum structure that we will compute herein based on inherited divisors.  Fortunately, a peculiarity of the KKLT construction obviates computing all autochthonous contributions.  To see this, we consider a toy example, in which $X$ is a Calabi-Yau orientifold, $D_1$ and $D_2$ are inherited prime toric divisors on $X$, and $2D_1 - D_2$ is an autochthonous --- and hence, by definition, effective, and thus also calibrated --- divisor.   If we find a point in the K\"ahler cone $\mathcal{K}_X$ where $\mathrm{vol}(D_1) = \mathrm{vol}(D_2) \equiv T$, then  $\mathrm{vol}(2D_1 - D_2) = T$ as well, and a Euclidean D3-brane wrapping $\mathrm{vol}(2D_1 - D_2)$ makes at most a contribution comparable to those of the effective divisors, \emph{not} parametrically larger.

Now we recall that any divisor, including an autochthonous one, is always expressible in terms of an integer (but not necessarily positive integer) linear combination of prime toric divisors.  Moreover, in the vacua that we will find below, the (Einstein frame) volumes of a subset of $h^{1,1}(X)$ prime toric divisors take \textit{integer} values, up to an overall factor $\log(W_0^{-1})/2\pi$: those hosting $\mathfrak{so}(8)$ stacks of seven-branes have volumes $c_2(\mathfrak{so}(8)) =6$ times larger than those hosting Euclidean D3-branes. If, at this point in moduli space, the other four prime toric divisors also have integer volumes, then in fact \textit{all} divisors have integer volumes, again up to an overall factor.  In such a case, just as in the toy example, the volume of an autochthonous divisor in the vacuum is
$k\times \log(W_0^{-1})/2\pi$, with $k\in \mathbb{Z}$. As the Calabi-Yau threefold is smooth inside the K\"ahler cone we have $k>0$. Euclidean D3-branes on autochthonous divisors with $k>1$ are exponentially negligible.  Morevoer, it is easy to show that neglecting Euclidean D3-branes on autochthonous divisors with $k=1$ produces an $\mathcal{O}(1)$ error in the vevs of the K\"ahler moduli. As these are of order $\log(W_0^{-1})\gg 1$, omitting $k=1$ autochthonous divisors produces an error only at subleading order in $\log(W_0^{-1})^{-1}\ll 1$.

This very general argument shows that autochthonous divisors can never make parametrically large contributions to the superpotential in our vacua.  However, we have also constructed  a class of autochthonous divisors that can be found from polytope data \cite{autochthonousinprogress}:  all such divisors turn out to be \emph{very} large in our examples, no less than 100 times larger than the leading prime toric divisors, and so can be completely neglected.

\section{K\"ahler potential and K\"ahler coordinates}\label{sec:control}

In the preceding sections, we have detailed a process for constructing flux compactifications on Calabi-Yau orientifolds in which the superpotential takes the form \eqref{kkltw}, with at least $h^{1,1}$ nonperturbative superpotential terms, all with constant Pfaffians.  We will call such a configuration a \emph{compactification with KKLT superpotential}.

A nontrivial question is whether such a compactification actually admits a supersymmetric $\mathrm{AdS}_4$ vacuum: specifically, does there exist a point in the K\"ahler moduli space of $X$ where the F-terms of all $h^{1,1}$ K\"ahler moduli vanish and the $\alpha^{\prime}$ expansion is well-controlled?
This question hinges on the form of the K\"ahler potential $\mathcal{K}_{K,\tau}$
for the K\"ahler moduli $T_i$ and the axiodilaton $\tau$,
to which we now turn.

At tree level we have that
\begin{equation}\label{eq:treelevel_Kahler_and_T}
\mathrm{exp}\bigl(-\mathcal{K}_{K,\tau}/2\bigr)\bigr|_{\text{tree}}\propto e^{-2\phi}\mathcal{V}_{\text{st}}(t)\, ,\quad \text{Re}(T_i)|_{\text{tree}}=e^{-\phi}\frac{1}{2}\kappa_{ijk}t^jt^k\, ,
\end{equation}
where
\begin{equation}\label{eq:volclass_def}
\mathcal{V}_{\text{st}} := \frac{1}{6}\kappa_{ijk}t^it^jt^k = \text{Im}(\tau)^{-\frac{3}{2}} \mathcal{V}_E
\end{equation}
is the string-frame volume of the Calabi-Yau threefold $X$, cf.~\eqref{eq:kis}, and $T_i$ are the holomorphic K\"ahler coordinates.
Here, $e^{-2\phi}\mathcal{V}_{\text{st}}$ is simply the four-dimensional dilaton obtained by dimensionally reducing the ten-dimensional Einstein-Hilbert term in string frame, and
$e^{\phi}\equiv g_s$.

The K\"ahler potential $\mathcal{K}_{K,\tau}$ receives perturbative and nonperturbative corrections in the $\alpha'$ and $g_s$ expansions.
In particular, nonperturbative corrections arise from Euclidean D(-1)-branes; worldsheet instantons, and more generally Euclidean $(p,q)$ strings, wrapped on two-cycles; Euclidean D3-branes wrapped on four-cycles; and Euclidean $(p,q)$ fivebranes
wrapped on the Calabi-Yau threefold $X$.  We can write
\begin{align}
&\mathrm{exp}\bigl(-\mathcal{K}_{K,\tau}/2\bigr)=\frac{1}{g_s^2} \sum_{k=0}^\infty g_s^k\,\mathcal{V}^{[k]}(t,z)+\mathcal{O}\left(e^{-2\pi/g_s}\right)\, ,\label{eq:kexp} \\
&T_i=-i\left(\int_X C_4\wedge \omega_i-\frac{\chi(D_i)}{24}C_0\right)+\frac{1}{g_s}\sum_{k=0}^\infty g_s^k \mathcal{T}^{[k]}_i(t,z)+\mathcal{O}\left(e^{-2\pi/g_s}\right)\, ,\label{eq:texp}
\end{align}
where each of the $\mathcal{V}^{[k]}$ and $\mathcal{T}^{[k]}_i$ enjoys a separate $\alpha'$ expansion.\footnote{For a related discussion of the perturbative expansion in $g_s$ and $\alpha'$, see \cite{Burgess:2020qsc}.}
In particular, we have
\begin{equation}\label{eq:volclass}
\mathcal{V}^{[0]} = \mathcal{V}_{\text{st}}  + \Delta \mathcal{V}^{[0]}\,,
\end{equation} where $\Delta \mathcal{V}^{[0]}$ encodes perturbative corrections in the $\alpha'$ expansion, as well as nonperturbative corrections from worldsheet instantons, all at \emph{string tree level}, which will be given in \eqref{eq:Kahlerpotential_coordinates} below.

\subsection{Corrections at string tree level}

In our class of vacua the string coupling $g_s$ is parametrically small,
\begin{equation}
g_s =  c_\tau^{-1} \cdot \frac{2\pi}{\log \left(W_0^{-1}\right)} \ll 1\,,
\end{equation}
where $c_\tau>1$ is a model-dependent number defined in \eqref{eq:ctau} that is determined by the overall magnitude of the racetrack coefficients $\mathbf{p}\cdot \mathbf{q}_i$ in \eqref{eq:W0_formula}, and is usually $\mathcal{O}(1)$.
Therefore, at least for sufficiently small $W_0$, we may restrict ourselves to the $k=0$ terms in \eqref{eq:kexp} and \eqref{eq:texp}. However, as \textit{Einstein frame} four-cycle volumes will turn out to also scale as $\log \left(W_0^{-1}\right)/2\pi$, the \textit{string frame} four-cycle volumes do not become large in the limit of small $W_0$.
Thus the $\alpha'$ expansion of $\mathcal{V}^{[0]}$ is not in general well-approximated by the lowest-order term $\mathcal{V}_{\text{st}}$, and likewise for $\text{Re}(T_i)$. This, however, does not pose an insurmountable problem, as we now explain.

The perturbative contributions to $\mathcal{V}^{[0]}$ come from $\alpha'$ corrections to the ten-dimensional effective action, more specifically the NS-NS sector,
\begin{equation}
S_{\mathrm{IIB}}^{\mathrm{NS-NS}}=\frac{2\pi}{\ell_s^8}\int d^{10}x\sqrt{-g}\, e^{-2\phi}\left(R+4(\del \phi)^2 -\frac{1}{2}|H_3|^2+...\right)\, ,
\end{equation}
where $...$ stands for terms with more than two derivatives in the metric, dilaton and two-form. Corrections to the effective action from brane sources (i.e.~open strings)
and from the R-R sector are dressed by a relative suppression factor of $g_s$, so these effects can contribute only to $\mathcal{V}^{[k]}(t,z)$ with $k\geq 1$. To see this one recalls that the K\"ahler potential at closed string tree level can be computed from a worldsheet CFT two-point function on the sphere, which is dressed by $g_s^{-\chi(S^2)}=g_s^{-2}$, while orientifolding introduces open strings whose tree level contribution to the K\"ahler potential comes from a disk amplitude, dressed by a factor $g_s^{-\chi(\text{disk})}=g_s^{-1}$: see~e.g.~\cite{Polchinski:1998rr,Berg:2004ek,Berg:2005ja}. Equivalently, one recovers this from the fact that both D-brane and O-plane tensions in string frame are proportional to $g_s^{-1}$. As usual, string loop corrections are even more suppressed, as they come from torus, annulus, Klein bottle and M\"obius strip amplitudes, all of which have $\chi=0$.

In our solutions the dilaton is constant, $\del \phi = 0$, so no correction proportional to $\del \phi$ contributes to the K\"ahler potential.
Finally, in our solutions we also have that \cite{Giddings:2001yu}
\begin{equation}
\star H_3=g_s F_3\, ,\quad \Rightarrow \quad \int_X d^6 y\sqrt{g}\,\frac{1}{2}|H_3|^2=\frac{g_s}{2}\int_X H_3\wedge F_3=g_sN_{\mathrm{D3}}^{\text{flux}}\, ,
\end{equation}
and as a consequence corrections from fluxes can be neglected if $g_sN_{\mathrm{D3}}^{\text{flux}}$ is suitably small.

In conclusion, at $\mathcal{O}(g_s^{-2})$ we are left with pure curvature corrections, as the effect of fluxes and orientifolding (open strings) are invisible at this order. Thus, all sources of breaking from $\mathcal{N}=2$ to $\mathcal{N}=1$ supersymmetry contribute only to $\mathcal{V}^{[k]}$ with $k\geq 1$, and we can therefore accurately compute the curvature corrections assuming eight unbroken supercharges!
In fact, these corrections are known \textit{exactly}, at least in principle, by virtue of mirror symmetry. A quick way to get to the result is to compare with the mirror dual type IIA O6 orientifold \cite{Grimm:2004ua,DeWolfe:2005uu} of the mirror $\widetilde{X}$, which is well-known to geometrize worldsheet instantons of $X$. We have
\begin{equation}\label{eq:KahlerV0}
\mathcal{V}^{[0]}(t,z)\equiv \mathcal{V}^{[0]}(t)=-\frac{i}{8}\int_{\widetilde{X}}\tilde{\Omega} \wedge \overline{\tilde{\Omega}}\, ,
\end{equation}
where $\widetilde{X}$ is the mirror threefold, and $\tilde{\Omega}$ is the holomorphic three-form of $\widetilde{X}$ normalized such that the fundamental period around LCS is equal to unity.\footnote{Note that the part of the $\mathcal{N}=1$ K\"ahler potential for the K\"ahler moduli that is inherited from the $\mathcal{N}=2$ parent theory is related to the hyper-K\"ahler potential of the hypermultiplet sector of type IIB on $X$, which via the c-map is related to the K\"ahler potential for the vector multiplets of type IIA on $X$ \cite{Rocek:2005ij,Robles-Llana:2006vct,Neitzke:2007ke}.  This is of course consistent with \eqref{eq:KahlerV0}.} Moreover, also by comparing to the mirror dual O6 orientifold one learns that the $\mathcal{T}_i:=\mathcal{T}_i^{[0]}$ are identified with appropriate periods of $\tilde{\Omega}$.\footnote{The appropriate $\mathbb{Z}$-linear combination of periods is straightforward to identify by matching the polynomial corrections of the periods at LCS with the $\alpha'$-corrected action \cite{Dasgupta:1997cd,Bachas:1999um,Fotopoulos:2001pt} for Euclidean D3-branes wrapped on the basis divisors at large volume.} We may write $\mathcal{V}^{[0]}(t)$ using a prepotential $\tilde{\mathcal{F}}(\tilde{z}^i)$ of the form \eqref{eq:prepotential}, with all geometric quantities of $\widetilde{X}$ replaced by those of $X$, i.e.
\begin{equation}
z^a\rightarrow \tilde{z}^i\equiv it^i+b^i \, ,\quad \tilde{\kappa}_{abc}\rightarrow \kappa_{ijk}\, ,\quad \ldots
\end{equation}
Here, $b^i=\frac{1}{2}\gamma^i$ is the half-integral $B_2$-field in the type IIB duality frame, as defined in \eqref{eq:Bfield}. Specifically, we have\footnote{Strictly speaking the formula \eqref{eq:Kahlerpotential_coordinates} for $\mathcal{T}_i$ holds only if the basis divisors can host Euclidean D3-branes with vanishing gauge-invariant worldvolume flux in our $B_2$ field background, i.e.~if $c_1(D_i)/2-\imath^*_{D_i}B_2$ is integer-valued for all $i$. In our examples we have checked that this is true.}
\begin{align}\label{eq:Kahlerpotential_coordinates}
\mathcal{V}^{[0]}=&\frac{1}{6}\kappa_{ijk}t^it^jt^k-\frac{\zeta(3)\chi(X)}{4(2\pi)^3}\nonumber\\
&+\frac{1}{2(2\pi)^3}
\sum_{\mathbf{q}\in \mathcal{M}(X)}\, \mathscr{N}_{\mathbf{q}}\,\Biggl( \text{Li}_3\Bigl((-1)^{\mathbf{\gamma}\cdot \mathbf{q}}e^{-2\pi \mathbf{q}\cdot \mathbf{t}}\Bigr)+ 2\pi \mathbf{q}\cdot \mathbf{t}\,\,\text{Li}_2\Bigl((-1)^{\mathbf{\gamma}\cdot \mathbf{q}}e^{-2\pi \mathbf{q} \cdot \mathbf{t}}\Bigr)\Biggr)\,, \\
\mathcal{T}_i=&\frac{1}{2}\kappa_{ijk}t^jt^k-\frac{\chi(D_i)}{24}+\frac{1}{(2\pi)^2}\sum_{\mathbf{q}\in \mathcal{M}(X)}q_i\, \mathscr{N}_{\mathbf{q}} \,\text{Li}_2\Bigl((-1)^{\mathbf{\gamma}\cdot \mathbf{q}}e^{-2\pi \mathbf{q}\cdot \mathbf{t}}\Bigr)\, .
\end{align}
The perturbative expansion of $\mathcal{V}^{[0]}$ consists only of the classical term, namely $\mathcal{V}_{\text{st}}$, and the famous BBHL correction \cite{Becker:2002nn} at order $\alpha'^3$.  However, there are infinitely many instanton corrections whose amplitudes are given by the genus-zero Gopakumar-Vafa invariants of $X$ \cite{Gopakumar:1998ii,Gopakumar:1998jq}, which can be computed using mirror symmetry \cite{Candelas:1990rm,Candelas:1993dm,Hosono:1993qy,Candelas:1994hw,Hosono:1994ax}. A comment is in order regarding the range of validity of \eqref{eq:Kahlerpotential_coordinates}. It is tempting to continue the expressions $\mathcal{V}^{[0]}$ and $\mathcal{T}_i$ through flop transitions between topologically distinct phases.
At the transition locus an effective curve $\mathcal{C}$ shrinks to zero volume and one has to distinguish between two qualitatively distinct cases.

First, if $\mathcal{C}$ does \textit{not} intersect any O7-planes, we have $\int_\mathcal{C}B_2=0$,
so one encounters logarithmic branch cuts
\begin{equation}
\frac{\text{Li}_2(e^{-2\pi t})}{(2\pi)^2}= \frac{t}{2\pi}\log(t)+\text{hol.}\, ,\quad \frac{\text{Li}_3(e^{-2\pi t})}{(2\pi)^3}= -\frac{t^2}{4\pi}\log(t)+\text{hol.}\, ,
\end{equation}
emerging at zero curve volume $t$. Thus, upon continuing to negative $t$ one na\"ively picks up a non-vanishing imaginary part, which is puzzling because the $\mathcal{T}_i$ were constructed to give the \textit{real} parts of $g_s T_i$. However, no matter how small the string coupling $g_s$ is, before the point $t=0$ is reached an infinite tower of instanton corrections from Euclidean $(p,q)$-strings with arbitrary $(p,q)$ become unsuppressed, invalidating \eqref{eq:Kahlerpotential_coordinates}.\footnote{One should be able to likewise determine these from $\mathcal{N}=2$ data, i.e.~the hyper-K\"ahler potential of the hypermultiplet sector \cite{Rocek:2005ij}, because fluxes remain dilute and, by assumption, orientifold planes do not intersect $\mathcal{C}$, but we will not do so here.} As an aside,
one often finds that $h^{1,1}_-(D)$ of a rigid divisor $D$ jumps across flop transitions of the above type. This suggests that the dilaton dependence of the Pfaffian $\mathcal{A}_D(z,\tau)$ can likewise jump. We speculate that upon interpolating from one phase to the next, one needs to resum Euclidean $(p,q)$ string corrections to the definition of the K\"ahler coordinates, along the lines of \cite{Robles-Llana:2007bbv}, in such a way that the Euclidean D(-1)-brane corrections to \eqref{eq:Kahlerpotential_coordinates} are modified, and such that
$e^{\mathcal{K}/2}|\mathcal{A}_D(\tau,z)e^{-2\pi T_D}|$ can be evaluated in either phase, with agreement at the flop transition locus.\footnote{Alternatively, one might conclude that the $\mathcal{N}=1$ quasi-moduli space actually \textit{ends} at the flop transition locus, fracturing the Calabi-Yau moduli space into disconnected components. This appears unlikely to us, because it certainly does not occur with $\mathcal{N}=2$ supersymmetry, and the $\mathcal{N}=2 \rightarrow \mathcal{N}=1$ supersymmetry breaking from fluxes and O-planes becomes arbitrarily weak in the conifold region in the limit that the curve shrinks.}  In any event, our analysis does not depend on the resolution of this puzzle.

Second, if $\mathcal{C}$ \textit{does} intersect an O7-plane transversely in a single point, or is contained in the O7-plane with intersection number $-1$, one has $\int_{\mathcal{C}}B_2=\frac{1}{2}$, and contributions from wrapped $(p,q)$-strings are parametrically suppressed at small $g_s$ \textit{except} for $(p,q)=(1,0)$, i.e.~worldsheet instantons. Indeed, in this case there are no branch cuts,\footnote{For an early related observation, see \cite{Aspinwall:1993xz}.}
\begin{equation}
\frac{\text{Li}_2(-e^{-2\pi t})}{(2\pi)^2}= -\frac{1}{48} +\frac{\log(2)t}{2\pi}+\mathcal{O}(t^2)\, ,\quad \frac{\text{Li}_3(-e^{-2\pi t})}{(2\pi)^3}= -\frac{3\zeta(3)}{32\pi^3}+ \frac{t}{48} +\mathcal{O}(t^2) \, ,
\end{equation}
so one can continue \eqref{eq:Kahlerpotential_coordinates} to negative $t$. Moreover, such transitions make sense physically: the divisor wrapped by the O7-plane intersecting $\mathcal{C}$ undergoes a blow-up/blow-down transition and an O3-plane gets absorbed/emitted in a way that preserves D3-brane charge \cite{Denef:2005mm,Carta:2020ohw}. Moreover, upon continuing to large negative $t$ one arrives again at an expression of the form \eqref{eq:Kahlerpotential_coordinates}, but with all geometric quantities replaced by those of the flopped phase, as needed for consistency.  This follows immediately from the polylogarithm identity
\begin{equation}
\frac{\text{Li}_2(-e^{-2\pi t})}{(2\pi)^2}=-\frac{\text{Li}_2(-e^{-2\pi (-t)})}{(2\pi)^2}+\frac{1}{2}t^2-\frac{1}{24}\, ,
\end{equation}
and likewise for $\text{Li}_3(-e^{-2\pi t})$, in beautiful agreement with the well-known transformation properties of $\chi(D_i)$ and $\kappa_{ijk}$ under flop transitions. In accordance with the above, we have \textit{not} found examples where $h^{1,1}_-(D)$ of a rigid divisor jumps across a flop transition of this type.

Let us pause to stress an important point. Suppose we are faced with a series of corrections in the $\alpha'$ expansion, and we seek to know whether their contributions  to the K\"ahler potential ruin the vacuum structure that was computed at leading order.  If the corrections have \emph{unknown} coefficients, then a conservative requirement is that all effective curves should have large volumes in string units: a curve of volume, say, $2\ell_{s}^2$ might be problematic, and it might not.  But when we can actually compute the coefficients of the leading series of corrections, a weaker --- and at the same time, much more precise --- condition suffices: the curve volumes need only lie within the radius of convergence of the series.  Because the leading corrections in \eqref{eq:Kahlerpotential_coordinates} are determined by GV invariants, which we can compute (see \S\ref{ss:GV}), we will be able to ensure control of the $\alpha'$ expansion in this sharper manner: see \S\ref{ss:convergence}.

\subsection{Corrections beyond string tree level}

The leading \emph{additional} correction to \eqref{eq:Kahlerpotential_coordinates}, which is suppressed by a further power of the string coupling, comes from the backreaction of three-form fluxes $F_3$ and $H_3$, and its magnitude is proportional to $g_s^{-1}N_{\mathrm{D3}}^{\text{flux}}$,
as explained above. Similarly, the corrections from D-brane sources are expected to be of order $g_s^{-1}Q$ where $Q$ is the corresponding (locally uncancelled) brane charge. As seven-brane charges are cancelled locally in our vacua, the only relevant corrections come from fluxes and D3-branes, and from the induced D3-brane charge on seven-branes and O3-planes, so the leading correction induced by fluxes is suppressed in comparison to the tree level result by a factor \cite{Giddings:2001yu,Giddings:2005ff}
\begin{equation}
g_{\mathcal{N}=1}:=g_s|N_{\mathrm{D3}}|\, .
\end{equation}
Here $N_{\mathrm{D3}}$ is the D3-brane tadpole, and we have assumed that string frame volumes are all $\mathcal{O}(1)$. More precisely, one should evaluate the D3-brane charge densities along four-cycles
\begin{equation}\label{eq:warping_controlI}
g_{\mathcal{N}=1}^{\omega_i}:=g_s \frac{|N_{\mathrm{D3}}^{\omega_i}|}{\text{Vol}(\omega_i)}= \frac{|N_{\mathrm{D3}}^{\omega_i}|}{\text{Vol}_E(\omega_i)}\, ,
\end{equation}
where $\text{Vol}(\omega_i)$ and $\text{Vol}_E(\omega_i)$ are the string frame and Einstein frame volumes, respectively, of divisors $\omega_i$, and
$N_{\mathrm{D3}}^{\omega_i}$ are the D3-brane charges on $\omega_i$.  The corrections suppressed by the $g_{\mathcal{N}=1}^{\omega_i}$ are the corrections from warping in the ten-dimensional solution \cite{Giddings:2001yu,Giddings:2005ff}, which indeed become negligible when all Einstein frame volumes are large in comparison to the locally uncancelled D3-brane charges.\footnote{See Appendix \ref{app:ds} for further analysis of this point.}
We will impose that $g_{\mathcal{N}=1}^{\omega_i}<1$ in our solutions, and also that
\begin{equation}\label{eq:warping_controlII}
g_{\mathcal{N}=1}^X:=g_s  \frac{|N_{\mathrm{D3}}|}{\mathcal{V}^{\frac{2}{3}} }<1\, ,
\end{equation}
to control the \textit{overall} importance of warping throughout $X$. For fixed $N_{\mathrm{D3}}$ and $N_{\mathrm{D3}}^{\omega_i}$ the control factors $g_{\mathcal{N}=1}^{X,\omega_i}$ scale as $\log(W_0)^{-1}$, so they become parametrically small in the limit of small $W_0$. However, as $N_{\mathrm{D3}}=\mathcal{O}(100)$  in some of our examples, the conditions $g_{\mathcal{N}=1}^{X,\omega_i}<1$ become nontrivial constraints nevertheless, and we will carefully check them.

The fact that the control parameters $g_s$ and $g_{\mathcal{N}=1}^{X,\omega_i}$ are very small in our examples provides strong evidence that our vacua are well-controlled.
One could check this more explicitly by computing the leading string loop corrections to the K\"ahler potential.
In carrying out such a computation --- which is beyond the scope of this work --- special attention should be paid to corrections from curves $\mathcal{C} \cong \mathbb{P}^1$ that are close to undergoing a flop transition, and thus have small volumes in string units.  As explained in the previous section, such curves come in two different classes.

In the first class, $\mathcal{C}$ does not intersect any seven-branes.  In the limit that such a $\mathcal{C}$ shrinks to a point, the local neighborhood of the singular geometry contains no brane sources.  For this reason, and because fluxes are negligible at short distances, the breaking of supersymmetry from $\mathcal{N}=2$ to $\mathcal{N}=1$ becomes arbitrarily weak in the limit of vanishing curve volume.  Thus, corrections to the K\"ahler potential coming from a small curve of this type are simply inherited from the hypermultiplet moduli space metric of the $\mathcal{N}=2$ parent theory.  Such corrections are captured by \eqref{eq:Kahlerpotential_coordinates} at string tree level, while the corrections beyond string tree level are known \cite{Ooguri:1996me,Robles-Llana:2006vct,Robles-Llana:2006hby,Robles-Llana:2007bbv} and can be shown to be negligible in our examples.

In the second class, a stack of seven-branes intersects the shrinking curve $\mathcal{C}$.  Cancellation of Freed-Witten anomalies in such a background requires the existence of a discrete B-field, and the presence of this B-field ensures that the limit of vanishing curve volume is a non-singular locus in the $\mathcal{N}=2$ moduli space.  Thus, there are no important corrections to the K\"ahler potential at string tree level. One easily shows that this result extends to all orders in $g_s$ in the parent $\mathcal{N}=2$ Calabi-Yau compactification.
	
We conclude that for both classes of curves, the string loop corrections to the K\"ahler potential that are \emph{inherited} from  the $\mathcal{N}=2$ parent theory can be neglected in our examples, to all orders in $g_s$.  It remains to consider genuine $\mathcal{N}=1$ corrections to the K\"ahler potential for curves $\mathcal{C}$ that intersect seven-branes.  If such corrections were to diverge at small curve volume, then the vacuum structure that we have described thus far would be endangered.
However, such a divergence would be quite remarkable: as discussed in the previous section, transitioning through the locus of vanishing curve volume appears to make perfect sense even in the $\mathcal{N}=1$ theory with O-planes and D-branes \cite{Denef:2005mm,Carta:2020ohw}, while divergent corrections at small curve volume would remove the transition locus to infinite distance in moduli space.  Nevertheless, an actual proof of the absence of such divergent corrections, say at order $g_s^0$, would be desirable.  We leave this interesting task for future work.

\section{Computational methods}\label{sec:comp}

Having determined the leading data of the effective $\mathcal{N}=1$ supergravity in our ensemble of compactifications, we now discuss the search for supersymmetric vacua therein.

\subsection{Iterative solution}\label{sec:control_Fterms}

We have just established that in our vacua, where $g_s \ll 1$ but string-frame volumes are of order unity,
the K\"ahler potential for the holomorphic K\"ahler moduli $T_i$ is determined by the $k=0$ terms of \eqref{eq:kexp}, which are given in \eqref{eq:Kahlerpotential_coordinates}, and which incorporates perturbative and worldsheet instanton corrections in the $\alpha'$ expansion, at string tree level.
In particular, the dependence of $\mathcal{V}^{[0]}$ on $\mathrm{Re}(T_i)$ is not as simple as in \eqref{eq:treelevel_Kahler_and_T}, which includes only the leading term in \emph{both} the $g_s$ and $\alpha'$ expansions.

Fortunately, the vacuum conditions that arise from the superpotential of \eqref{kkltw} are rather insensitive to the precise form of $\mathcal{V}^{[0]}(\text{Re}\,T)$, and we will be able to iteratively incorporate the small effects of the $\alpha'$ corrections in \eqref{eq:Kahlerpotential_coordinates}.
This is seen as follows.
We have the F-flatness conditions
\begin{equation}\label{eq:Kahler_Fterms}
D_{T_i}W(T)=-\frac{2\pi}{c_i}\mathcal{A}_i\, e^{-\frac{2\pi}{c_i}T_i}-g_s \,\frac{t^i}{2\mathcal{V}^{[0]}}\left(W_0+\sum_j\mathcal{A}_j e^{-\frac{2\pi}{c_j}T_j}\right)\, ,
\end{equation}
where
we have used the fact that the basis of $H_4(X,\mathbb{Z})$ is chosen to be a set of $h^{1,1}$ divisors $D_i$ that contribute to the superpotential via Euclidean D3-branes or gaugino condensation, and we neglect, for now, commensurate contributions from further rigid divisors (cf.~\S\ref{sec:autochthonous}).

Let us define
\begin{equation}
\epsilon^i:=- g_s \mathcal{A}_i^{-1} \frac{t^i}{2\mathcal{V}^{[0]}} \frac{c_i}{2\pi}\,.
\end{equation}
Then, using
\begin{equation}
|\epsilon^i| \lesssim g_s\sim \log(W_0)^{-1}\ll 1\,,
\end{equation}
we can iteratively solve \eqref{eq:Kahler_Fterms} to obtain
\begin{equation}\label{eq:F-term-solution}
T_i=\frac{c_i}{2\pi}\log(W_0^{-1})-\frac{c_i}{2\pi}\log\Biggl[\epsilon^i\Biggl(1+\sum_{j}\mathcal{A}_j \epsilon^j
+\sum_{k, j }\mathcal{A}_{j}\mathcal{A}_k\epsilon^j\epsilon^k+\ldots\Biggr)\Biggr]\, ,
\end{equation}
and one finds a solution\footnote{Note that the axion vevs $\text{Im}(T_i)$ are determined by the complex phases of the $\mathcal{A}_i$ and thus cannot be determined without precise knowledge of the $\mathcal{A}_i$.}
\begin{equation}
T_i=T_i^{(0)}+\delta T_i\, ,\quad T_i^{(0)}:=\frac{c_i}{2\pi}\log(W_0^{-1})
\end{equation}
with a relative correction
\begin{equation}\label{eq:relcorr}
\delta T_i/T_i^{(0)}=\mathcal{O}\left(\frac{\log\left[\log(W_0^{-1})\right]}{\log (W_0^{-1})}\right)\ll 1\, ,
\end{equation}
that vanishes in the limit $W_0\rightarrow 0$, and is of order a few percent in our examples. It is straightforward to perturb \eqref{eq:Kahler_Fterms} by a further commensurate instanton, e.g.~from an autochthonous divisor (cf.~\S\ref{sec:autochthonous}) or another prime toric divisor, to see that the vevs of the $T_i$ get perturbed by at most an $\mathcal{O}(1)$ factor that likewise becomes negligible for small $W_0$.

In the above argument we have imagined following a discrete series of flux vacua leading to smaller $W_0$, and we have implicitly assumed that there is no conspiracy that causes $\frac{t^i}{2\mathcal{V}^{[0]}}$ to scale as
$\log(W_0^{-1})$ in the limit $W_0 \to 0$.
This is well-justified: the values $\frac{t^i}{2\mathcal{V}^{[0]}}$ in a series of vacua are independent of the choice of background fluxes, except through the effects of
the (computable) $\mathcal{O}(1)$ changes in the coefficients $c_\tau$ defined in \eqref{eq:ctau}.
We expect such changes in $c_\tau$ to be independent of the scaling of $W_0$ along a series of vacua, and this is indeed borne out in our examples.

We conclude that a full solution of the F-flatness conditions \eqref{eq:Kahler_Fterms} should exist near the candidate point $T_i\approx \frac{c_i}{2\pi}\log(W_0^{-1})$,
absent a conspiracy in the moduli space metric.\footnote{One further possible conspiracy is that quantum effects might become strong enough to `cap off' moduli space before a candidate point is reached. While conceivable, we find it hard to envision a scenario where no nearby solution with similar properties would exist.}
Even so, we would much prefer to \textit{prove} the existence of a vacuum --- and to compute the vacuum energy and the moduli masses --- by
means of a reliable computation of the K\"ahler potential near such a point.
For this reason, we will select vacua at points where we can compute
the worldsheet instanton corrections to $\mathcal{V}^{[0]}$ and $\text{Re}(T_i)$
rather systematically, and thus accurately compute the K\"ahler potential: see \S\ref{ss:convergence}.

\subsection{Algorithm for F-flat solutions}\label{ss:fflat}

As explained in \S\ref{sec:control_Fterms},
the F-flatness conditions for the K\"ahler moduli are solved for
\begin{equation}\label{fflatsol}
\mathrm{Re}(T_i) \approx \frac{c_i}{2\pi} \mathrm{log}\bigl(W_0^{-1}\bigr)~~\forall i\,.
\end{equation}
We now turn to finding solutions of this form and verifying that they lie in a region of parametric control, where the assumptions that led to \eqref{fflatsol} are justified \textit{a posteriori}.

As a first step, we consider solving \eqref{fflatsol} using the tree-level expression $T_i\rightarrow \frac{1}{g_s}\frac{1}{2}\kappa_{ijk}t^jt^k$ of \eqref{eq:treelevel_Kahler_and_T}. When $h^{1,1} \gg 1$, the K\"ahler cone $\mathcal{K}_X$ typically has exponentially many subcones, or chambers:
for example, flopping a suitable curve in $X$ leads one to a new chamber, where new possibilities for flops may arise.\footnote{See e.g.~\cite{Demirtas:2020dbm} for a recent exploration at large $h^{1,1}$.}
Given a compactification with KKLT superpotential, a randomly chosen triangulation of $\Delta^{\circ}$ will typically correspond to a chamber of $\mathcal{K}_X$ in which there does \emph{not} exist a  supersymmetric $\mathrm{AdS}_4$ vacuum.  We will therefore need to search through the secondary fan to find a triangulation in which such a vacuum does exist.  Because the number of chambers is exponentially large at large $h^{1,1}$, a brute force search would be ineffective.

We will now describe an effective algorithm for finding points in the extended K\"ahler cone $\mathcal{K}_X$ where the basis divisors have the desired values.
First, as both $\log(W_0)$ and $g_s$ enter only as overall factors in the F-term equation \eqref{fflatsol}, we may instead solve the equation $\frac{1}{2}\kappa_{ijk}t^jt^k=c_i$, which is independent of the choice of fluxes. Second, we will illustrate the algorithm in a simplified case in which all dual Coxeter numbers $c_i$ are set to one, but the generalization is immediate.

We wish to find a point in the extended K\"ahler cone where a basis set of $h^{1,1}$ linearly independent prime toric divisors ${ D_i }$, $i=1, \dots, h^{1,1}$, have unit volumes, while the remaining four divisors have larger (or equal) volumes. The first challenge is to identify choices of basis divisors with the property that once their volumes are set to unity, the volumes of the remaining four divisors are strictly positive. This is equivalent to requiring that the constant vector $\tau_\star = (1, \dots, 1)$ is contained in the dual of the cone of effective divisors, $\mathcal{E}(X)^\circ$. The number of possible basis choices --- chosen from subsets of $h^{1,1}$ prime toric divisors that all contribute to the superpotential --- is finite, and is often small enough to allow for a brute force search.

Once an appropriate basis is identified, the next task is to find the K\"ahler parameters $t_\star$ that result in unit divisor volumes $\tau_\star$. One might be tempted to parameterize the extended K\"ahler cone by the divisor volumes $\tau^i$ and aim to prove that it contains $\tau_\star$. However, to the best of our knowledge, there does not exist an algorithm to \textit{construct} the corresponding phase of the Calabi-Yau hypersurface given only $\tau^i$.
In contrast, $t_\star$ naturally corresponds to a point in the secondary fan, from which one can obtain a fine, regular, star triangulation (FRST), defining a toric fan and a Calabi-Yau hypersurface.

We start by picking a random point $h_{\text{init}}$ in the subset of the secondary fan of FRSTs, which we denote by $\mathcal{G}$. Such a point is naturally associated to a point in the extended K\"ahler cone, $t_{\text{init}}$, with basis divisor volumes $\tau_{\text{init}}$. Consider any point on the straight line between $\tau_{\text{init}}$ and $\tau_\star$,
\begin{equation}
\tau_\alpha =  (1-\alpha) \tau_{\text{init}} + \alpha \tau_\star
\end{equation}
parameterized by $0 \leq \alpha \leq 1$. Since $\mathcal{E}^\circ (X)$ is convex, $\tau_{\text{init}} \in \mathcal{E}(X)^\circ$ and $\tau_\star \in \mathcal{E}(X)^\circ$ implies that $\tau_\alpha \in \mathcal{E}(X)^\circ$. Our strategy is to start from a randomly chosen $\tau_{\text{init}}$ and move towards $\tau_\star$ on this line.\footnote{Note that this algorithm can fail to converge in some examples, e.g.~if there is an unknown autochthonous divisor that has negative volume at the candidate point. Conversely, if the algorithm succeeds, all possible autochthonous divisors have positive volume.}

The corresponding path between $t_{\text{init}}$ and $t_\star$ is not straight, since the divisor volumes $\tau (t)$ are quadratic functions of the K\"ahler parameters $t^i$ with coefficients $\kappa_{ijk}$ that jump across phases. However, $\tau (t)$ are continuous and once differentiable, giving rise to a path that is continuous, with no cusps. This enables us to follow the path efficiently.

Then, our final task is to devise a numerical algorithm that starts from $t_{\text{init}}$ and follows the continuous path towards $t_\star$. We first divide the path into $N \gg 1$ small sections, by considering the set of points defined by $\alpha = m/N$, $m = 0, \dots ,N$, denoting the corresponding points in $\mathcal{E}(X)^\circ$ and $\mathcal{K}(X)$ by $\tau_m$ and $t_m$, respectively. Following the path is then reduced to moving from $t_m$ to $t_{m+1}$. Let $t_{m+1} = t_m + \eps$. Then,
\begin{align}
\tau_m^i &= \frac{1}{2} \kappa_{ijk} t_m^j t_m^k , \\
\tau_{m+1}^i &= \frac{1}{2} \kappa_{ijk} (t_m^j + \eps^j) (t_m^k + \eps^k) = \tau_m^i + \kappa_{ijk} t_m^j \eps^k + \mathcal{O} (\eps^2)\,.
\end{align}
Determining $\eps$ then requires solving the linear system
\begin{equation}
\kappa_{ijk} t^j \eps^k = \tau_{m+1}^i -  \tau_m^i .
\end{equation}

Once a point $\frac{1}{2}\kappa_{ijk}t^j t^k=c_i$ has been found, the solution for the $t^i$ must be scaled by a factor
$c_{\tau}^{-\frac{1}{2}}=(\mathbf{p}\cdot \tilde{\mathbf{q}}_2)^{1/2}$ --- see \eqref{eq:ctau} --- and the perturbative and non-perturbative corrections in \eqref{eq:Kahlerpotential_coordinates} need to be incorporated systematically. Clearly, this can only be done inside the radius of convergence of the type IIB worldsheet instanton expansion, where at most finitely many curves contribute sizeable corrections. Assuming a solution exists\footnote{Verifying this assumption is the subject of \S\ref{ss:convergence}.} within the radius of convergence, it can be found iteratively as follows. We start with the zeroth-order solution
\begin{equation}
\frac{1}{2}\kappa_{ijk}t^j_{(0)}t^k_{(0)}=\frac{c_i}{c_\tau}\, .
\end{equation}
Then, we define $t^i_{(n)}$ for $n>0$ recursively as the solution to the quadratic equation
\begin{equation}\label{eq:algorithm_incorporate_instantons}
\frac{1}{2}\kappa_{ijk}t^j_{(n)}t^k_{(n)}=\frac{c_i}{c_\tau}+\frac{\chi(D_i)}{24}-\frac{1}{(2\pi)^2}\sum_{\mathbf{q}\in \mathcal{M}(X)}q_i\, \mathscr{N}_{\mathbf{q}} \,\text{Li}_2\Bigl((-1)^{\mathbf{\gamma}\cdot \mathbf{q}}e^{-2\pi \mathbf{q}\cdot \mathbf{t}_{(n-1)}}\Bigr)\, ,
\end{equation}
as a function of the $t_{(n-1)}^i$. At each $n$ one may approximate the instanton sum by keeping only the terms that make a contribution larger than a fixed small threshold. If a solution exists, one should be able to find it this way, to arbitrary precision, by iterating to high enough $n$.

\subsection{Gopakumar-Vafa invariants}\label{ss:GV}

In order to carry out the algorithm that we have just described,
we need to compute the genus-zero Gopakumar-Vafa invariants $\mathscr{N}$ of $X$.
For a general
threefold $X$ these
can be obtained via mirror symmetry, by computing the expansion of the period vector of the mirror threefold $\widetilde{X}$ around LCS \cite{Candelas:1990rm,Candelas:1993dm,Candelas:1994hw}. The results of \cite{Hosono:1993qy,Hosono:1994ax} can \textit{in principle} be used to do so algorithmically, using publicly-available code \cite{Instanton}, once the triple intersection form of $X$ and a simplicial cone containing the Mori cone $\mathcal{M}(X)$ are in hand.

However, with presently-available software it is not feasible to systematically compute GV invariants in threefolds with $h^{1,1} \gg 10$, and in particular to do so at the high degrees needed for our purposes.  In \cite{compmirror} we will present improved methods that allowed us to compute GV invariants in the regime of interest.
Here we will restrict ourselves to reviewing
a few facts that are relevant for the present work.

First, along certain rational rays $\mathbf{r}$ in the Mori cone, the
GV invariants come in infinite families: for $k \mathbf{q} \in \mathbf{r}$, there are infinitely many $k \in \mathbb{N}$ for which $\mathscr{N}_{k\mathbf{q}} \neq 0$.
We call such a curve class $\mathbf{q}$ a \emph{potent curve}, and we call $\mathbf{r}$ a \emph{potent ray}.
Along such rays, the GV invariants typically grow exponentially: see \S\ref{ss:convergence}.
We denote by $\mathcal{M}_{\infty}(X)$ the closure of the cone over all potent rays.

Second, along special rays, often outside of $\mathcal{M}_{\infty}(X)$, the GV invariants come in isolated sets associated with a curve class $\mathbf{q}$ and a \textit{finite} number of its multiples: we have $\mathscr{N}_{k\mathbf{q}} \neq 0$ for finitely many $k \in \mathbb{N}$ (and often, for only one such $k$).
We call such a curve class $\mathbf{q}$ a \emph{nilpotent curve}, and we call $\mathbf{r}$ a \emph{nilpotent ray}.
Nilpotent curve classes that lie outside of $\mathcal{M}_{\infty}(X)$ are collapsible $\mathbb{P}^1$'s.

The dual of $\mathcal{M}_{\infty}(X)$ contains the extended K\"ahler cone \cite{WGV}, and sufficiently far out in this cone the instanton expansion converges, even if a number of collapsible curves are \textit{arbitrarily} small.  If a candidate solution $\mathbf{t}\in \mathcal{K}_X$ of the F-flatness conditions lies at a point in the (extended) K\"ahler cone where some number of collapsible curves are small, we can simply evaluate the di-logarithms in \eqref{eq:algorithm_incorporate_instantons} to account for these.

Although small collapsible curves are relatively innocuous, small curves in $\mathcal{M}_{\infty}(X)$ are not.  We will need to check that all curves in $\mathcal{M}_{\infty}(X)$ are large: at least, large enough so that at most a few contribute appreciably to the right-hand side of \eqref{eq:algorithm_incorporate_instantons}.  In other words, we will need to ensure that there exists a solution to the F-flatness conditions within the radius of convergence of the instanton expansion.   We now turn to this final check.

\subsection{Radius of convergence}\label{ss:convergence}

The LCS singularity is never the only
singularity in moduli space, so the worldsheet instanton expansion generally has a finite radius of convergence around any point.
Along any fixed direction $\mathbf{t}=\lambda\mathbf{t}_0$ in the K\"ahler cone $\mathcal{K}_X$, with $\lambda >0$, there should exist $\lambda_c>0$ such that
the expansion converges for all $\lambda>\lambda_c$ and diverges for all $\lambda<\lambda_c$.

The coefficients of the instanton terms in the prepotential are governed by the GV invariants $\mathscr{N}_{\mathbf{q}}$,
\begin{equation}
\mathcal{F}_{\text{inst}}(\lambda)\propto \sum_{\mathbf{q}\in \mathcal{M}(X)}\, \mathscr{N}_{\mathbf{q}}\, \text{Li}_3\left(e^{-2\pi\lambda\,  \mathbf{q}\cdot \mathbf{t}_0}\right)\, ,
\end{equation}
and the arguments of the polylogarithm become arbitrarily small far out in the Mori cone. To analyze the asymptotic behavior
we first normalize $\mathbf{t}_0$ such that $d_{\mathbf{q}}:=\mathbf{q}\cdot \mathbf{t}_0\in \mathbb{N}$,
and for $k \in \mathbb{N}$ we define
\begin{equation}
\mathscr{N}_k:= \sum_{\mathbf{q}:\,d_{\mathbf{q}}=k} \mathscr{N}_{\mathbf{q}}\,.
\end{equation}
Then we consider
\begin{equation}
\mathcal{F}_{\Sigma}(\lambda):=\sum_{k=1}^\infty \mathscr{N}_k\, (e^{-2\pi \lambda})^k\,.
\end{equation}
By the ratio test we have
\begin{equation}
e^{2\pi \lambda_c}=\lim_{k\rightarrow \infty}\frac{\mathscr{N}_k}{\mathscr{N}_{k-1}} \, ,
\end{equation}
and thus, GV invariants grow \textit{exponentially} at large degree,
\begin{equation}
\mathscr{N}_k\sim e^{2\pi  \lambda_c k}\, ,\quad \text{as}\quad k\rightarrow \infty\, .
\end{equation} This growth is a consequence of our assumption that the radius of convergence is finite.

Conversely, one can estimate the radius of convergence of the instanton expansion by computing GV invariants.
Although this approach does not give a formal proof of control, being reliant on extrapolation to curves of arbitrarily large degree, it is still rather powerful.
The growth rate of GV invariants as a function of degree has been observed\footnote{See e.g.~\cite{Candelas:1990rm,Klemm:1999gm,Huang:2007sb,Couso-Santamaria:2016vcc}.}
to asymptote very quickly to an exponential rate, which then gives a reliable estimate of the radius of convergence.  For example, in the case of the quintic, the leading estimate is $\lambda_{c}^{(1)} = \frac{1}{2\pi}\mathrm{log}(\mathscr{N}_1)= \frac{1}{2\pi}\mathrm{log}(2875) \approx 1.27$, whereas the actual radius of convergence is $\lambda_c \approx 1.208$ \cite{Candelas:1990rm}.

Though it is in general not feasible to compute GV invariants systematically to high degree at large $h^{1,1}$, due to the sheer number of curve classes, it is possible to compute to very high degree inside low-dimensional faces of the Mori cone. Moreover, by finding an appropriate phase where a given face of $\mathcal{M}_{\infty}(X)$ is also a face of $\mathcal{M}(X)$ one can compute GV invariants in many low-dimensional faces of $\mathcal{M}_{\infty}(X)$.  In this way one can in principle compute the GV invariants along a large number of rays in $\mathcal{M}_{\infty}(X)$, forming a full-dimensional cone, and we can test whether at a candidate point in moduli space the worldsheet expansion truncated to that sector converges.
This approach can never fully \emph{prove} control over the instanton expansion, as in principle there could exist a ray somewhere in the interior of $\mathcal{M}_\infty(X)$ with rapidly growing GV invariants.  However, in examples one usually observes that the growth rate of GV invariants in the interior of $\mathcal{M}_{\infty}(X)$ is a simple interpolation without extrema between the growth rates of the generators of $\mathcal{M}_{\infty}(X)$. Therefore, we \textit{do} expect to be able to estimate control over the instanton expansion by inspecting the curve classes of low-dimensional faces, as we will do in our examples.

Although the above approach will allow us to estimate the contributions of \emph{potent} curves, we will also need to incorporate nilpotent curves, to which we now turn.
Finding lattice points in the Mori cone at large $h^{1,1}$ is a difficult task by itself, and when further restricting the search to curves with non-vanishing GV invariants it becomes seemingly insurmountable.
However, we have devised a method that begins by finding
curves inherited from the toric ambient variety, which by a slight abuse of terminology we refer to as
\emph{toric curves}.  Many such curves turn out to have non-vanishing GV invariants.  In fact, the set of toric curves generally contains the Hilbert basis of the Mori cone in the examples with small $h^{1,1}$ where a fully systematic comparison is possible.  This is extremely helpful for our purposes, because the Hilbert basis contains the smallest (and hence most important) effective curves.

Our approach is then as follows.
From the set of toric curves we pick those with volumes\footnote{For this purpose we scale the K\"ahler parameters homogeneously, corresponding to $c_{\tau}=1$, so that the details of the flux vacuum are immaterial.}
less than, say, 2, which gives us a few hundred curves. We remove the curves that can be written as sums of others, and so are not Hilbert basis elements, and can then compute the GV invariants of the remainder.
In examples at small $h^{1,1}$ we have found that the curves found in this way account for the great majority of curves with non-vanishing GV invariants below the volume threshold in question.

In the examples described below, we are able to systematically compute GV invariants of \emph{all} effective curves with volumes $\lesssim 0.1$, and, among the $\mathcal{O}(100,000)$ curves included, the $\mathcal{O}(10)$ curves with non-vanishing GV invariants are none other than the toric curves!  Along with the fact that \emph{all} of the small toric curves have $\mathcal{O}(1)$ GV invariants, we expect to have captured the most important contributions to the instanton expansion. Thus, we can find the leading few hundred terms instead of the $\mathcal{O}(10)$ that we would have been able to obtain with a more direct approach.

\section{Examples}\label{sec:examples}

Our procedure for constructing vacua can be applied to a very large number of geometries.
In principle the approach is valid for a fair fraction of all the threefolds resulting from the Kreuzer-Skarke list.
However, with present tools the search for flux vacua becomes costly for $h^{2,1} \gtrsim 10$, as the flux lattice dimension is then at least 20.\footnote{Approaches such as those of~\cite{Cole:2019enn,Bena:2021wyr} might aid in finding flux vacua at larger $h^{2,1}$.}
At the same time, explicitly checking convergence of the worldsheet instanton corrections to the K\"ahler potential by computing genus-zero Gopakumar-Vafa invariants to high degrees
becomes expensive for $h^{1,1} \gtrsim 50$, and requires special methods for $h^{1,1} \gg 100$.

In this work we have restricted our attention to polytopes that admit simple orientifolds in which there are at least $h^{1,1}$ pure rigid prime toric divisors.
For $h^{2,1} \le 4$ the search over flux quanta is inexpensive, and one can find hundreds of supersymmetric $\mathrm{AdS}_4$ vacua with $W_0 \lesssim 10^{-10}$ in minutes on a laptop.
Most of the polytopes that we have checked do in fact admit such vacua.

In a few polytopes one can easily find extremely small values, $W_0 \lesssim 10^{-50}$.
In a larger class of polytopes, such values emerge after a more determined search, while in other polytopes we have not yet found such enormous hierarchies.

In this section we present a few illustrative vacua.  Each example is defined by a pair of reflexive polytopes $(\Delta^\circ,\Delta)$ and triangulations defining toric varieties $(V,\widetilde{V})$ and their Calabi-Yau hypersurfaces $(X,\widetilde{X})$, chosen such that our flux vacua lie in the K\"ahler cone of $\widetilde{X}$ and the K\"ahler moduli are stabilized at a point in the K\"ahler cone of $X$.
Each orientifold is defined by negating a toric coordinate $x_i\rightarrow -x_i$, and in all cases $h^{1,1}_-(X)=h^{2,1}_+(X)=0$, so the D3-brane tadpole is equal to $\frac{1}{2}\bigl(h^{1,1}(X)+h^{2,1}(X)\bigr)+1$.
Key data such as Hodge numbers and $W_0$ values are listed below, but as the K\"ahler moduli spaces are high-dimensional, it would be impractical to list full polytope data, intersection numbers, K\"ahler moduli vevs, curve volumes, etc.  These data are all available, in {\tt{CYTools}} format, as supplemental materials associated to the arXiv e-print.

\subsection{Vacuum with $(h^{2,1},h^{1,1})=(5,113)$}\label{sec:main_examples}

We begin with the reflexive polytope $\Delta$ whose vertices are the columns of
\begin{equation}\label{eq:Delta_vertices_main}
\begin{pmatrix}
1  & -3 & -3 & 0  & 0  & 0 &-5 &-2  \\
0 & -2  & -1 & 0  & 0  & 1 &-3 &-1  \\
0 & 0   & -1 & 0  & 1  & 0 & 0 & 1 \\
0 & 0   & 0  & 1  & 0  & 0 &-1 &-1
\end{pmatrix}\, .
\end{equation}
Besides the origin and points interior to facets, $\Delta$ contains one further point interior to a 1-face. The polar dual of $\Delta$, denoted $\Delta^\circ$, has vertices
\begin{equation}\label{eq:DeltaCirc_vertices_main}
\begin{pmatrix}
1  & -1 & -1 & -1 & -1 & -1 & -1 & -1 & -1\\
-1 & -1 &  2 & -1 & -1 & -1 &  2 &  2 &  2\\
-1 & 5  & -1 & -1 & -1 &  5 & -1 &  2 &  2\\
-1 & 9  &  0 & -1 &  3 & -1 & -1 & -1 &  0
\end{pmatrix}\, ,
\end{equation}
and has $108$ further integer points interior to 1-faces and 2-faces. Partial FRST's of $\Delta^\circ$ and $\Delta$ define toric varieties $V$ and $\widetilde{V}$, respectively, and the corresponding generic anti-canonical hypersurfaces define a mirror pair of smooth Calabi-Yau threefolds $X$ and $\widetilde{X}$
with
\begin{equation}
h^{1,1}(X)=h^{2,1}(\widetilde{X})=113\, ,\quad h^{2,1}(X)=h^{1,1}(\widetilde{X})=5\, .
\end{equation}
The threefold $X$ is both $\Delta$-favorable and $\Delta^\circ$-favorable, and so is $\tilde{X}$. We denote by $D_I$, $I=1,\ldots,117$, the prime toric divisors of $X$, with $D_{1},\ldots,D_{9}$ corresponding to the vertices listed in \eqref{eq:DeltaCirc_vertices_main}. Likewise, $\tilde{D}_{\tilde{I}}$, $\tilde{I}=1,...,8$, will denote the prime toric divisors of $\widetilde{X}$ corresponding to the vertices in \eqref{eq:Delta_vertices_main}.

We consider a type IIB O3/O7 orientifold of $X$ defined by the involution of $V$,
\begin{equation}
\widehat{\mathcal{I}}:\,x_1\mapsto -x_1\,.
\end{equation}
A few key properties of this orientifold, independent of the choice of FRST, are:
\begin{itemize}
	\item $h^{1,1}_-(X)=h^{2,1}_+(X)=0$, and thus no geometric moduli are projected out.
	\item There is an O7-plane on the divisor $D_1$ with $h^\bullet(D_1,\mathcal{O}_{D_1})=(1,0,2)$.
	\item There are 25 O7-planes wrapping other prime toric divisors, \textit{all} of which are rigid.
	\item There are 48 O3-planes at the triple intersections of certain prime toric divisors.
	\item The D3-brane tadpole is equal to $\frac{\chi_f}{4}=\frac{1}{2}\bigl(h^{1,1}(X)+h^{2,1}(X)\bigr)+1=60$.
\end{itemize}
As stated in \S\ref{sec:setup}, we cancel the D7-brane tapole locally, so each of the rigid divisors hosting an O7-plane actually hosts a confining $\mathcal{N}=1$ pure Yang-Mills theory with gauge algebra $\mathfrak{so}(8)$, and in the absence of fluxes the divisor $D_1$ hosts an $\mathcal{N}=1$ Yang-Mills theory with the same gauge algebra and two adjoint chiral multiplets.

Our first task will be to find flux vacua of the form described in \S\ref{sec:setup}. The prime toric divisors $\{\tilde{D}_1,\tilde{D}_2,\tilde{D}_3,\tilde{D}_4,\tilde{D}_5\}$ of $\widetilde{X}$ will be our chosen basis of $H_4(\widetilde{X})$, and our basis of curves will be its dual basis. One can now search for flux vacua in any of the LCS cones defined by triangulations of $\Delta$. In a suitable triangulation the triple intersection numbers and second Chern classes are
\begin{align}
&\tilde{\kappa}_{1ab}=\begin{pmatrix}
89 &0 &16 &12 &7  \\
& 0& 0& 0& 0\\
& & 0&3 &0  \\
& & & 0& 3 \\
& & & &  -3\\
\end{pmatrix}\, ,\quad \tilde{\kappa}_{2ab}=\begin{pmatrix}
8& -2& -2& -2  \\
& 0&  1& 0  \\
& &    0& 1 \\
& & &    0\\
\end{pmatrix}\, ,\\
&\tilde{\kappa}_{3ab}=\begin{pmatrix}
0& 0& 0  \\
&  0& 0 \\
& &   0\\
\end{pmatrix}\, ,\quad
\tilde{\kappa}_{4ab}=\begin{pmatrix}
0 & 0 \\
&   0\\
\end{pmatrix}\, ,\quad \tilde{\kappa}_{555}=-1\, ,\quad \tilde{c}_a =\begin{pmatrix}
146 \\
-4  \\
24 \\
24 \\
14
\end{pmatrix}\, ,
\end{align}
where for fixed $a'=1,...,5$ we display only the $\tilde{\kappa}_{a' ab}$ with $a'\leq a \leq b$.

One readily verifies that the flux choice\footnote{This example is also presented in the companion paper \cite{shortpaper}.}
\begin{equation}\label{5113flux}
\mathbf{M}=\begin{pmatrix}
0 & 2 & 4 & 11 & -8
\end{pmatrix}^T\, ,\quad \mathbf{K}=\begin{pmatrix}
8 & -15 & 11 & -2 & 13
\end{pmatrix}^T\, ,
\end{equation}
satisfies the conditions for a perturbatively flat vacuum, along which the dilaton is related to the complex structure moduli via
\begin{equation}\label{5113flat}
\mathbf{z}=\mathbf{p}\,\tau\, ,\quad \mathbf{p}=\begin{pmatrix}
\frac{7}{58} & \frac{15}{58} & \frac{101}{116} & \frac{151}{58} &  \frac{-13}{116}
\end{pmatrix}\, .
\end{equation}
The D3-brane charge in fluxes is equal to $-\frac{1}{2}\mathbf{M}\cdot \mathbf{K}=56$, so there are four mobile D3-branes.

We have computed the GV invariants of $\widetilde{X}$ systematically, and the leading instantons along the perturbatively flat valley have charges $\tilde{\mathbf{q}}_i$ equal to the columns of
\begin{equation}
\begin{pmatrix}
0 & 3 \\
-2 & 0 \\
1 & 0 \\
0 & 0 \\
1 & 1
\end{pmatrix}\,,
\end{equation}
and the corresponding GV invariants are
\begin{equation}\label{5113gv}
\mathscr{N}_{\tilde{\mathbf{q}}_i}=\begin{pmatrix}
-2 & 252
\end{pmatrix}\, .
\end{equation}
The resulting flux superpotential is
\begin{equation}\label{5113wf}
W_{\text{flux}}(\tau)=8\,\zeta \left(-2\,e^{2\pi i \tau \cdot \frac{7}{29}}+252\,e^{2\pi i \tau \cdot \frac{7}{28}}\right)+\mathcal{O}\left(e^{2\pi i \tau \cdot \frac{43}{116}}\right) \, ,
\end{equation} where the constant $\zeta$ was defined in \eqref{eq:zeta}.
The effective K\"ahler potential for the flat direction parameterized by $\tau$ can be approximated near LCS by
\begin{align}
\mathcal{K}_{\text{eff}}\Bigl(\text{Im}(\tau)\Bigr)=&-\log\Bigl(\,\text{Im}(\tau)\Bigr)-\log\left(-i\int_X \Omega\wedge \overline{\Omega}\right)\nonumber\\
=& -4\log \Bigl(\text{Im}(\tau)\Bigr)+\mathcal{K}_0+\mathcal{O}\left(\text{Im}(\tau)^{-3}\right)\, ,
\end{align}
with constant $e^{\mathcal{K}_0}:=\left(\frac{4}{3}\tilde{\kappa}_{abc}p^ap^bp^c\right)^{-1}=1170672/12843563$.
One then finds
\begin{equation}
g_s  \approx \frac{2\pi }{116\log(261/2)} \approx 0.011\,.
\end{equation}
The vev of the flux superpotential is
\begin{equation}\label{5113w02}
W_0\approx  0.526 \times \left(\frac{2}{252}\right)^{29} \approx 6.46\times 10^{-62}\,.
\end{equation}
The exponential smallness of $W_0$ is manifestly a result of the hierarchy in the GV invariants appearing in \eqref{5113gv}, raised to the power of the racetrack exponent $29$ appearing in \eqref{5113wf}.  The GV invariants are of course intrinsic to the geometry, while the $7/28:7/29$ racetrack results from the choice of fluxes in \eqref{5113flux}, through the perturbatively flat direction \eqref{5113flat} that these fluxes leave open.

Our next task is to stabilize the K\"ahler moduli. For this purpose, we first note that the 25 rigid O7-planes support $\mathfrak{so}(8)$ stacks that contribute superpotential terms with dual Coxeter numbers $c_i\equiv c_2(\mathfrak{so}(8))=6.$
There are an additional 91 rigid prime toric divisors, of which 83 are pure regardless of triangulation, while 8 have the property that $h^{2,1}(\widehat{D})$ depends on the triangulation.\footnote{This issue will not arise in the further examples presented below: the leading prime toric divisors there will be pure and rigid in all triangulations.} These 8 require careful examination.

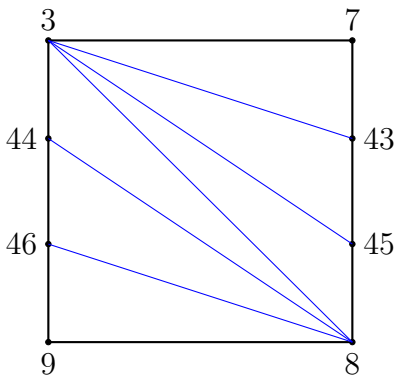
\begin{figure}
\centering
\begin{tikzpicture}
\draw[thick] (-2,2)--(2,2);
\draw[thick] (-2,-2)--(2,-2);
\draw[thick] (-2,-2)--(-2,2);
\draw[thick] (2,-2)--(2,2);
\filldraw[black] (-2,2) circle (1pt) node[left,above] {3};
\filldraw[black] (-2,-2) circle (1pt) node[left,below] {9};
\filldraw[black] (2,2) circle (1pt) node[right,above] {7};
\filldraw[black] (2,-2) circle (1pt) node[right,below] {8};
\filldraw[black] (2,0.7) circle (1pt) node[right] {43};
\filldraw[black] (2,-0.7) circle (1pt) node[right] {45};
\filldraw[black] (-2,0.7) circle (1pt) node[left] {44};
\filldraw[black] (-2,-0.7) circle (1pt) node[left] {46};

\draw[blue] (-2,2)--(2,-2);
\draw[blue] (-2,2)--(2,0.7);
\draw[blue] (-2,2)--(2,-0.7);
\draw[blue] (-2,0.7)--(2,-2);
\draw[blue] (-2,-0.7)--(2,-2);
\end{tikzpicture}
\caption{A triangulation of $\Theta^{(2)}.$}\label{fig:triang}
\end{figure}

The divisors in question are $D_3$, $D_7$, $D_8$, $D_9$, $D_{43}$, $D_{44}$, $D_{45}$, and $D_{46}$, and correspond to points $(3,7,8,9,43,44,45,46)$
in a 2-face $\Theta^{(2)} \subset \Delta^{\circ}$ with $g(\Theta^{(2)})=1$.
The points $(3,7,8,9)$ are vertices of $\Delta^\circ$, while $(43,44,45,46)$ are interior to 1-faces: see Figure \ref{fig:triang}.

Let us define $n(i)$ to be the number of lines interior to $\Theta^{(2)}$ that are connected to the point $i$. Then we compute \cite{hodge},
\begin{equation}\label{eq:vertform}
h^{2,1}(\widehat{D}_3)=1+n(3)\,,~h^{2,1}(\widehat{D}_7)=n(7)\,,~h^{2,1}(\widehat{D}_8)=n(8)\,,~h^{2,1}(\widehat{D}_9)=n(9)\,,
\end{equation}
and for any point $i \in(43,44,45,46)$ we obtain
\begin{equation}\label{eq:edgeform}
h^{2,1}(\widehat{D}_{i})=n(i)-1\,.
\end{equation}

The triangulation depicted in Figure \ref{fig:triang} corresponds to a phase in which we have found a solution to the F-flatness conditions for the K\"ahler moduli.  In this phase, of the eight divisors corresponding to points in $\Theta^{(2)}$, only $D_7$,  $D_9$, $D_{43}$, $D_{44}$, $D_{45}$, and $D_{46}$ support \emph{leading} contributions to the nonperturbative superpotential.
In particular, in this phase the volumes
of $D_3$ and $D_8$ exceed the volumes of the leading contributors by a factor $\sim 30$, so any potential instantons from $D_3$ or $D_8$ would be completely negligible.
Comparing Figure \ref{fig:triang} to \eqref{eq:vertform} and \eqref{eq:vertform}, we see that $D_7$,  $D_9$, $D_{43}$, $D_{44}$, $D_{45}$, and $D_{46}$ are all \emph{pure} in this phase.
Thus we have $25+83+6=114$ superpotential terms with constant Pfaffians, all of which make commensurate contributions to the potential for the K\"ahler moduli, and omitting all divisors that are not pure and rigid is self-consistent.

In sum, taking the triangulation of $\Theta^{(2)}$ shown in Figure \ref{fig:triang}, we have specified a \emph{compactification with KKLT superpotential}, as defined in \S\ref{sec:control}, and have found a supersymmetric AdS$_4$ vacuum therein.

At the corresponding point $\mathbf{t}_{\star}$ in K\"ahler moduli space,
the volume of $X$ in string units is $\mathcal{V}^{[0]}_{\text{st}}\approx 945.18$,
while the Einstein-frame volume of $X$ is $\mathcal{V}_E = \mathcal{V}^{[0]}_{\text{st}} g_s^{-3/2} \approx 8.1 \times 10^5$. The divisors supporting the leading Euclidean D3-branes have Einstein-frame volumes $\approx 22$, while the divisors hosting gaugino condensates are six times larger.

\begin{figure}
	\centering
	\includegraphics[keepaspectratio,width=17cm]{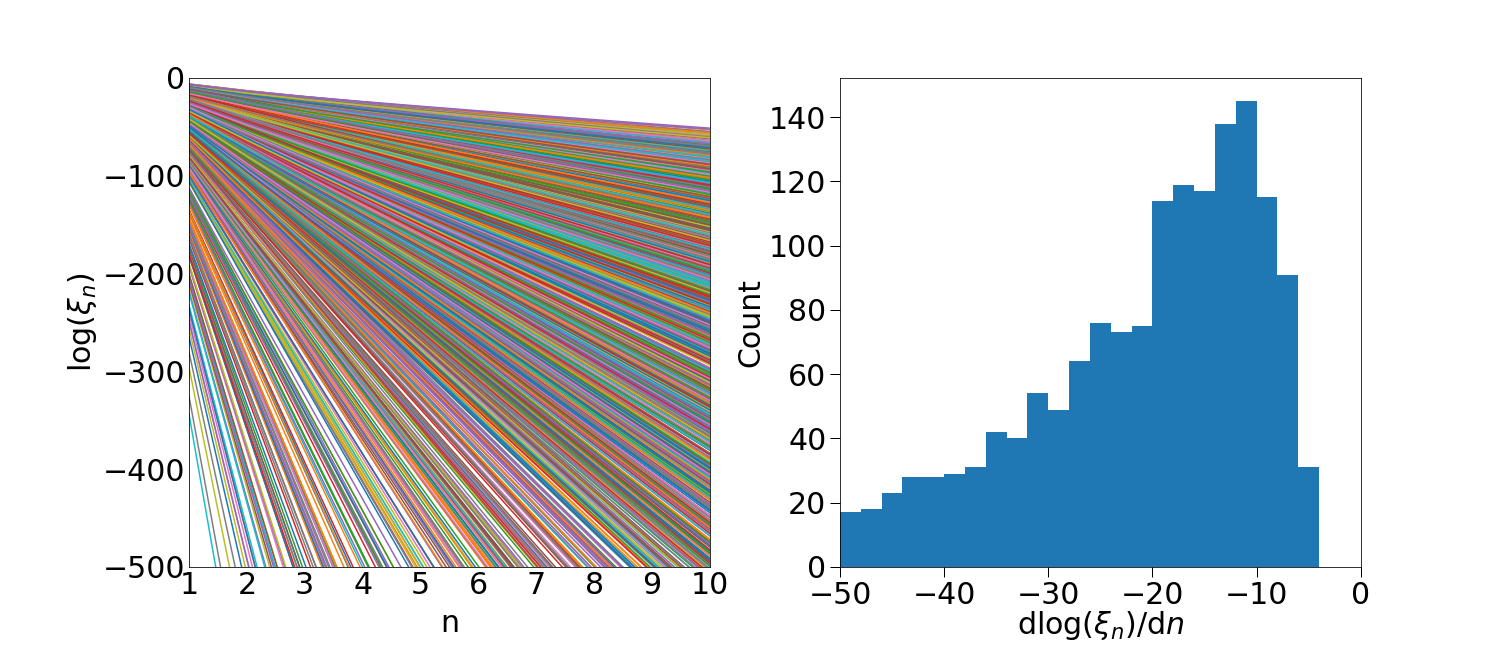}
	\caption{Convergence of worldsheet instanton sum for $(h^{2,1},h^{1,1})=(5,113)$.
Left: We plot the log-magnitude $\mathrm{log}(\xi_n)$, cf.~\eqref{eq:xi}, of the $n$-th term in the instanton series associated with a sample of 1728 potent rays in $\mathcal{M}_\infty(X)$, spanning a 101-dimensional cone. Right: a histogram of the slopes of $\mathrm{log}(\xi_n)$ with respect to $n$ for the set of potent rays. It is apparent that the sum converges.}
	\label{fig:convergence_5-113-4627}
\end{figure}

We now examine the volumes of curves at $\mathbf{t}_{\star}$.  There are 238 curves that are complete intersections of toric divisors and have volumes $\leq 1$, and we have incorporated these curves
in \eqref{eq:algorithm_incorporate_instantons}.
Moreover, by computing GV invariants systematically we have determined that these 238 curves include \textit{all} the effective curves with volume $\leq 0.05$
that contribute to the K\"ahler potential and the definition of the holomorphic coordinates \eqref{eq:Kahlerpotential_coordinates}. Based on the distribution of curve volumes, we expect not to have missed curves with volumes $\lesssim 0.5$.
As
\begin{equation}
\frac{\text{Li}_2(e^{-\pi })}{(2\pi)^2}\approx  0.0011\, ,
\end{equation}
we thus understand all relevant contributions to \eqref{eq:Kahlerpotential_coordinates} from worldsheet instantons, provided that our solution point is in fact inside the radius of convergence of the instanton expansion. While we cannot compute the GV invariants along \textit{all} rays in $\mathcal{M}_\infty(X)$ in a completely systematic manner, we have found $1728$ random rays inside low-dimensional faces of $\mathcal{M}_\infty(X)$, spanning a $101$-dimensional cone, and computed their GV invariants to very high degree. For each such ray, we clearly see that the associated series of worldsheet instanton corrections \textit{converges} and is negligible overall. This is shown in Figure \ref{fig:convergence_5-113-4627}, where we plot the quantity
\begin{equation}\label{eq:xi}
\xi_n := \mathscr{N}_{n\mathbf{q}}\, e^{-2\pi n\, \mathbf{q}\cdot \mathbf{t}}\,,
\end{equation} on a log scale.

Indeed, the smallest \textit{potent} curve $\mathcal{C}_{\mathrm{min}}$ in $\mathcal{M}_{\infty}(X)$ has
\begin{equation}
t_{\text{min}}\approx 1.19\, ,\quad \mathscr{N}=3\,~~~~\text{and contributes} \quad  3\cdot\frac{\text{Li}_2(e^{-2\pi \cdot 1.19})}{(2\pi)^2}\approx  4.3 \times 10^{-5}\, .
\end{equation}

To illustrate the asymptotic behavior we select a potent curve $\mathcal{C}'$ and compute GV invariants along the corresponding ray.\footnote{For reference, the corrected volume of $\mathcal{C}'$ is $2.01$.}
The GV invariants of $\mathcal{C}',2\mathcal{C}',\ldots 10\mathcal{C}'$ are

\begin{tabular}{l}
		$\phantom{-}3$\\
		$-6$\\
		$\phantom{-}27$\\
		$-192$\\
		$\phantom{-}1695$\\
		$-17064$\\
		$\phantom{-}188454$\\
		$-2228160$\\
		$\phantom{-}27748899$\\
		$-360012150$\,.
\end{tabular}

\noindent Skipping ahead, the GV invariant of $100\,\mathcal{C}'$ is

$~~\seqsplit{-91461158123783137122697397476857357418750633461367914322579026697369512751047337367692277761351484717813209296148860000}\,.$

\FloatBarrier

The growth rate with degree is evidently exponential, and the computation out to $100\,\mathcal{C}'$ shows that the rate is very stable.
We therefore have high confidence in assessing the impact of the curves in our sample.

We also note that perturbative corrections in $\alpha'$, and worldsheet instanton corrections, have negligible effect on the F-term of the dilaton $D_{\tau}W$, because
\begin{equation}
\del_{\tau}K=ig_s\times \left(2-\frac{\mathcal{T}_it^i}{2\mathcal{V}^{[0]}}\right)\approx 0.0056i\, ,
\end{equation}
scales as $g_s\approx 0.01 \ll 1$. Thus, even after accounting for nontrivial $\alpha'$ corrections to the K\"ahler potential, we may approximate $D_{\tau}W$ by $\del_{\tau}W$. Overall, worldsheet instantons affect the K\"ahler potential marginally,
\begin{equation}
\mathcal{V}^{[0]}\equiv \mathcal{V}^{[0]}_{\text{st}}+\delta \mathcal{V}^{[0]}\, ,\quad \mathcal{V}^{[0]}_{\text{st}}\approx 945.18\, ,\quad \delta \mathcal{V}^{[0]}\approx-0.23\, ,
\end{equation}
at the solution of the F-term equations. Furthermore, the parameters $g_{\mathcal{N}=1}^{X,\omega_i}$ defined in \eqref{eq:warping_controlI} and \eqref{eq:warping_controlII}, which measure the strength of unknown $\mathcal{N}=1$ corrections to the K\"ahler potential, are indeed small,
\begin{equation}
g_{\mathcal{N}=1}^X\approx  0.0069\,, \quad \max_i\left( g_{\mathcal{N}=1}^{\omega_i}\right)\approx 0.014\,.
\end{equation}

Because $W_0$ is very small, the arguably largest sub-leading correction to our computation of the K\"ahler moduli expectation values, cf.~\eqref{eq:relcorr}, is also small,
\begin{equation}
\frac{\log\left[\log(W_0^{-1})\right]}{\log (W_0^{-1})}\approx 0.04\, .
\end{equation}
Thus, at last, we have found a controlled supersymmetric $\mathrm{AdS}_4$ vacuum, with vacuum energy
\begin{equation}
V_0=-3e^{\mathcal{K}}|W|^2\approx -3e^{\mathcal{K}}W_0^2\approx -3e^{\mathcal{K}_0} \frac{g_s^7}{(4\mathcal{V}^{0})^2}\cdot W_0^2\approx -1.68 \times 10^{-144}M_{\mathrm{pl}}^4\, .
\end{equation}
\medskip

\newpage
\noindent {\textbf{A second flux vacuum with $(h^{2,1},h^{1,1})=(5,113)$}}\\

\begin{figure}
	\centering
	\includegraphics[keepaspectratio,width=17cm]{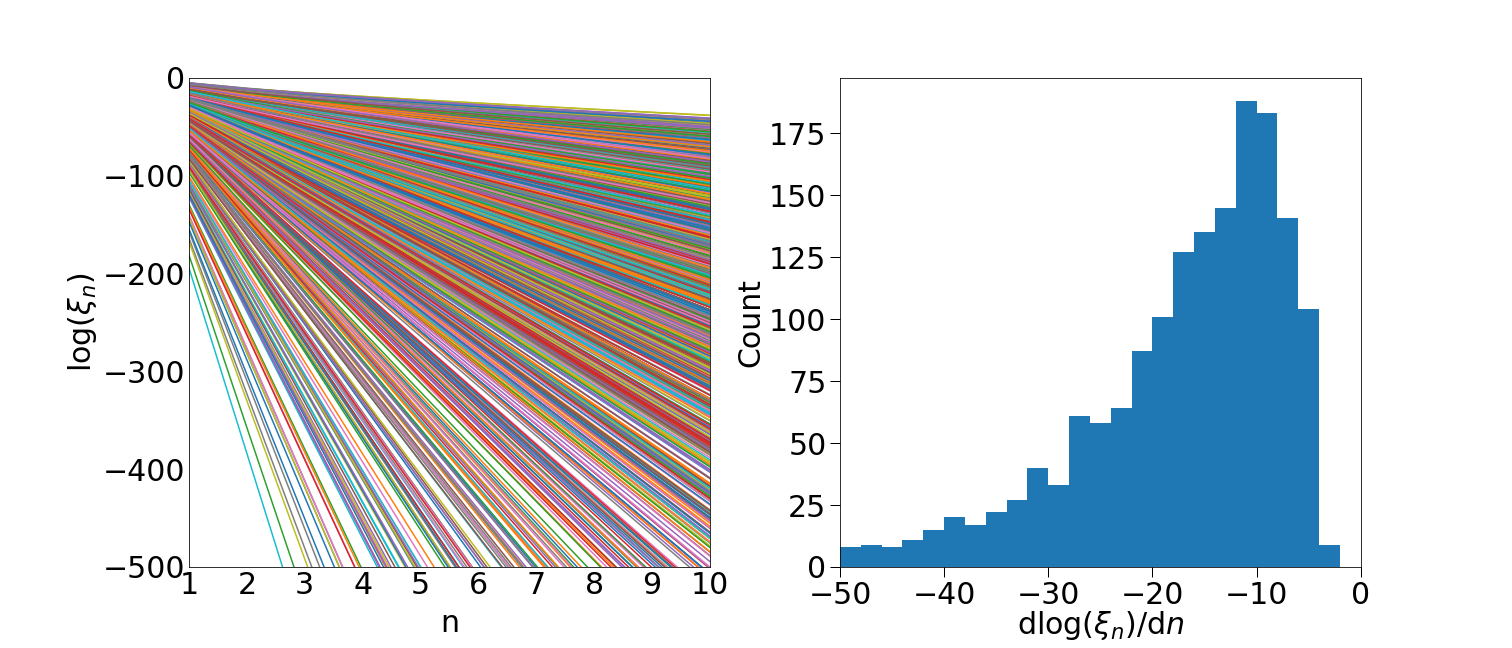}
	\caption{Convergence of worldsheet instanton sum for the second vacuum in $(h^{2,1},h^{1,1})=(5,113)$.
Left: We plot the log-magnitude $\mathrm{log}(\xi_n)$, cf.~\eqref{eq:xi}, of the $n$-th term in the instanton series associated with a sample of 1728 potent rays in $\mathcal{M}_\infty(X)$, spanning a 101-dimensional cone. Right: a histogram of the slopes of $\mathrm{log}(\xi_n)$ with respect to $n$ for the set of potent rays. It is apparent that the sum converges, but the instanton series decays more slowly towards large degree in comparison to the first flux vacuum in $(h^{2,1},h^{1,1})=(5,113)$, cf.~Figure \ref{fig:convergence_5-113-4627}.}
	\label{fig:convergence_5-113-4627_smallerW0}
\end{figure}

Let us now consider a different choice of flux vectors in the same geometry,
\begin{equation}
\mathbf{M}=\begin{pmatrix}
0 & 2 & 4 & 13 & -8
\end{pmatrix}^T\, ,\quad \mathbf{K}=\begin{pmatrix}
0 & -14 & 9 & -1 & 10
\end{pmatrix}^T\, ,
\end{equation}
which again satisfy the conditions for a perturbatively flat vacuum. For this new choice we obtain
\begin{equation}
\mathbf{p}=\begin{pmatrix}
\frac{9}{70} & -\frac{1}{140} & \frac{141}{280} & \frac{81}{40} &  -\frac{73}{280}
\end{pmatrix}\,,\quad -\frac{1}{2}\mathbf{M}\cdot \mathbf{K}=\frac{83}{2}\, ,
\end{equation}
so there are 17 mobile D3-branes and a single `half' D3-brane. The leading instantons along the perturbatively flat valley have charges $\tilde{\mathbf{q}}_i$ equal to the columns of
\begin{equation}
\begin{pmatrix}
1 & 3\\
1 & 0\\
0 & 0\\
0 & 0\\
0 & 1
\end{pmatrix}
\end{equation}
and their GV invariants are
\begin{equation}
\mathscr{N}_{\tilde{\mathbf{q}}_i}=\begin{pmatrix}
-2 & 252
\end{pmatrix}\, .
\end{equation}
The resulting flux superpotential is
\begin{equation}
W_{\text{flux}}(\tau)=\zeta \left(4\,e^{2\pi i \tau \cdot \frac{34}{280}}+2016\,e^{2\pi i \tau \cdot \frac{35}{280}}\right)+\mathcal{O}\left(e^{2\pi i \tau \cdot \frac{9}{70}}\right) \, ,
\end{equation}
and one finds $e^{\mathcal{K}_0}=5488000/20186543$ and
\begin{equation}
g_s \approx 0.0036 \, .
\end{equation}
The vev of the flux superpotential is
\begin{equation}
W_0\approx  \frac{1008}{17}\times \zeta \times \left(\frac{8820}{17}\right)^{-35}   \approx 1.13\times 10^{-95}\, .
\end{equation}
In going from the previous flux vacuum to this one, all that changes in the superpotential is the value of $W_0$. Moreover, string frame volumes are stabilized at different values because the value of $c_\tau$ has slightly increased. We find
\begin{equation}
\mathcal{V}^{[0]}\equiv \mathcal{V}^{[0]}_{\text{st}}+\delta \mathcal{V}^{[0]}\, ,\quad \mathcal{V}^{[0]}_{\text{st}}\approx 388.70\, ,\quad \delta \mathcal{V}^{[0]}\approx-0.25\, ,
\end{equation}
and the Einstein-frame volume of $X$ is $\mathcal{V}_E \approx 1.8 \times 10^6$. Convergence of the instanton sum can be seen in Figure \ref{fig:convergence_5-113-4627_smallerW0}.  Finally, the vacuum energy is
\begin{equation}
V_0=-3e^{\mathcal{K}}|W|^2\approx -3.31 \times 10^{-214} M_{\mathrm{pl}}^4\, .
\end{equation}

\subsection{Vacuum with $(h^{2,1},h^{1,1})=(7,51)$}

The vertices of $\Delta$ are the columns of
\begin{equation}\label{eq:Delta_vertices_7-51}
\begin{pmatrix}
1 & 1   & -2 & -2 & 0  &-2 & 0 & 0  \\
0 & 0   & -1 & -1 & 1  & 1 & 0 & 0  \\
0 & 1   & -1 & 1  & 0  &-1 & 0 & 1 \\
0 & 1   & 1  & -1 & 0  &-1 & 1 & 0
\end{pmatrix}\, .
\end{equation}
In this example there are $h^{1,1}+2=53$ rigid prime toric divisors $D_I\subset X$ with $h^{2,1}(\widehat{D}_I)=0$.

The D3-brane tadpole is $30$, and a suitable flux choice is
\begin{equation}
\mathbf{M}=\begin{pmatrix}
4 & 4 & 0 & -3 & 2 & 0 & -2
\end{pmatrix}^T\, ,\quad \mathbf{K}=\begin{pmatrix}
-4 & -4 & -3 & 2 & -3 & 3 & 3
\end{pmatrix}^T\, ,
\end{equation}
leading to a perturbatively flat vacuum where
\begin{equation}
\mathbf{z}=\mathbf{p}\,\tau\, ,\quad \mathbf{p}=\begin{pmatrix}
\frac{13}{6} & \frac{1}{3} & -\frac{2}{3} & 1 &  \frac{7}{10} & \frac{8}{5} & \frac{11}{10}
\end{pmatrix}\, .
\end{equation}
The D3-brane charge in fluxes is $-\frac{1}{2}\mathbf{M}\cdot \mathbf{K}=25$, so there are five mobile D3-branes.
The leading instantons along the perturbatively flat valley have charges corresponding to the columns of
\begin{equation}
\begin{pmatrix}
0 & 1 & 0 & 1\\
-1 & 1 & 0 & 0 \\
0 & 0 & 0 & 1\\
-1& 1 & 1 & 1 \\
0 & 0 & -1 & 0 \\
1 & -2 & 0 & 0\\
0 & 0 & 0 & -2
\end{pmatrix}\, ,
\end{equation}
and their
GV invariants are
\begin{equation}
\mathscr{N}_{\tilde{\mathbf{q}}}=\begin{pmatrix}
-2 & -4 & 56 & -4
\end{pmatrix}\, .
\end{equation}
The remaining flux superpotential is
\begin{equation}
W_{\text{flux}}(\tau)=\zeta \left(2\,e^{2\pi i \tau \cdot \frac{8}{30}}+320\,e^{2\pi i \tau \cdot \frac{9}{30}}\right)+\mathcal{O}\left(e^{2\pi i \tau \cdot \frac{1}{3}}\right) \, ,
\end{equation}
which stabilizes the dilaton with vev
\begin{equation}
g_s  \approx  \frac{2\pi }{30\log(180)} \approx 0.040\, ,
\end{equation}
and the resulting vev of the flux superpotential is
\begin{equation}
W_0\approx  40\times \zeta \times \left(180\right)^{-9}   \approx 4.1\times 10^{-21}\, .
\end{equation}
We find a solution to the F-flatness conditions for the K\"ahler moduli with $\mathcal{V}^{[0]}\approx 141.4$, and the Einstein-frame volume of $X$ is $\mathcal{V}_E \approx 2.4 \times 10^5$.

\begin{figure}
	\centering
	\includegraphics[keepaspectratio,width=17cm]{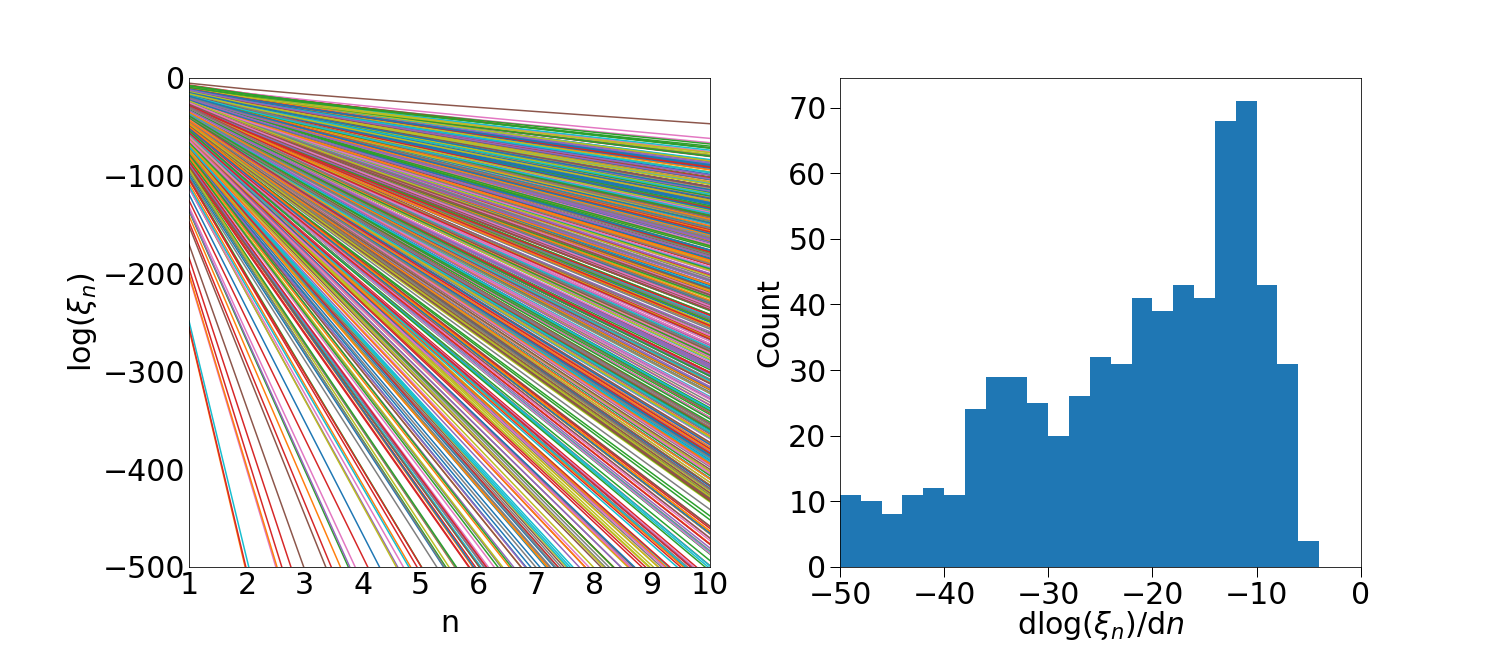}
	\caption{Convergence of worldsheet instanton sum for $(h^{2,1},h^{1,1})=(7,51)$.
Left: We plot the log-magnitude $\mathrm{log}(\xi_n)$, cf.~\eqref{eq:xi}, of the $n$-th term in the instanton series associated with a sample of 758 potent rays in $\mathcal{M}_\infty(X)$, spanning a 48-dimensional cone.  Right: a histogram of the slopes of $\mathrm{log}(\xi_n)$ with respect to $n$ for the set of potent rays. It is apparent that the sum converges.}
	\label{fig:convergence_7-51-13590}
\end{figure}

Convergence of the worldsheet instanton expansion at this point in K\"ahler moduli space is shown in Figure \ref{fig:convergence_7-51-13590}, and the instanton corrections shift $\mathcal{V}^{[0]}$ by $\delta \mathcal{V}^{[0]}\approx -0.1$.
The control parameters defined in \eqref{eq:warping_controlI} and \eqref{eq:warping_controlII} are
\begin{equation} 
g_{\mathcal{N}=1}^X\approx  0.045\,, \quad \max_i\left( g_{\mathcal{N}=1}^{\omega_i}\right)\approx 0.011\,.
\end{equation}

Thus we have a controlled supersymmetric AdS$_4$ vacuum with vacuum energy
\begin{equation}
V_0=-3e^{\mathcal{K}}|W|^2\approx -3.1 \times 10^{-57}M_{\mathrm{pl}}^4\, .
\end{equation}

\subsection{Vacuum with $(h^{2,1},h^{1,1})=(5,81)$}

\begin{figure}
	\centering
	\includegraphics[keepaspectratio,width=17cm]{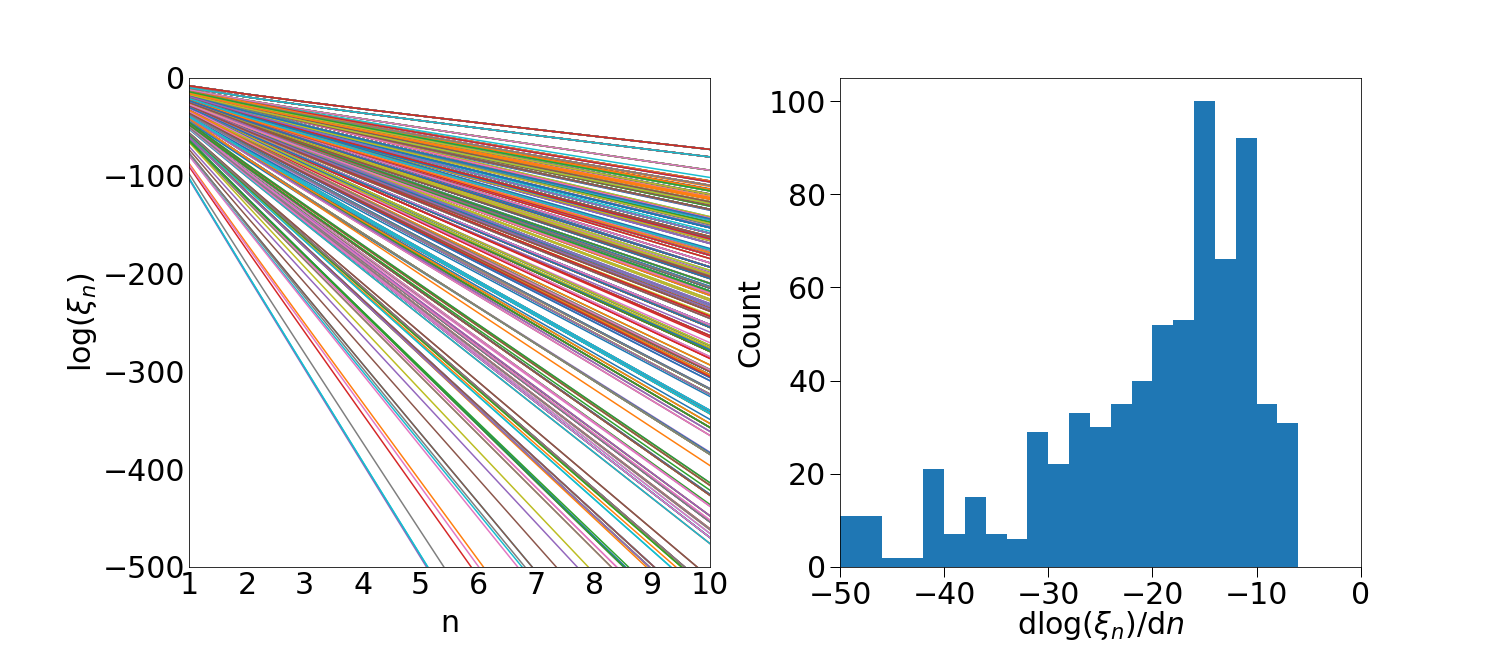}
	\caption{Convergence of worldsheet instanton sum for $(h^{2,1},h^{1,1})=(5,81)$.
Left: We plot the log-magnitude $\mathrm{log}(\xi_n)$, cf.~\eqref{eq:xi}, of the $n$-th term in the instanton series associated with a sample of 727 potent rays in $\mathcal{M}_\infty(X)$, spanning a 76-dimensional cone.  Right: a histogram of the slopes of $\mathrm{log}(\xi_n)$ with respect to $n$ for the set of potent rays. It is apparent that the sum converges.}
	\label{fig:convergence_5-81-3213}
\end{figure}

The vertices of $\Delta$ are the columns of
\begin{equation}\label{eq:Delta_vertices_5-81}
\begin{pmatrix}
1  & -2 & -2 & -2 & -2 & 0 & 0 & 0 & 0 \\
0 & -1  & -1 & 0  & 0  & 0 & 0 & 1 & 1 \\
0 & -1  & 0  & -1 & 0  & 0 & 1 & 0 & 1\\
0 & 1   & 0  & 0  & -1 & 1 & 0 & 0 & -1
\end{pmatrix}\, .
\end{equation}
In this example there are $h^{1,1}+3=84$ rigid prime toric divisors $D_I\subset X$ with $h^{2,1}(\widehat{D}_I)=0$.
The D3-brane tadpole is $44$, and a suitable flux choice is
\begin{equation}
\mathbf{M}=\begin{pmatrix}
3 & -5 & 2 & -2 & -5
\end{pmatrix}^T\, ,\quad \mathbf{K}=\begin{pmatrix}
-5 & 5 & -4 & -1 & 5
\end{pmatrix}^T\, ,
\end{equation}
leading to a perturbatively flat vacuum where
\begin{equation}
\mathbf{z}=\mathbf{p}\,\tau\, ,\quad \mathbf{p}=\begin{pmatrix}
\frac{13}{8} & \frac{59}{24} & \frac{5}{4} & \frac{5}{4} &  \frac{5}{12}
\end{pmatrix}\, .
\end{equation}
The D3-brane charge in fluxes is $-\frac{1}{2}\mathbf{M}\cdot \mathbf{K}=\frac{71}{2}$, so there are eight mobile D3-branes and a single `half' D3-brane.
The leading instantons along the perturbatively flat valley have charges given by the columns of
\begin{equation}
\begin{pmatrix}
1 & 0 & 0 & 1\\
0 & 0 & 0 & -1\\
-1 & 1 & 0 & 0\\
0 & -1 & 0 & 1\\
0 & 1 & 1 & 0
\end{pmatrix}\, ,
\end{equation}
with
GV invariants
\begin{equation}
\mathscr{N}_{\tilde{\mathbf{q}}}=\begin{pmatrix}
2 & 2 & 2 & 56
\end{pmatrix}\, .
\end{equation}
The remaining flux superpotential is
\begin{equation}
W_{\text{flux}}(\tau)=-\zeta \left(2\,e^{2\pi i \tau \cdot \frac{9}{24}}+324\,e^{2\pi i \tau \cdot \frac{10}{24}}\right)+\mathcal{O}\left(e^{2\pi i \tau \cdot \frac{5}{6}}\right) \, ,
\end{equation}
which stabilizes the dilaton with vev
\begin{equation}
\langle\tau \rangle \approx i \frac{24\log(-180)}{2\pi } \approx 19.84i-12 \, ,
\end{equation}
and the resulting vev of the flux superpotential is
\begin{equation}
W_0\approx  36\times \zeta \times 180^{-10}   \approx 2.04\times 10^{-23}\, .
\end{equation}
We find a solution to the F-flatness conditions for the K\"ahler moduli with $\mathcal{V}^{[0]}\approx 198.1$, and the Einstein-frame volume of $X$ is $\mathcal{V}_E \approx 1.7 \times 10^5$. Convergence of the worldsheet instanton expansion at this point in K\"ahler moduli space is shown in Figure \ref{fig:convergence_5-81-3213}, and instanton corrections shift $\mathcal{V}^{[0]}$ by an amount $\delta \mathcal{V}^{[0]}\approx -0.2$.
The control parameters defined in \eqref{eq:warping_controlI} and \eqref{eq:warping_controlII} are
\begin{equation} 
g_{\mathcal{N}=1}^X\approx  0.0065\,, \quad \max_i\left( g_{\mathcal{N}=1}^{\omega_i}\right)\approx 0.0071\,.
\end{equation}

Thus we have a controlled supersymmetric AdS$_4$ vacuum with vacuum energy
\begin{equation}
V_0=-3e^{\mathcal{K}}|W|^2\approx    -8.6 \times 10^{-63} M_{\mathrm{pl}}^4\, .
\end{equation}

\subsection{Vacuum with $(h^{2,1},h^{1,1})=(4,214)$}\label{sec:ex:4:214}
The vertices of $\Delta$ are given by
\begin{equation}
\begin{pmatrix}
-1 & 1 & -1 & -1 & -1 & -1 & -1\\
2  & -1& -1 & -1 & -1 & -1 & -1\\
-1 & 0 & 1  &  1 & 2  & 2  &  3\\
-1 & 0 &  1 & 2  & 1  &  3 &  2
\end{pmatrix}
\end{equation}
There are $h^{1,1}+2=216$ rigid prime toric divisors $D_I\subset X$, all of which have $h^{2,1}(\widehat{D}_I)=0$.
The D3-brane tadpole is 110.  We choose fluxes
\begin{equation}
\mathbf{M} = \begin{pmatrix}
10&11&-11&-5
\end{pmatrix}^T\, ,\quad \mathbf{K} = \begin{pmatrix}
-3&-5&8&6
\end{pmatrix}^T\, ,
\end{equation}
such that the corresponding perturbatively flat vacuum satisfies
\begin{equation}
\mathbf{z}=\mathbf{p}\tau\, ,\quad \mathbf{p}=\begin{pmatrix}
\frac{293}{110}& \frac{163}{110} & \frac{163}{110}, \frac{13}{22}
\end{pmatrix}\, .
\end{equation}
The D3-brane charge in fluxes is $-\frac{1}{2}\mathbf{M}\cdot \mathbf{K}=\frac{203}{2}$, so there are eight mobile D3-branes and a single `half' D3-brane.
The leading instantons along the perturbatively flat valley have charges given by the columns of
\begin{equation}
\begin{pmatrix}
1 & 0 & -1 & 0\\
-1 & 0 & 1 & 1\\
-1 & 1 & 1 & 0\\
1 & -2 & 0 & -2
\end{pmatrix}
\end{equation}
and the
GV invariants are
\begin{equation}
\mathscr{N}_{\tilde{\mathbf{q}}}=\begin{pmatrix}
1 &-2& & 252 & -2
\end{pmatrix}\, .
\end{equation}
The remaining flux superpotential is
\begin{equation}
W_{\text{flux}}(\tau)=5\,\zeta \left(-\,e^{2\pi i \tau \cdot \frac{32}{110}}+512\,e^{2\pi i \tau \cdot \frac{33}{110}}\right)+\mathcal{O}\left(e^{2\pi i \tau \cdot \frac{13}{22}}\right) \, ,
\end{equation}
which stabilizes the dilaton with vev
\begin{equation}
g_s \approx \frac{2\pi }{110\log(528)} \approx 0.009\, ,
\end{equation}
and the resulting vev of the flux superpotential is
\begin{equation}
W_0\approx  80\times \zeta \times 528^{-33}   \approx 2.3\times 10^{-90}\, .
\end{equation}
We find a solution to the F-flatness conditions for the K\"ahler moduli with $\mathcal{V}^{[0]}\approx 4711$, and the Einstein-frame volume of $X$ is $\mathcal{V}_E \approx 5.4 \times 10^6$.

\begin{figure}
	\centering
	\includegraphics[keepaspectratio,width=17cm]{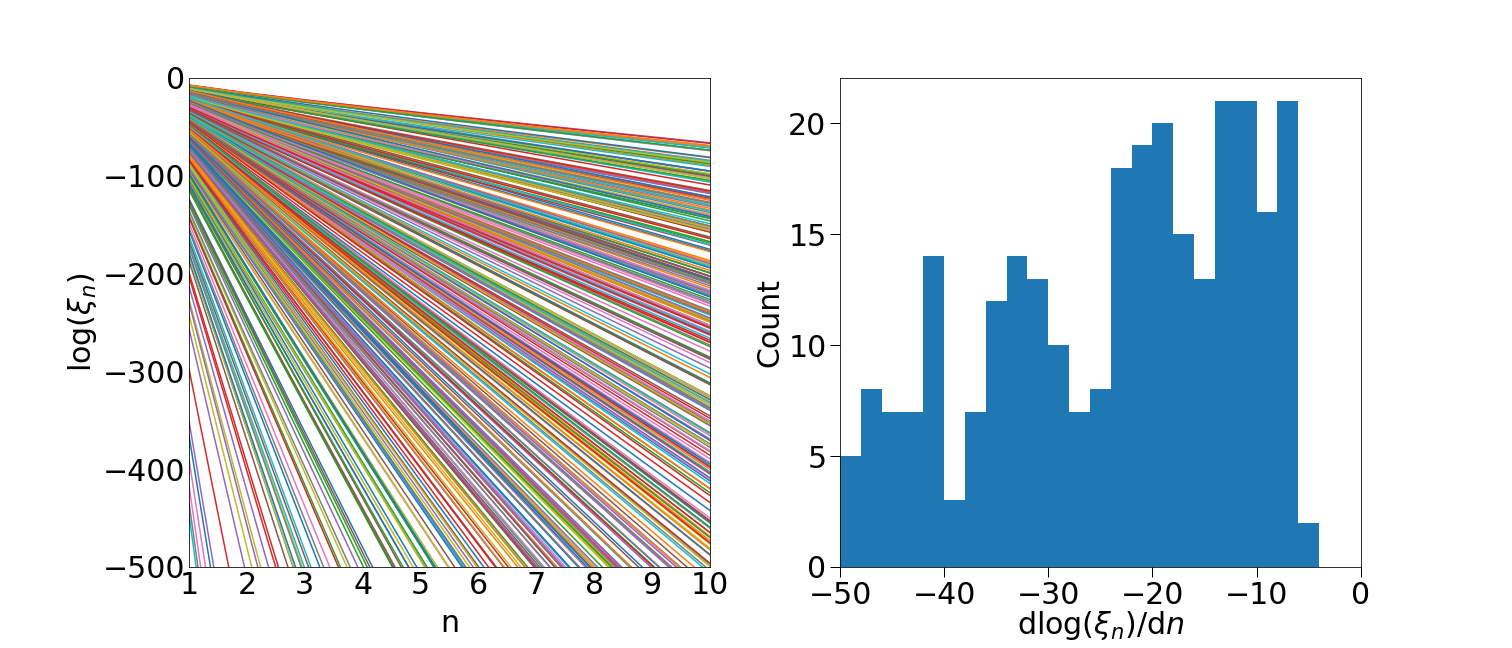}
\caption{Convergence of worldsheet instanton sum for $(h^{2,1},h^{1,1})=(4,214)$.
Left: We plot the log-magnitude $\mathrm{log}(\xi_n)$, cf.~\eqref{eq:xi}, of the $n$-th term in the instanton series associated with a sample of 411 potent rays in $\mathcal{M}_\infty(X)$, spanning a 118-dimensional cone.  Right: a histogram of the slopes of $\mathrm{log}(\xi_n)$ with respect to $n$ for the set of potent rays. It is apparent that the sum converges.}	
	\label{fig:convergence_4-214-647}
\end{figure}

Convergence of the worldsheet instanton expansion at this point in K\"ahler moduli space is shown in Figure \ref{fig:convergence_4-214-647}, and the instanton corrections shift $\mathcal{V}^{[0]}$ by $\delta \mathcal{V}^{[0]}\approx -0.4$.
The control parameters defined in \eqref{eq:warping_controlI} and \eqref{eq:warping_controlII} are
\begin{equation} 
g_{\mathcal{N}=1}^X\approx  0.0036\,, \quad \max_i\left( g_{\mathcal{N}=1}^{\omega_i}\right)\approx 0.0022\,.
\end{equation}

Finally, the supersymmetric AdS$_4$ vacuum has vacuum energy
\begin{equation}
V_0=-3e^{\mathcal{K}}|W|^2 \approx -5.5\times 10^{-203} M_{\mathrm{pl}}^4\, .
\end{equation}

\section{Discussion}\label{sec:discussion}

The vacua that we have constructed are novel incarnations of the ideas of Kachru, Kallosh, Linde, and Trivedi, with one important distinction: the mechanism of \cite{Demirtas:2019sip}\footnote{For related earlier work, see \cite{Giryavets:2003vd,Denef:2004dm}.} for producing an exponentially small flux superpotential leaves an imprint in the pattern of moduli expectation values.

Recall that we began by finding quantized three-form fluxes $\vec{f},\vec{h}$ for which there exists an exactly flat direction in the joint axiodilaton and complex structure moduli space of a Calabi-Yau threefold $X$, at the level of the \emph{perturbative} prepotential for these moduli.  We termed such a configuration a \emph{perturbatively flat vacuum}.  The true prepotential includes nonperturbative corrections that can be understood as worldsheet instantons of type IIA string theory on the mirror threefold $\widetilde{X}$.  For the type IIB theory these are, of course, classical effects, and they affect the classical Gukov-Vafa-Witten flux superpotential via \eqref{eq:finst}.  Taking fluxes $\vec{f},\vec{h}$ that yield a perturbatively flat vacuum and evaluating the true flux superpotential along the flat direction and near large complex structure, the result is then exponentially small.  Typically such a configuration is a runaway, but for suitably restricted $\vec{f},\vec{h}$ the worldsheet instanton terms form a racetrack that stabilizes the moduli along the flat direction.

One feature of this mechanism is that the dilaton is stabilized near weak coupling and the complex structure moduli are stabilized near large complex structure.  In particular, $g_s \propto 1/\mathrm{log}(W_0^{-1})$.  Because the F-flatness conditions for the K\"ahler moduli stabilize the divisors at Einstein-frame volumes $\mathrm{Re}(T_i) \propto \mathrm{log}(W_0^{-1})$, we find a solution in which the \emph{string-frame} volumes of divisors and curves are not parametrically large or small, even though their Einstein-frame volumes are large.  As we carefully explained in \S\ref{sec:control}, control of the $\alpha'$ expansion  then depends on whether the smallest effective curves in $X$ happen to be large enough for the worldsheet instanton series to converge.
Specifically, these worldsheet instanton contributions to the K\"ahler potential are automatically accounted for by the classical K\"ahler potential of the mirror O6 orientifold in type IIA, and can thus be computed accurately by computing the periods of the mirror threefold.\footnote{Note that we are studying type IIB worldsheet instantons on $X$ and, separately, type IIA worldsheet instantons on $\widetilde{X}$, the former as corrections to the K\"ahler potential for the K\"ahler moduli, and the latter as corrections to the flux superpotential, via the prepotential for the complex structure moduli, in the type IIB compactification of interest.} Because the radius of convergence can be inferred from the asymptotic growth of genus-zero Gopakumar-Vafa invariants, we were able to establish control in examples by computing these invariants.

Another feature is that one complex direction in the axiodilaton and complex structure moduli space --- the perturbatively flat direction --- receives a mass of order $W_0$, which is also the mass scale of the K\"ahler moduli.

Neither of these features is required by the KKLT mechanism per se, nor were they foreseen for other reasons, but they are characteristic of our class of constructions.

The statistics of the cosmological constants in our vacua deserve some comment.  We have found solutions with vacuum energy of magnitude $10^{-200}$ in Planck units, without a search of commensurate cost.\footnote{For discussions of the complexity of related problems, see e.g.~\cite{Halverson:2018cio}, as well as the analysis in \cite{Bao:2017thx} of the simpler landscape of \cite{Arkani-Hamed:2005zuc}.}  The \emph{methods} that we have developed to construct orientifolds, identify rigid divisors, find F-flat solutions, and compute Gopakumar-Vafa invariants, all at large $h^{1,1}$, are fairly novel, and we believe they could be of use in the future.  However, these are pieces of technology for studying compactifications in general, and not specifically for finding vacua with small vacuum energy.  Indeed, it is almost incidental that $h^{1,1} \gg 1$ in our examples: except for considerations of the D3-brane charge tadpole, increasing $h^{1,1}$ has no evident benefit in our constructions.

But the core problem in searching for small vacuum energy in a landscape of flux vacua is (expected to be) that of choosing the right fluxes.  One might naively anticipate that to find a flux superpotential of order $10^{-100}$, one will have to search in a very high-dimensional lattice, say of dimension $\sim \mathcal{O}(100)$, and explore a vast number of choices.  In our work this is not the case.  We have $h^{2,1}=4$ in an example with $W_0 \approx 10^{-90}$, so the lattice is eight-dimensional, and the search for flux vectors takes just minutes on a laptop.

An underlying reason is that by finding fluxes that allow for perturbatively flat vacua, we have arranged that the sum over all perturbative --- and hence, possibly large --- contributions to the superpotential actually \emph{vanishes}, and what remains is suppressed by exponentials in the mirror worldsheet instanton expansion around large complex structure.
Thus, our construction includes a fine-tuned and \textit{exact} cancellation of a vast array of order-unity perturbative contributions to the superpotential.
The possibility of such an exact cancellation hinges on the quantization of parameters in string theory: the superpotential, in particular, is determined by essentially integer data.
Because of this cancellation, everything appearing in the final expression for the vacuum energy is proportional to a nonperturbative effect, either a Euclidean D3-brane or strong gauge dynamics on a four-cycle in $X$, or a worldsheet instanton of type IIA wrapping a curve in $\widetilde{X}$.

In this sense, our construction of exponentially small flux superpotentials, and exponentially small vacuum energy, is \emph{natural}, in the sense of dimensional transmutation.\footnote{Of course, solving the cosmological constant problem would require exponentially small vacuum energy after supersymmetry breaking, which we have certainly not achieved!}  One might wonder if a similar mechanism is at work in our universe, perfectly cancelling perturbative contributions to the vacuum energy and lifting it to the observed value in a nonperturbative fashion.

There is of course some tuning of discrete data in our solutions: we had to choose $X$ with suitable patterns of Gopakumar-Vafa invariants in order to support a racetrack of worldsheet instantons, and find fluxes allowing compatible perturbatively flat vacua.  But a polynomial degree of tuning of such integers leads to exponential hierarchies in the vacuum energy: for example, in a threefold with $(h^{2,1},h^{1,1})=(5,113)$ we found
\begin{equation}\label{theratio}
W_0\propto  \left(\frac{2}{252}\right)^{29} \approx 10^{-61}\,.
\end{equation}  In this example the numbers $2$ and $252$ arise as the Gopakumar-Vafa invariants of the two leading curves, while the exponent $29$ results from the $7/28:7/29$ racetrack \eqref{5113wf} of worldsheet instantons on these curves, which is a consequence of the flux choice \eqref{5113flux}.\footnote{In the other four examples we presented, the racetracks took the form $34/280:35/280$, $8/30:9/30$, $9/24:10/24$, and $32/110:33/110$.}

Our results suggest a new perspective on the abundance of vacua with small $W_0$ in the type IIB flux landscape.
The classic statistical treatment of \cite{Denef:2004ze} relied on approximating the fluxes as continuous.  In this approximation, applied to a model where $\mathcal{N}$ is the total number of flux vacua, the smallest value of $W_0$ that one expects to find is of order $1/\sqrt{\mathcal{N}}$ \cite{Denef:2004ze}.  However, we have exhibited solutions in which $W_0$ is hierarchically smaller than this prediction.  The resolution of this mismatch is that our solutions critically rely on the values of flux \emph{integers}: the conditions for a perturbatively flat vacuum, which are equations over the integers, are fulfilled in a set of measure zero within the space of continuous fluxes.
Thus, our vacua are not captured in the statistics of \cite{Denef:2004ze}.  A systematic treatment of the statistics of small $W_0$ is an interesting task for the future.

The alert reader will have recognized that the solutions presented here are completely unrealistic: the cosmological constant is negative and $\mathcal{N}=1$ supersymmetry is preserved.
While it is possible that solutions in the class given here could be uplifted to de Sitter vacua, in order to exhibit maximal parametric control and maximal scale-separation we have focused on examples in which the magnitude of the superpotential is extremely small.  Thus, the gravitino is far too light, as are the K\"ahler moduli and the previously-perturbatively-flat complex structure direction.  Even if the cosmological constant were small and positive after uplifting, the degree of supersymmetry breaking would be unrealistically small, and moreover the moduli problem would almost surely be fatal for cosmology.

Nevertheless, we view these solutions as stepping stones to realistic vacua.  In this work we have restricted our attention to configurations in which we could establish control of corrections in the $\alpha'$ expansion with our \emph{present} knowledge of these corrections, and with our present ability to compute Gopakumar-Vafa invariants at large $h^{1,1}$.  With improved capabilities one could doubtless explore a much wider parameter space, including examples in which $W_0 \ll 1$ but, say, $W_0 \gtrsim 10^{-10}$.  We have found hundreds of examples of this form, but sifting out those among them that are best-controlled is a task for the future.  Uplifts of such vacua could in principle allow for realistic cosmology and particle physics.

At the same time, our solutions are instructive in their own right, because they present a slightly different perspective on the cosmological constant problem in string theory than one finds following \cite{Bousso:2000xa,Ashok:2003gk,Denef:2004ze}.

\section{Conclusions}\label{sec:conclusions}

We have demonstrated that supersymmetric $\mathrm{AdS}_4$ vacua
with exponentially small vacuum energy
can be constructed in large numbers in orientifolds of Calabi-Yau hypersurfaces in toric varieties.

The geometry, orientifolding, quantized fluxes, and D-brane configurations in our constructions are all totally explicit.
We enumerated nonperturbative superpotential terms that suffice to stabilize all the K\"ahler moduli,
and we ensured that all Pfaffian prefactors $\mathcal{A}_D$ of Euclidean D3-brane superpotential terms are constants, with no dependence on the moduli.  Lacking a theory of the Pfaffians, we were not able to compute these numbers, but we nevertheless established that well-controlled vacua exist for a wide range of values of the $\mathcal{A}_D$.

Our analysis relied on novel techniques that we have developed for constructing orientifold configurations and computing Gopakumar-Vafa invariants at large $h^{1,1}$, as well as for finding F-theory uplifts and computing the Hodge numbers of divisors therein.  We hope to present more details of these methods in the near future \cite{orientifoldKS,compmirror,trilayer,hodge}.

There are several directions for future work.  Computing the Pfaffian numbers $\mathcal{A}_D$ would be valuable.
It would be interesting to extend our construction beyond hypersurfaces in toric varieties, and to develop dual descriptions of similar vacua, in compactifications of F-theory, M-theory, or type IIA string theory.  Exploring constraints on the conformal field theories dual to our solutions would also be worthwhile.
Perhaps the most pressing question is whether some of our solutions can be uplifted to de Sitter vacua of string theory.

\section*{Acknowledgements}
We are grateful to Xin Gao, Yuval Grossman, Tom Hartman, Arthur Hebecker and Daniel Junghans for useful discussions, and to Daniel Harlow for encouraging comments.
We thank
Naomi Gendler, Ben Heidenreich, Tom Rudelius, and Mike Stillman for collaborations on related topics.
The work of M.D., M.K., L.M., and A.R.-T.~was supported in part by NSF grant PHY-1719877, and the work of L.M.~and J.M.~was supported in part by the Simons Foundation Origins of the Universe Initiative.   This research was conducted with support from the Cornell University Center for Advanced Computing.

\appendix

\section{Comments on de Sitter vacua}\label{app:ds}

The supersymmetric AdS$_4$ solutions  that we have constructed clearly serve as stepping stones towards de Sitter vacua in type IIB compactifications.
However, in the body of the paper we have not confronted the question of an uplift to de Sitter space: our intention was to first achieve optimal control in supersymmetric solutions.

In this Appendix we briefly describe an observation that may be relevant
for the question of control over the backreaction from seven-branes, as discussed in the recent literature \cite{Carta:2019rhx,Gao:2020xqh,Carta:2021lqg}. The potential problem observed in \cite{Carta:2019rhx} is that an anti-D3-brane uplift requires a tuning of a throat hierarchy \cite{Klebanov:2000hb,Giddings:2001yu}
\begin{equation}
a_0^4\sim \mathrm{exp}\Biggl({-\frac{8\pi}{3}\frac{N_{\text{D3}}^{\text{throat}}}{R^4_{\text{throat}}}}\Biggr)\,,
\end{equation}
where $N_{\text{D3}}^{\text{throat}}$ is the D3-brane charge hosted in the throat, $R_{\text{throat}}$ is the Einstein-frame curvature radius at the bottom of the throat, and $a_0$ is the warp factor at the tip of the throat.
For supergravity control of the infrared region of the throat one needs $R^4_{\text{throat}}\gtrsim 1$ \cite{Kachru:2002gs}. For the uplift to compete with the F-term potential of the supersymmetric AdS$_4$ vacuum one further needs
\begin{equation}
a_0^4\sim |W_0|^2\, \quad \Rightarrow \quad \frac{N_{\text{D3}}^{\text{throat}}}{R_{\text{throat}}^4}\approx \frac{3}{2}\frac{\log(W_0^{-1})}{2\pi}\approx \frac{3}{2}\frac{\text{Re}(T_i)}{c_i}\, ,
\end{equation}
and one would thus require $N_{\text{D3}}^{\text{throat}}\gtrsim \frac{\text{Re}(T_i)}{c_i} $. Moreover, if the dual Coxeter numbers $c_i$ are not very large\footnote{Recall that $c_i \in \{1,6\}$ in our examples, and $6$ counts as not very large for present purposes.} one might expect that the overall volume $\mathcal{V}_E$ of the threefold $X$ is stabilized at
\begin{equation}\label{eq:volguess}
\mathcal{V}_E \stackrel{?}{\sim} \bigl(\text{Re}(T)_i\bigr)^{\frac{3}{2}}\,.
\end{equation}
Now \emph{if} \eqref{eq:volguess} holds, one is forced into the regime
\begin{equation}\label{eq:KKLTregime?}
N_{\text{D3}}^{\text{throat}}\gtrsim \mathcal{V}_E^{\frac{2}{3}}\, .
\end{equation}
However, as $\left(N_{\text{D3}}^{\text{throat}}\right)^{\frac{1}{4}}$ also sets the transverse size of the throat,  it would follow that one cannot consistently glue in the warped throat into a weakly-warped larger bulk threefold $X$. Attempting to shrink $\mathcal{V}_E$ to the required small size then causes warp factor singularities, which are otherwise localized exponentially close to the seven-branes, to move into the bulk \cite{Carta:2019rhx,Gao:2020xqh}. These singularities were discussed further in \cite{Carta:2021lqg}, where it was shown that nonperturbative effects in the $\alpha'$ expansion can resolve the singularities, leaving behind a strongly-curved but non-singular region in $X$.  Although it then follows that the bulk physics is regular, computing the K\"ahler potential in such a regime is a formidable task.

The results of \cite{Carta:2021lqg} lead to a slight puzzle: at least in the simple example where the seven-brane singularity emerges from a D7-brane stack wrapped on K3, one finds that the transverse distance between the classically singular locus and the position of the seven-branes is of order
\begin{equation}
r_0\sim R_{\text{CY}}\,\mathrm{exp}\Biggl(-\frac{2\pi \text{Vol}(K3)}{|N_{\text{D3,K3}}|}\Biggr)\, ,
\end{equation}
where $N_{\text{D3,K3}}$ is the D3-brane charge hosted on the seven-brane stack and $R_{\text{CY}}$ is the radius of $X$.
This result immediately generalizes to other seven-brane configurations. Thus, for a more general collection of seven-branes one would expect that singularities are \textit{exponentially} controlled if all Einstein frame divisor volumes are larger than the D3-brane charges hosted on those divisors. In fact, this is precisely the constraint we have imposed in \eqref{eq:warping_controlI}. However, more singular outcomes are possible in some cases,
as a large number of sources of small amounts of D3-brane charge can source a macroscopic singularity via their collective charge seen at long distances, as discussed for the case of a gas of O3-planes in \cite{Gao:2020xqh}, but such singularities are generally evaded if the overall volume satisfies
\begin{equation}\label{eq:sugra_control}
\mathcal{V}_E^{\frac{2}{3}}\gg N_{\text{D3}}^{\text{total}}>N_{\text{D3}}^{\text{throat}}\, ,
\end{equation}
which is precisely our constraint \eqref{eq:warping_controlII}.
Now we have come full circle and recovered again the tension between supergravity control \eqref{eq:sugra_control} and the KKLT regime \eqref{eq:KKLTregime?}.

Up to this point we have been reviewing the recent literature, but let us add a new observation.
In the models that we have constructed, \emph{the volume $\mathcal{V}_E$ is much larger than predicted by} \eqref{eq:volguess}.
For example, in the vacuum detailed in \S\ref{sec:main_examples} we found
$\text{Re}(T)_i \approx 22 $, so the naive guess \eqref{eq:volguess} would predict $\mathcal{V}_E \approx 103$,
whereas we find $\mathcal{V}_E \approx 8.1 \times 10^5$.
In this case the relation \eqref{eq:volguess} underestimates  $\mathcal{V}_E$ by a factor of $\sim 8000$.

This finding has nothing to do with the smallness of $W_0$, but is simply a generic, purely geometric property of Calabi-Yau threefolds at large $h^{1,1}$: when a full-dimensional collection of effective divisors have volumes of order unity, the overall volume can become quite large \cite{Demirtas:2018akl}.

To be concrete, in our example of \S\ref{sec:ex:4:214} one can imagine that the entire D3-brane charge $N_{\text{D3}}=110$ allowed by the D3-brane tadpole contributes to the formation of a warped throat, that a perturbatively flat vacuum arises with $c_{\tau}= \frac{216}{110}$, and that the $F_3$ flux on the conifold $S^3$ is set to its critical value $M=12$, thus marginally ensuring stability of the anti-D3-brane \cite{Kachru:2002gs}. With the above parameters we have secretly guaranteed the KKLT fine-tuning
\begin{equation}
a_0^4\sim \mathrm{exp}\Biggl(-\frac{8\pi}{3}\frac{N_{\text{D3}}}{g_sM^2}\Biggr)= W_0^{c_\tau \frac{2}{3}\frac{110}{144}}= W_0^2\,,
\end{equation}
independent of the actual value of $W_0$. In order to ensure that $R^4_{\text{throat}}>1$ we must have $g_s>\frac{1}{144}$, so the smallest allowed value for $W_0$ would be\footnote{Note that a more stringent constraint $g_sM^2\gtrsim 50$ has been claimed \cite{Bena:2018fqc}, though it is not entirely clear to us that the effective field theory employed there is reliable in the relevant regime. Even so, using their constraint one still finds $W_0^{\text{min}}\approx 10^{-4}$, which is quite small nevertheless. We note further that the radius of curvature  in string units
is $R^2\sim g_s M$, which is less than unity for $M=12$ and $g_s=1/144$, but in this case the physics is controlled by the Klebanov-Strassler gauge theory instead \cite{Klebanov:2000hb}, and we see no reason why metastable supersymmetry breaking should disappear in this regime.}
\begin{equation}\label{eq:minW0}
W_0^{\text{min}}=\mathrm{exp}\Bigl(-\frac{2\pi}{c_\tau g_s} \Bigr) \approx 4\times 10^{-198}\, .
\end{equation}
At a solution to the F-term equations we obtain
\begin{equation}
\mathcal{V}_E\approx 4711\times \left(\frac{\log(1/W_0)}{2\pi}\right)^{\frac{3}{2}}< 4711\times \left(\frac{\log(1/W_0^\text{min})}{2\pi}\right)^{\frac{3}{2}}\, ,
\end{equation}
where the large prefactor $4711$ is a concrete manifestation of the fact that setting volumes of low-dimensional cycles to moderate values can result in very large volumes of higher-dimensional cycles when $h^{1,1}$ is large \cite{Demirtas:2018akl}.

Thus, instead of \eqref{eq:KKLTregime?} one then finds
\begin{equation}\label{eq:largevolume}
\mathcal{V}_E^{\frac{2}{3}}\approx 281 \times \frac{\log(1/W_0)}{2\pi} \gg N_{\text{D3}}=110\, ,
\end{equation}
even for rather  modest values of $W_0$, and the warping control criteria of \eqref{eq:warping_controlI} and \eqref{eq:warping_controlII} are comfortably satisfied. We stress again that this has nothing to do with
the smallness of the
flux superpotential --- even for the smallest allowed superpotential of \eqref{eq:minW0}, the ratio $\frac{\log(1/W_0)}{2\pi N_{D3}}$
only reaches the modest value $0.67$ --- but instead results  from the  scaling of cycle volumes with $h^{1,1}$. While many curves in turn have small string frame volumes at this point in moduli space, we have argued that, in this instance, the K\"ahler potential receives no dangerous corrections from such curves: indeed, we showed explicitly that a large set of infinite towers of worldsheet instantons yield negligible corrections.

In this work we have not actually constructed a warped throat in such an example,\footnote{With the above parameters such a throat would have flux quanta $(M,K)\sim (12,9)$, which does not seem unreasonable at all.} but  the above parameter values do not appear out of reach. We conclude that there is no evident obstacle to
circumventing
the problems pointed out in \cite{Carta:2021lqg,Gao:2020xqh}, by finding an appropriate Calabi-Yau compactification that unifies the above scaling of \eqref{eq:largevolume} with an appropriately warped throat, even without realizing the contrived O3-plane configurations suggested in \cite{Gao:2020xqh}. The tools to engineer such models have been developed in \cite{Alvarez-Garcia:2020pxd,Demirtas:2020ffz},
but using them to build a controlled KKLT  de Sitter vacuum is a task for the future.

\bibliography{refs}
\bibliographystyle{JHEP}
\end{document}